\documentclass[a4paper,fleqn,1p]{cas-sc}
\usepackage[authoryear,longnamesfirst,sort&compress]{natbib}

\usepackage{graphicx} % Required for inserting images
\usepackage{amsthm}
\usepackage{bbm} % Extended set of blackboard-bold symbols
\usepackage{mathtools}
\usepackage{physics}
\usepackage{xcolor}
\usepackage{enumerate}
\usepackage[uncertainty-mode=separate]{siunitx}
\usepackage[normalem]{ulem} %to strikethrough text using \sout{....}
\graphicspath{{Graphics/}}

\def\tsc#1{\csdef{#1}{\textsc{\lowercase{#1}}\xspace}}
\tsc{WGM}
\tsc{QE}
\newcommand{\bs}[1]{{\mathbf #1}}
\newcommand{\mc}[1]{\mathcal{#1}}
\newcommand{\cvec}[1]{\vectorbold{#1}}
\newcommand{\nvec}[1]{{\texttt{#1}}}

\newcommand{\R}{\mathbb{R}}

\renewcommand{\r}{\cvec{r}}
\renewcommand{\k}{\cvec{k}}
\newcommand{\y}{\nvec{y}}
\newcommand{\s}{\cvec{s}}
\newcommand{\id}{\mathbbm{1}}
\newcommand{\dt}{\dd{t}}
\newcommand{\disc}[1]{\mc{D}_{\bs{#1}}}

\newcommand{\cll}{{\mc{I}_{\cvec{0},-\r}}}
\newcommand{\clg}{{\mc{C}_{\cvec{0},-\r}}}
\newcommand{\cgl}{{\mc{C}_{-\r,\cvec{0}}}}
\newcommand{\cgg}{{\bar\Pi_{\cvec{0},-\r}}}

\newcommand{\hll}{{\mc{I}_{\r,\cvec{0}}}}
\newcommand{\hlg}{{\mc{C}_{\r,\cvec{0}}}}
\newcommand{\hgl}{{\mc{C}_{\cvec{0},\r}}}
\newcommand{\hgg}{{\bar\Pi_{\r,\cvec{0}}}}

\newcommand{\rhotch}{\rho (\nvec{T} \times_3 \nvec{h})\times_2 \nvec{c}}
\newcommand{\chired}{\chi_{\text{red}}}
\newcommand{\ichired}{\chi^{-1}_\text{red}}
\newcommand{\nod}{N_{0d}}
\newcommand{\ndl}{N_{d\infty}}
\newcommand{\ndisc}{N_\text{disk}}
\newcommand{\ldinf}{l_{d\infty}}

\newcommand{\critpointthree}{
	(\num{0.31691(1)}, \num{1.11436(1)})
	}
\newcommand{\critpointtwo}{
	(\num{0.45244(3)}, \num{5.3263588(2)})
}
\newcommand{\etathree}{-0.006(10)}
\newcommand{\gammathree}{1.989(15)}
\newcommand{\deltathree}{5.04(6)}
\newcommand{\betathree}{0.493(8)}
\newcommand{\etatwo}{0.36(13)}
\newcommand{\gammatwo}{1.0000(12)}
\newcommand{\deltatwo}{10(6)}

\newcommand{\Int}[1]{{\texttt{Int}_{#1}}}
\newcommand{\Interp}[1]{{\texttt{Interp}_{#1}}}

\usepackage{booktabs}
\newcommand{\tabfigure}[2]{\raisebox{#2}{\includegraphics[scale = 0.15]{#1}}}

\date{\today}

\begin{document}	

%\printinunitsof{in}\prntlen{\textwidth}
%\expandafter\string\the\font

\let\WriteBookmarks\relax
\def\floatpagepagefraction{1}
\def\textpagefraction{.001}

\shorttitle{}    

\shortauthors{}  

\title [mode = title]{Pseudospectral Methods and Critical Phenomena}  



\author[edi]{J.C. Skelton}[orcid=0009-0009-9693-9699]
\cormark[1]
\ead{jake.skelton@ed.ac.uk}

\author[fri]{J. M. Brader}[orcid=0000-0002-4101-6522]
%\ead{}
	
\author[fri]{S. M. Tschopp}[orcid=0000-0003-3259-052X]
	
\author[edi]{B. D. Goddard}[orcid=0000-0002-8781-014X]

\affiliation[edi]{organization={The School of Mathematics and Maxwell Institute for Mathematical Sciences, University of Edinburgh},
	%	addressline={Peter Guthrie Tait Road}, 
	city={Edinburgh},
	%          citysep={}, % Uncomment if no comma needed between city and postcode
	%	postcode={EH9 3FD}, 
	%	state={},
	country={UK}}
\affiliation[fri]{organization={Department of Physics, University of Fribourg},
	%	addressline={}, 
	city={Fribourg},
	%          citysep={}, % Uncomment if no comma needed between city and postcode
	postcode={CH-1700}, 
	%	state={},
	country={Switzerland}}

\cortext[1]{Corresponding author}



\begin{abstract}
We employ pseudospectral methods to solve the homogeneous Ornstein-Zernike (OZ) equation for a model fluid in the vicinity of the critical point. Focusing on the Mean-Spherical  Approximation (MSA) as a closure to the OZ equation, we obtain numerical estimates for the critical exponents $\eta$, $\delta$ and $\gamma$ for a system of hard-core Yukawa particles both in two and three dimensions. The three-dimensional MSA exponents are already well-known from an analytic solution and are recovered by our numerical methods. The two-dimensional exponents are a new output of this work.  The pseudospectral method allows for rapid and highly accurate solution of liquid-state integral equation theories, and enables calculations on truely infinite domains, as needed for highly correlated states. In addition, we analyse the standard Picard iteration scheme and propose a variation of it which provides increased stability and speed of convergence.
\end{abstract}


% \begin{highlights}
% 	\item 
% 	\item 
% 	\item 
% \end{highlights}

% Keywords

\maketitle

\section{Introduction}

Two common numerical problems that arise from physical applications are the solution of partial differential equations (PDEs) and of integral equations.  Sometimes, the solution of both types of equations are required in the same application.  A concrete example, which partially motivates the approach taken in this work, comes from the theoretical study of inhomogeneous fluids.  The dynamics of such systems can be described by Dynamical Density Functional Theory (DDFT) (see, e.g., the review~\cite{teVrugtReview}) and its generalisations such as Superadiabatic Dynamical Density Functional Theory (SDDFT)~\cite{TschoppSuperadiabatic,TschoppSuperadiabatic2,TschoppSuperadiabatic3}.  In these applications, in order to understand the time-evolution of a system, one needs to solve a PDE in time and space.  However, these are often not in closed form and require the solution of an auxiliary integral equation.  In DDFT this can include the need to solve an equilibrium Density Functional Theory (DFT) problem for an initial condition, which can be formulated as an integral equation, or the solution of the Ornstein-Zernike (OZ) equation (as studied here) to calculate hydrodynamic interaction terms for use in DDFT~\cite{RexHI1,RexHI2,GoddardPRL,GoddardHI,GoddardHIConfined,DuranHI} or for any implementation of SDDFT.

It seems reasonable to ask whether the same, or similar, numerical approach can be used to solve both the PDEs and integral equations in a given application. Independently of this, it is instructive to investigate whether approaches that are successful for one problem may be applied to the other.  For the numerical solution of PDEs, (pseudo)spectral methods ((P)SMs) are one of the three main numerical approaches; the other two being finite difference (FDM) and finite element (FEM) methods.  Returning to our examples of DDFT and SDDFT, pseudospectral methods have been shown to be highly successful in solving the PDEs~\cite{NoldPSM, GoddardPRL, GoddardHI, GoddardHIConfined, RodenSedimentation, RodenMultishape, MillsFlow, LutskoPSM}, some of which involve very similar integral terms to those in the applications studied here.  However, the use of PSM for multidimensional integral equations, especially those with non-smooth solutions, is much less well-established~\cite{ZakyIE}.  Our contribution here aims to provide an accurate, efficient, and robust PSM approach to such problems, the applications of which are much wider than the concrete examples given above, see, e.g.,~\cite{WazwarIEBook}.

Here we focus on what we call (following~\cite{Boyd}) \emph{pseudospectral} methods (sometimes called \emph{interpolating} spectral methods), rather than \emph{non-interpolating}, or simply \emph{spectral} methods.  For clarity, the latter do not have a grid of interpolation points.  Rather, functions are expanded in a (typically polynomial) basis.  Operations such as differentiation, integration, and interpolation can then be performed via known results for the basis functions with inputs being the expansion coefficients.  In contrast, pseudospectral methods represent a function by its values at some given set of points, often referred to as \emph{collocation} or \emph{interpolation} points (which can be thought of as a Lagrange basis, rather than an orthogonal polynomial basis as used in non-interpolating methods).  The analogous operations are then defined in terms of these function values.
For many problems, the accuracy of pseudospectral methods is almost as good as that of (non-interpolating) spectral methods, but with much simpler and more computationally efficient algorithms.

To understand the basics of PSM, it is instructive to compare them to the better-known FDM and FEM approaches.  The key, overarching difference is that PSM involve \emph{global} representations of functions and operators, whereas both FDM and FEM are `local'.  

A typical FEM represents a function on a (typically large) set of (often uniformly spaced) grid points. In contrast, an FEM divides up the domain into a (typically large) number of subdomains, and represents a function on each of these subdomains by a (typically low-order) polynomial.  One can increase the resolution of an FDM by increasing the number of grid points.  For an FEM one can increase both the number of subdomains (analogous to increasing the number of grid points) and/or increase the degree of the polynomial representation on each subdomain.  Both FDM~\cite{SmithFDM} and FEM~\cite{JohnsonFEM} are well-studied, relatively easy to implement, and have been applied to a huge range of (predominantly local) problems.  Additionally, FEM is well-suited to studying problems on complicated domains, e.g., through triangulation routines.  Whilst not directly relevant to the current work, it is helpful to point out one disadvantage of such triangulations in the context of studying spherical particles that is important for future work: For inhomogeneous systems, one would often be interested in a `test particle' approach where a single particle (typically a disk or sphere) is excluded from the domain.  One then wants to compute various correlation functions (some of which are described below), where the values at the boundary of the particle can be particularly crucial.  In triangulation approaches, this boundary is not necessarily approximated to a sufficiently high degree of accuracy, and also breaks the rotational symmetry of the problem.  PSM can overcome these issues, as will be demonstrated below.

Both FDM and FEM suffer from some deficiencies when the number of gridpoints is increased.  For example, the well-known Runge phenomenon when using polynomial interpolation on a uniform grid, or the ill-conditioned and ill-behaved nature of FEM when the degree of the polynomials is increased~\cite{Trefethen,Boyd}.  Before discussing the benefits of PSM for the problem at hand, it is also worthwhile considering how interpolation, integration, and convolution behave in FDM and FEM.  Typically, interpolation and integration would be performed using a low order polynomial approximation, and so the accuracy would be expected to behave like $\dd{x}^p$, where $\dd{x}$ is a typical grid spacing and $p$ is a small integer.  Thus, in order to obtain high accuracy, it is necessary to choose a very small $\dd{x}$, which can lead to prohibitively large problems in multiple dimensions, even utilising the sparsity of the interpolation and integration operators.  For convolutions there is an additional issue in that the operators are no longer sparse; each point (or subdomain) is connected to each other point (or subdomain).  This destroys one of the key advantages of FDM and FEM.  Finally, it is worth noting that neither FDM or FEM can treat infinite domains, and so problems such as those tackled here must be studied on truncated domains, which can lead to large, unphysical effects.

In contrast to FDM and FEM, a PSM typically uses a small number of collocation points, but the matrices representing operations such as differentiation and interpolation are dense.  The key reason that this is not prohibitively computationally expensive is that, assuming the function being considered is sufficiently nice (e.g., smooth or analytic), then the error of PSM typically decreases as $\exp(-k N)$ where $N$ is the number of collocation points and $k$ is a constant.  This allows a very small number of (carefully chosen) points to be used whilst still achieving better accuracy than is typically possible using FDM or FEM.  Of course, the regularity requirements on the function mean that PSM are not universally useful.  Fortunately, the question of how to choose suitable collocation points has been well-studied and is easy to answer for many problems.

PSM are also more restricted in their application to complicated domains than, for example, FEM.  In order to tackle a problem on a domain other than a finite box, one must find a sufficiently `nice' map between (e.g.) the unit box and the domain in question.  However, for a number of `simple' domains, this problem is well-understood; this includes how to tackle infinite intervals.  We shall demonstrate here that the necessary domains to tackle the integral equations of interest are `nice enough' to admit a suitable mapping.  An additional approach, which we also implement here, is the spectral element method which, like FEM, divides the domain into a number of subdomains.  It then employs a PSM on each of these domains.  Typically, when one is solving a PDE using such a method it is necessary to consider matching conditions between the elements so that the solution has the correct global regularity (typically, for problems that apply PSM the requirement is that the solution is smooth).  We note that such a matching procedure is not necessary in our case, as there is no direct interaction between the elements, and the expected global solution is not even continuous.

A fundamental challenge in the physics of fluids, which enables the advantages of PSM to be exploited, is the description of phase transitions and the associated critical phenomena. Phase transitions present one of the most striking and nontrivial collective phenomena exhibited by interacting many-body systems. When one or more first derivatives of the
relevant thermodynamic potentials change discontinuously (e.g., when a liquid solidifies to a crystal), then the transition, typically associated with symmetry breaking, is referred to as first-order. 
Despite the often impressive nature of first-order transitions, deeper and more universal physical mechanisms are revealed by the continuous transitions in which the first derivatives of the thermodynamic potential remain continuous while only higher-order derivatives (e.g., the compressibility or the specific heat) are divergent or change discontinuously. 

As the critical point of a continuous transition is approached, by variation of the thermodynamic parameters, local density fluctuations can become macroscopic in extent, leading to a rapid growth of the related second-order thermodynamic derivatives (e.g., the compressibility). Moreover, the development 
of large-scale density fluctuations causes the scattering intensity of impinging electromagnetic or particle waves
to become large at small wavevectors, as can be readily observed from the turbid appearance (`critical opalescence') of near-critical binary fluid mixtures. 
The most remarkable feature of these various divergent behaviours is that the power-law exponents which characterize them are the same for a wide range of seemingly 
disparate physical systems. This \textit{universal} behaviour ultimately arises from the scale invariance of critical systems, and does not depend on their specific details; the values of the critical exponents 
depend only on general features, such as spatial dimensionality and symmetry. 
There also exist scaling laws which provide simple mathematical relations between the critical exponents. 
Good accounts of the statistical mechanics of critical phenomena can be found in the books by Binney \textit{et al.} \cite{Binney} and Goldenfeld \cite{Goldenfeld}. 

When investigating the critical behavior of a given system, either experimentally or theoretically, it is often useful to focus on the correlation functions which characterize the average spatial arrangement of the particles. The most important correlation functions are the two-body functions, which quantify the correlation of density fluctuations at 
two different points in space. Not only are these directly accessible in scattering experiments but, for systems interacting via (the most commonly studied) pairwise additive interactions, also provide 
full information about the thermodynamics. Calculation of the two-body correlations thus gives access to the full set of equilibrium critical exponents. 

While this is all very well in principle, a central and long-standing problem in the theory of liquids is how to actually calculate the 
two-body correlation functions from first-principles, i.e.,~directly from the underlying interparticle interactions. 
For interacting systems in either two or three spatial dimensions (usually the cases of interest) approximations are certainly required, since exact solutions can only be obtained 
for strongly idealized one-dimensional systems. 
The most successful route to obtain useful approximations has proven to be the method of integral equations, based on approximate closures of the OZ equation. 
Among the many available closure schemes (see e.g., \cite{Janssen,caccamo_review}) the most well-known are the hypernetted-chain (HNC) \cite{HNCoriginal1,HNCoriginal2}, the Percus-Yevick (PY) \cite{PercusOriginal} and the mean-spherical approximation (MSA) 
\cite{MSAoriginal}. 
While HNC and PY usually provide reliable predictions for soft and strongly-repulsive interparticle potentials, respectively, they both fail to describe the liquid-gas phase transition which emerges for attractive systems as the temperature is reduced. In contrast, the MSA has the appealing property that it provides a realistic description of the phase transition, with a critical 
point at which the isothermal compressibility diverges with non-classical (i.e.,~non-mean-field) critical exponents. 
A further positive feature of the MSA is that it can be solved analytically for the special case of a three-dimensional system of particles interacting via a hard-core repulsion plus an attractive Yukawa tail, namely the hard-core Yukawa (HCY) model \cite{Waisman1973,Hoye1977}. 
The MSA for the HCY model thus provides a convenient test-bed to investigate the performance of numerical algorithms aiming to accurately solve the OZ equation; if the analytical results can be well reproduced in three dimensions, then one can proceed with confidence when applying the same numerical method to systems/approximations for which no analytical solution is available.    

In the present work, 
after reviewing the OZ equation and the associated numerical challenges in Section~\ref{sec:OZ}, we will describe a novel PSM scheme to compute the integrals required to solve the OZ equation in Sections~\ref{sec:discretisation} and~\ref{sec:pseudospectral}. This requires a careful choice of numerical domain decomposition and choice of collocation points. In Section~\ref{sec:self-consistent} we describe how to solve the OZ equations via both the standard Picard scheme and a novel iterative approach.  We validate the numerical approaches in Section~\ref{sec:validation} by comparing to known solutions for a system of HCY particles in 3D, recovering both the correct phase boundary and critical exponents. Finally, in Section~\ref{sec:exponentreview} we consider the same system in two dimensions to obtain new information about the critical behaviour of the MSA, for which no results were previously available. The case of two-dimensions is of particular interest, since fluctuations are much more pronounced than in the three dimensional system and, consequently, the critical properties deviate strongly from the predictions of mean-field theory. In Section~\ref{sec:conclusions} we provide some conclusions and provide some suggestions for future work.

\section{The Ornstein-Zernike equation and closure approximations} \label{sec:OZ}

\begin{figure}
	\centering
	\includegraphics{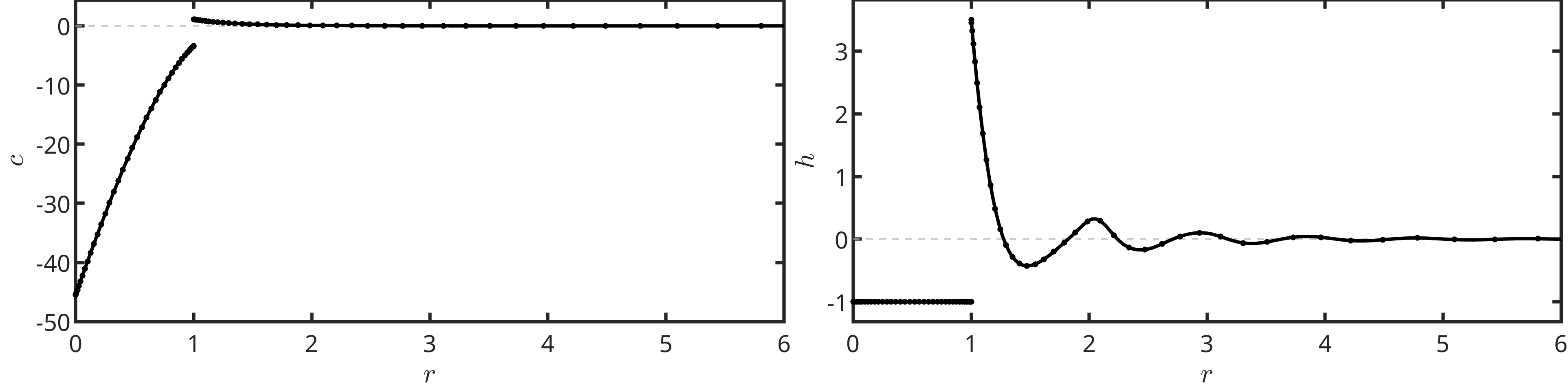}
	\caption{
	An example solution to the 3D OZ relation (in bulk) for a model with a hard core and a long-range attractive interparticle potential, illustrating the discontinuous nature of the correlation functions, $c$ and $h$.}
	\label{fig:ozprofile}
\end{figure}

The OZ equation connects the total correlation function, $h$, and the two-body direct correlation function, $c$, examples of which are shown in figure \ref{fig:ozprofile}. In $\R^D$, it can be expressed as
\begin{equation} \label{eqn:oz}
	h(\r,\r') = c(\r,\r') + \int_{\R^D} \dd{\r''} \rho(\r'') c(\r,\r'') h(\r'',\r').
\end{equation}
In bulk, the (one-body) density $\rho$ is constant and $h$ and $c$ depend on only a single, scalar variable (namely the separation between $\r$ and $\r'$). Defining $r = \abs{\r - \r'}$ and renaming the dummy integration variable as $\r'$, we have
\begin{equation} \label{eqn:ozbulk}
	h(r) = c(r) + \rho \int_{\R^D} \dd{\r'} c(\abs{\r-\r'}) h(\abs{\r'}).
\end{equation}

The OZ equation is a single equation in two unknowns ($h$ and $c$) and, therefore, to determine a unique solution, it must be augmented with a closure scheme.
Any such scheme must incorporate the fact that for hard spheres or hard disks, the radial distribution function $g(r)$ is zero for separations less than the hard-core diameter, $d$, i.e., within the exclusion sphere or exclusion disk of radius $d$, since the particles have a zero probability to overlap.  Using the definition $h = g - 1$ results in the condition $h(r) = -1$ for $r<d$.
Alongside this exact condition, we also require an approximate closure for the remainder of the domain, i.e., for $r\geq d$.
Many common closures are local functions on this domain, and we can write such a closure as
\begin{equation} \label{eqn:closure}
	\begin{aligned}
		h(r) + 1 &= 0 \quad &r<d, \\
		G\big(c(r),h(r),r\big) &= 0 \quad &r>d,
	\end{aligned}
\end{equation}
for some $G$. We note that closures are often expressed instead in terms of the indirect correlation function $\gamma=h-c$ and $c$, such that $G\rightarrow G(c(r),\gamma(r),r)$, but the two formulations are clearly equivalent.

In this work we will consider only the MSA closure, in which the direct correlation function is assumed proportional to the interparticle potential, $\phi(r)$, 
	\begin{equation}\label{eqn:msa}
		G(c,h,r) = c + \beta\phi = 0
	\end{equation}
though there are many other possibilities that have been considered in the literature \cite{caccamo_review,TschoppSuperadiabatic} and our approach is relatively agnostic to the choice of closure.

We note a key aspect that affects the numerical solution of the OZ equation: for hard particles, the closure schemes result in solutions $h$ and $c$ that are discontinuous at $r = d$, but smooth everywhere else; see figure~\ref{fig:ozprofile}.  It is well known~\cite{Hansen06} that the combination $\gamma = h-c$ is always smooth, which enables one to deal with only (at most) one discontinuous function, but this still imposes strong constraints on numerical approaches that aim to produce highly accurate solutions of the OZ equation.

\subsection{Numerical challenges of OZ}
\label{sec:numericalChallenges}

The main difficulty when solving the (inhomogeneous) OZ equation \eqref{eqn:oz} lies in the form of its integral, which is  complicated to implement numerically, not only due to its dimension, but also to the interdependency of the positions at which the functions $h$ and $c$ have to be evaluated. Fortunately, in the bulk case, since the density is constant, it can be removed from the integral, which then reduces to a convolution. The bulk OZ equation \eqref{eqn:ozbulk} is therefore historically solved in Fourier space, where convolutions become multiplications. Indeed, the Fourier-transformed bulk OZ is given by
\begin{equation} \label{eqn:ozbulkft}
	\tilde{h}(\abs{\k}) = \tilde{c}(\abs{\k}) + \rho \tilde{c}(\abs{\k}) \tilde{h}(\abs{\k}),
\end{equation}
where the tilde indicates a transformed quantity.
It is worth pointing out that the Fourier transformation not only removes the integral, but also disentangles the positions at which the correlations are evaluated, i.e.,
\begin{equation} 
	c(\abs{\r-\r'}) h(\abs{\r'}) \rightarrow \tilde{c}(\abs{\k}) \tilde{h}(\abs{\k}),
\end{equation}
thus curing the equation of its most undesirable real-space features, re-shaping it to a more manageable state.
Unfortunately, setting up a numerical scheme to iteratively solve the OZ equation in Fourier space involves the need for a cut-off value in real space, beyond which the correlations are assumed to vanish. This does not present serious issues in many cases, but shows limitations in regions close to thermodynamic state points which generate long-range correlations (e.g., the critical and triple points). The 3D and 2D Fourier grids are given by the zeros of the sine function and the zero-order Bessel function, respectively \cite{lado}. The grid spacings in real and transformed space depend on each other via the choice of both the number of grid points (that corresponds to the number of considered zeros in the transformed space) and the value of the real-space cut-off.
To ensure that the correlation functions vanish both in real and transformed space, the cut-off value has to be large enough. However, since the only choice available is the number of points, not their location, there is no access to resolve specific positions.
It is thus difficult to simultaneously resolve the correlations in both spaces, especially for strongly varying functions as is the case for the total correlation function and its transform, at high packing. The only possibility is to max-out the number of considered points for the smallest safe guess of spatial cut-off value.
Moreover, Fourier transformations are not well-suited to discontinuous functions, generating unwanted Gibbs phenomena when applied directly. Replacing the total correlation $h$ with the (always smooth and continuous) indirect correlation function, $\gamma$, is thus of great help. The transformed bulk OZ \eqref{eqn:ozbulkft} is then given by
\begin{equation}
   \tilde{\gamma}(\abs{\k}) = \rho \tilde{c}(\abs{\k}) \big( \tilde{c}(\abs{\k}) +  \tilde{\gamma}(\abs{\k}) \big).
\end{equation}
In addition, the (also often discontinuous) direct correlation, $c$, has to be rewritten as the sum of a step function, $f_{\text{step}}$, and a continuous (however not smooth) function, $c_{\text{cont}}$, and (back)transformed in two steps. Indeed, the step function can be analytically (back)transformed, while $c_{\text{cont}}$ can be straightforwardly numerically Fourier (back)transformed, according to
\begin{align}
    f_{\text{step}}(\abs{\r}) =    \begin{cases}
                                        c_{\text{jump}} \qquad \abs{\r}<d, \\
                                        0  \qquad \qquad \text{elsewise},
                                    \end{cases}
                                    & \overset{FT}{\longleftrightarrow} \quad    \tilde{f}_{\text{step}}(\abs{\k}) =    \begin{cases}
                                                4\pi c_{\text{jump}} \frac{\sin{\left(d \abs{\k}\right)} - d \abs{\k} \cos{\left(d \abs{\k}\right)}}{\abs{\k}^3} \qquad \text{in 3D}, \\
                                                2\pi c_{\text{jump}} \frac{d J_1(d\abs{\k})}{\abs{\k}} \hspace*{1.65cm}  \qquad \text{in 2D},
                                            \end{cases}\\
    c_{\text{cont}}(\abs{\r}) = c(\abs{\r}) - f_{\text{step}}(\abs{\r}) \quad &\overset{FT}{\longleftrightarrow} \quad \tilde{c}_{\text{cont}}(\abs{\k}) = \tilde{c}(\abs{\k}) - \tilde{f}_{\text{step}}(\abs{\k}),
\end{align}
where $c_{\text{jump}}$ is defined as the amplitude of the discontinuity at $\abs{\r}=d$, given by the difference of the right and left limits $c(\abs{\r}\rightarrow d^{+})-c(\abs{\r}\rightarrow d^{-})$, and where $J_1$ is the order one Bessel function.
Those numerical considerations are even more important given that the iteration scheme to solve the OZ equation needs back and forth transformations of the correlations at each step. Indeed, since the OZ equation has only one input, the one-body density, while having two unknowns, the correlation functions, it needs to be supplemented by a closure relation (such as the MSA given above), which is stated in the real space. One thus has to solve the OZ equation in the transformed space but come back to real space to implement the closure. This back and forth transformation scheme is computationally very demanding, being quite slow and requiring the storage of double the number of (large) matrices than would be required for a scheme based purely in real or transformed space.

To connect to the wider literature and demonstrate some further challenges involved in the numerical solution of the OZ equations, we note that there is some similarity with non-linear, multi-dimensional Volterra-Fredholm integral equations (NL-MD-VFIE) of the second kind.  There has been some recent work on the numerical solution of such systems using spectral methods, see, e.g.,~\cite{ZakyIE,AminIE1,AminIE2}. 

Here we find it helpful to consider the PY closure as an example, as it allows us to write down explicit equations for $h$ and $c$. The PY closure imposes $h(r) = -1$, for $r<d$, and $c(r) = 0$, for $r>d$.  We may then rewrite the bulk OZ \eqref{eqn:ozbulk} as the coupled pair of equations
\begin{equation} \label{eqn:ozwithdomains}
\begin{aligned}
    c(r) &\overset{r< d}{=} -1 - \rho \int_{\disc{\r}} \dd{\r'} c(\abs{\r-\r'}) h(\abs{\r'}),\\
	h(r) &\overset{r\geq d}{=} c(r) + \rho \int_{\disc{\r}} \dd{\r'} c(\abs{\r-\r'}) h(\abs{\r'}),
\end{aligned}
\end{equation}
where $\disc{\r}$ is the disk of radius $d$ centred at $\vectorbold{r}$.
If, for example, we knew $h$, then the first equation would be similar to a NL-MD-VFIE (but with a more complicated integration domain than is normally treated in such problems).  We would encounter a similar case if we knew $c$ in the second equation.  However, this is not the case here, and we must solve the coupled set of equations.
Under more complicated closure schemes, the integration domains in \eqref{eqn:ozwithdomains} remain infinite, and the equations for $h$ and $c$ both contain the local (i.e., un-integrated) value of the other unknown.  As far as we are aware, application of PSMs to this form of problem has not been studied before.

When treating a general closure, there is an additional complication since $h$ and $c$ are both discontinuous (at $d$) for systems of hard-core particles.  Hence, a na\"ive PSM is bound to fail, and one needs to apply a spectral element method on the real line for both $h$ and $c$.  As will become clear in the following section, the reality is much more complicated than this: one also needs to decompose the integration domain to ensure that both functions are smooth on each subdomain in order to obtain accurate interpolation and integration. Therefore, we do not expect any of the existing approaches in the literature to be directly applicable here, but they do provide some further motivation for our approach.

\section{Discretisation of the OZ integral} \label{sec:discretisation}

Henceforth, we consider bulk systems of hard-core particles in both 3 and 2 dimensions.
We begin by noting that we have two functions ($h$ and $c$) that are both defined on $\R^+$ and discontinuous at particle diameter $d$.  We also wish to solve the OZ equation \eqref{eqn:ozbulk}, which we find convenient to rewrite as 
\begin{equation} \label{eqn:ozbulkshift}
	h(r) = c(r) + \rho \int_{\R^D} \dd{\s} c(\abs{\s}) h(\abs{\s+\r}),
\end{equation}
by making the change of variables $\r' = \s + \r$ in the integral.

If one chooses the discretisation points of $\R^+$ to be the same for $h$ and $c$ then the local part of \eqref{eqn:ozbulkshift} is straightforward to address and does not need any interpolation.  The challenge therefore lies in computing the integral term.  To do so, we implement a domain decomposition strategy.  This will enable us to perform highly accurate interpolation and integration of smooth functions, without further concern for discontinuities.

\subsection{Domain decomposition}

\begin{table}[h]
	\centering
	\begin{tabular}{rcc}
		&					$\abs{\r+\s} < d$ & 	$\abs{\r+\s} \geq d$ \\
		\hline
		$\abs{\s} < d$ & 		$\cll$ &				$\clg$ \\
		$\abs{\s} \geq d$ & 	$\cgl$ &				$\cgg$ \\
		\hline
	\end{tabular}
	\caption{Notation for the four domains required to ensure smoothness of the two correlation functions.}
	\label{tab:subdoms}
\end{table}
	
\begin{figure}[pos=h]
	\centering
	\includegraphics{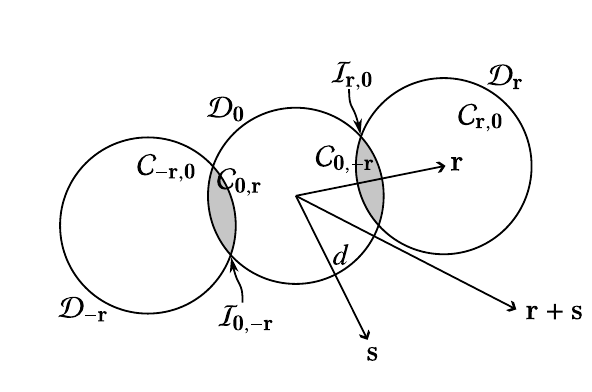}
	\caption{The subdomains of $\mathbb{R}^2$ relevant for integration and interpolation in our scheme.
	The two disks centred at the origin and $-\r$ are of primary relevance for decomposing \eqref{eqn:ozbulkshift} into equation \eqref{eqn:intbydomains}, while $h$ is evaluated  on these subdomains after translation by $\r$, leading to the disk centred at $\r$ in the figure.
	Note that $\cll$ is a subset of $\hgl$, and $\hll$ is a subset of $\clg$, but not vice-versa.
	The two remaining shapes not marked in the figure are the entire space $\mathbb{R}^2$ with $\disc{0}$ and $\disc{-\r}$ removed, and with $\disc{0}$ and $\disc{\r}$ removed.}
	\label{fig:subdomains}
\end{figure}

The domain of integration in equation \eqref{eqn:ozbulkshift} can be partitioned into four pieces according to the conditions $\abs{\s}<d$ and $\abs{\s+\r}<d$, which will be recognised as defining balls of radius $d$ centred at the origin and $-\r$, respectively.
We will denote these balls as $\disc{0}$ and $\disc{-\r}$ respectively, and consider them as sets of points obeying the above inequalities.
With this established, the four subdomains on which the correlation functions remain continuous are defined by
\begin{align} \label{eqn:subdomains1}
	\cll &= \disc{0} \cap \disc{-\r}, \\
	\clg &= \disc{0} \setminus \disc{-\r}, \\
	\cgl &= \disc{-\r} \setminus \disc{0}, \\
	\cgg &= \mathbb{R}^D \setminus \left(\disc{0} \cup \disc{-\r}\right).
\label{eqn:subdomains4}
\end{align}
For reference, we give the equivalent definition in terms of inequalities in table \ref{tab:subdoms}.
The convolution nature of \eqref{eqn:ozbulkshift} means that integration on any of the above subdomains entails evaluating one of the functions at a different set of points, related to the original domain through a translation by $\r$. 
Again using subscripts to denote the required set-theoretic operations, these translated domains are given by (in the same order as in equations \eqref{eqn:subdomains1}--\eqref{eqn:subdomains4}) $\hll, \hlg, \hgl, \hgg$, and are illustrated in figure \ref{fig:subdomains}.

As the next step, we rewrite the integral in equation \eqref{eqn:ozbulkshift} as a sum over these subdomains: 
\begin{align} 
	&\int_{\mathbb{R}^2}\dd{\s} c(\abs{\s}) h(\abs{\r+\s}) \notag \\
    &=
	\int_{\cgg}\dd{\s} c(\abs{\s}) h(\abs{\r+\s})
	+\int_{\cgl}\dd{\s} c(\abs{\s}) h(\abs{\r+\s}) 
	+\int_{\cll}\dd{\s} c(\abs{\s}) h(\abs{\r+\s}) 
    +\int_{\clg}\dd{\s} c(\abs{\s}) h(\abs{\r+\s}) 
    \notag \\
    &=
	\int_{\cgg}\dd{\s} c(\abs{\s}) h(\abs{\r+\s})
	-\int_{\cgl}\dd{\s} c(\abs{\s}) 
	-\int_{\cll}\dd{\s} c(\abs{\s}) 
    +\int_{\clg}\dd{\s} c(\abs{\s}) h(\abs{\r+\s}),
    \label{eqn:intbydomains}
\end{align}
where we have used the fact that, in the middle two integrals, $h$ is evaluated with an argument in $[0,d]$ and hence can be set to the known value of $-1$.  We reiterate that this is an exact condition for hard spheres/disks, and so we apply it throughout this work.  However, we note that the methodology we describe below can be easily generalised if $h$ is not a constant on $[0,d]$.

We will now describe, using this domain decomposition, how to evaluate the integral purely in real space, in a discretisation-agnostic manner.  Following that, in Section~\ref{sec:pseudospectral}, we will explain how we perform this integration to a very high degree of accuracy via a particular choice of pseudospectral discretisation.  

\subsection{Interpolation and integration} \label{sec:InterpInt}

Suppose that we have discretised each of the domains in table~\ref{tab:subdoms} such that each domain, $\mc{A}$, is represented by a set of $N_{\mc{A}}$ points, stored as a column vector \texttt{a}. We fix a discretisation of the non-negative real line (or, in most approaches, of a sufficiently long interval $[0,L]$), which we denote by $\texttt{r}$, a column vector of length $N$.

We now also assume that we have access to two types of operator for each of the domains, where the discretisation of a function $f$ on $\mc{A}$ is denoted by $\texttt{f} = f(\texttt{a})$, with $\texttt{f}$ again a column vector of length $N_{\mc{A}}$:
\begin{enumerate}

\item $\Interp{\mc{A}}$, a matrix of dimension $[N_{\mc{A}},N]$, performs numerical interpolation of a function $f(r)$ defined on $\R^+$ (or $[0,L]$), as represented by its values at \texttt{r}, onto the corresponding values $f\big(|\texttt{a}|\big)$, i.e.,
\[
    f(\abs{\texttt{a}}) \approx \Interp{\mc{A}} * \texttt{f},
\]
where the left-hand side should be interpreted as $f\big(|\texttt{a}|\big) = \big[f\big(|\texttt{a}_1|\big),f\big(|\texttt{a}_2|\big), \dots, f\big(|\texttt{a}_{N_\mathcal{A}}|\big)\big]^T$. 

\item $\Int{\mc{A}}$ performs numerical quadrature (integration) on the domain $\mc{A}$.  $\Int{\mc{A}}$ is a row vector of length $N_{\mc{A}}$ such that
\[
    \int_{\mc{A}} \dd{\s} f(\s) \approx \Int{\mc{A}} * \texttt{f}.
\]

\end{enumerate}
Suppose that we have access to $\nvec{c} = c(\texttt{r})$ and $\nvec{h} = h(\texttt{r})$.
Then, for a given integral in equation \eqref{eqn:intbydomains} of the form
\[
    I_{\mc{A}}(r) = \int_{\mc{A}} \dd{\s} c(\abs{\s}) h(\abs{\r+\s}),
\]
we may approximate it through the following steps:
\begin{enumerate}
    \item Approximate $c(\abs{\s})$ by $\nvec{c}_{\mc{A}} = \Interp{\mc{A}} * \nvec{c}$;
    \item Approximate $h(\abs{\r+\s})$ by $\nvec{h}_{\mc{A}_{\r}} = \Interp{\mc{A}_{\r}} * \nvec{h}$, where $\mc{A}_{\r}$ is $\mc{A}$ shifted by $\r$;
    \item Approximate 
    $I_{\mc{A}}(r)$ by $ \Int{\mc{A}} * ( \nvec{c}_{\mc{A}} \odot \nvec{h}_{\mc{A}_{\r}} )$, where $\odot$ denotes the pointwise, or Hadamard, product.
\end{enumerate}
Recall that, since we are in bulk, we may choose any $\r$ such that $\abs{\r} = r$.  We conventionally choose this to be $(r,0)^T$.

\subsection{Tensor formulation} \label{sec:tensor}

Having described how (given a discretisation, interpolation, and integration) one may numerically approximate the integrals in \eqref{eqn:intbydomains}, we now outline a method which implements this in a memory-efficient way.  We first highlight a potential issue with the method described in Section~\ref{sec:InterpInt}.  Suppose that we wish to solve equation \eqref{eqn:ozbulkshift} for each element $\texttt{r}_i$ of the vector $\texttt{r}$.   Each of the domains in \eqref{eqn:intbydomains} depends on $\texttt{r}_i$, and would typically be discretised using many more points than are necessary for the 1D interval on which $h$ and $c$ are defined.  This is typical behaviour since the domains in \eqref{eqn:intbydomains} are in 3D or 2D.  Hence, using the notation of Section~\ref{sec:InterpInt}, $N_{\mc{A}} \gg N$, and $\Interp{\mc{A}}$ has dimension $[N_{\mc{A}},N]$.

In contrast, fixing a point $r$, one can think of the integral in \eqref{eqn:ozbulkshift} as a bilinear operator in $c$ and $h$.  After discretisation, the approximate integration routine is also a bilinear operator, now in $\nvec{c}$ and $\nvec{h}$.  Hence, we can write the numerical approximation to \eqref{eqn:ozbulkshift} as 
\begin{equation} \label{eqn:ozbulk_discrete}
	h_i = c_i + \rho\sum_{j,k=1}^{n} T_{ijk} c_j h_k
\end{equation}
for some tensor $T_{ijk}$ of dimension $N^3$.  In other words, we have already contracted over the dimensions corresponding to the discretisation of the 3D or 2D domains, leading to a much more memory efficient approach.  See Appendix~\ref{app:tensor} for more details on the tensor formulation.

\section{Pseudospectral discretisation} \label{sec:pseudospectral}

Having described, for an arbitrary discretisation scheme, how to compute (the numerical approximation to) the integral in equation \eqref{eqn:ozbulkshift}, we now present an accurate and efficient approach using pseudospectral methods.

For an introduction to pseudospectral methods, we refer the reader to the textbooks~\cite{Boyd,Trefethen,Canuto2006Book,Canuto2007Book}, as well as to previous literature concerning the particular implementation that forms the basis of the approach here~\cite{NoldPSM, GoddardPRL, GoddardHI, GoddardHIConfined, RodenSedimentation, RodenMultishape, MillsFlow}.  Note that many of these approaches have successfully tackled (D)DFT problems which have some very similar challenges to the problem studied here.  A significant difference between previously-studied problems and the approach we take here is that the former did not make use of the OZ equation, but instead used simpler (and less accurate) approaches such as fundamental measure theory. As described above, the key idea of PSM is that numerical operations, such as differentiation and integration, can be performed to a very high accuracy with a small (compared, e.g., to FDM or FEM) number of collocation points, at least when the functions involved are smooth.  This the main motivator for the domain decomposition in the previous section; $c$ and $h$ are now smooth on each of the domains in \eqref{eqn:intbydomains}.  

For each domain $\mc{A}$, we require the following:
\begin{enumerate}[(1.)]
\item A set of collocation points that accurately describes the domain.
\item The corresponding integration weights.
\item An accurate method for interpolation.
\end{enumerate}

\subsection{Collocation points, integration, and interpolation} \label{sec:points_weights_interp}

For (1.) and (2.), collocation points and integration, we first demonstrate the approach in 2D, i.e., for hard disks as we believe this is much easier to understand visually.  We will then explain the simple approach employed to go from 2D to 3D.

We begin by describing the 1D discretisations which are used to construct those in higher dimensions (essentially through a tensor product approach). We utilise two standard pseudospectral discretisations.  For periodic domains, such as the angular coordinate in a disk or annulus, we use Fourier collocation points.  For non-periodic domains, we use Chebyshev collocation points.  For $N+1$ points, these are given explicitly as $x_k=k/(N+1), k\in\{0,\dots,N\}$ in the Fourier case, and in the Chebyshev case as
\begin{equation} \label{eqn:cglpoints}
	x_k = -\cos\left(\frac{\pi k}{N}\right), \quad k\in\{0,\dots,N\}
\end{equation}
Note that we include a non-traditional minus sign in the definition of the Chebyshev points.  This has no effect on, for example, integration as the discretisation is symmetric.  The reason for this choice is that it preserves the `left-right' order of points.  In other words, the first point in the list is $-1$ and the last is $1$; cf.~the standard approach, e.g.,~\cite{Trefethen}.

Alongside the explicit collocation points, the integration weights (traditionally denoted by $w$) are also known: 
\begin{itemize}
    \item Fourier: $w_j = 1/(N+1)$.
    \item Chebyshev: 
    \begin{align}
        w_j = \frac{2d_j}{N} \left\{
        \begin{array}{ll}
         1 - \sum_{k=1}^{(N-2)/2} \frac{ 2\cos 2 k \theta_j }{4k^2-1} - \frac{\cos \pi j}{N^2-1}  &\qquad \text{for } N \text{ even}\\
         1 - \sum_{k=1}^{(N-1)/2} \frac{ 2\cos 2k \theta_j}{4k^2-1}&\qquad \text{for } N \text{ odd}
        \end{array}\right.,
    \end{align}
where $\theta_n = n \pi /N$ and $d_n = 1/2$ for $n \in \{0,N\}$ and $1$ otherwise.

\end{itemize}

For the Fourier case, we note that the integration vector is constant.  For the Chebyshev case we note that there are a number of other choices of representation, e.g., through the inverse of the Chebyshev differentiation matrix.  We have chosen to use Clenshaw-Curtis quadrature weights, which are simple to implement numerically; see, e.g.,~\cite[Chapter 12]{Trefethen},~\cite{NoldPSM}.

It turns out that (3.), interpolation, can be understood as a global question on how to perform accurate (polynomial) interpolation.  The approach that we choose here is based on barycentric interpolation~\cite{Barycentric}, which is fast, stable, and accurate.  While all of these properties are desirable for any numerical scheme, a key property for PSMs is the very high accuracy; there is no point being able to perform high accuracy integration if the chosen interpolation scheme is inaccurate.
For a function $f$ determined by values at $N$ points $x_k$, we interpolate to determine the value of $f$ at an arbitrary point $x$ (different from all of the $x_k$) via
\begin{align}
f(x) = \frac{\sum_{k=0}^N \frac{v_k}{x - x_k}f(x_k) }{\sum_{k=0}^N  \frac{v_k}{x-x_k}},
\end{align}
with weights $v_0 = 1/2$ , $v_{k \notin \{0,N\}}= (-1)^k$ and $v_N = (-1)^N/2$.  Note that the weights, $v_k$ are normally denoted by $w_k$, but we use $v_k$ here to avoid any confusion with the integration weights.
We also note that, for a given set of `output' points to be interpolated onto, one can write the interpolation as a matrix multiplication with matrix \texttt{Interp}, as used in Section~\ref{sec:InterpInt}.

\subsection{More general 1D domains}
\label{sec:1DMaps}

In the above, we have explained how to integrate on the intervals $[0,1]$ and $[-1,1]$ for the Fourier and Chebyshev cases, respectively.  However, in practice, we would need to be able to interpolate and integrate on arbitrary domains $[a,b]$, where $a, b \in [-\infty,\infty]$. Note that the infinite cases are only applicable to the Chebyshev discretisation.

We explain the process for the Chebyshev case with interval $[-1,1]$, but the methodology for the Fourier case is analogous. Let us fix an interval $[a,b]$ and two maps, $P:[-1,1] \to [a,b]$ and $C = P^{-1}:[a,b] \to [-1,1]$; we have assumed that $P$ is invertible.  We call $[-1,1]$ the \emph{computational} domain and $[a,b]$ the \emph{physical} domain.  Hence, $P$ maps the computational domain to the physical domain, and $C$ maps the physical domain to the computational domain.  In the below, we use the convention that the computational domain variable is $x$, and the physical space variable is $y$.

In order to perform interpolation within the physical domain $[a,b]$, we map the points to be interpolated at into the computational domain and then use the standard Barycentric interpolation.  Note that the mapping affects only the positions of the collocation points and not the value of any function to be interpolated.

In order to perform integration in the physical domain, one needs only to account for the change of metric due to the map $P$, which can be done using its Jacobian $J_P$.  Suppose that $\texttt{J}_P$ is a \emph{column} vector representing $\dv*{P}{x}$ at each $\texttt{x}_j$ and, as above $\texttt{w}$ is the vector of integration weights in the computational domain.  Then the integration weights in the physical space are given by $\texttt{w}_P = \texttt{w} \odot \texttt{J}_P$, where once again $\odot$ denotes the pointwise or Hadamard product.

\begin{figure}
\includegraphics[width=\textwidth]{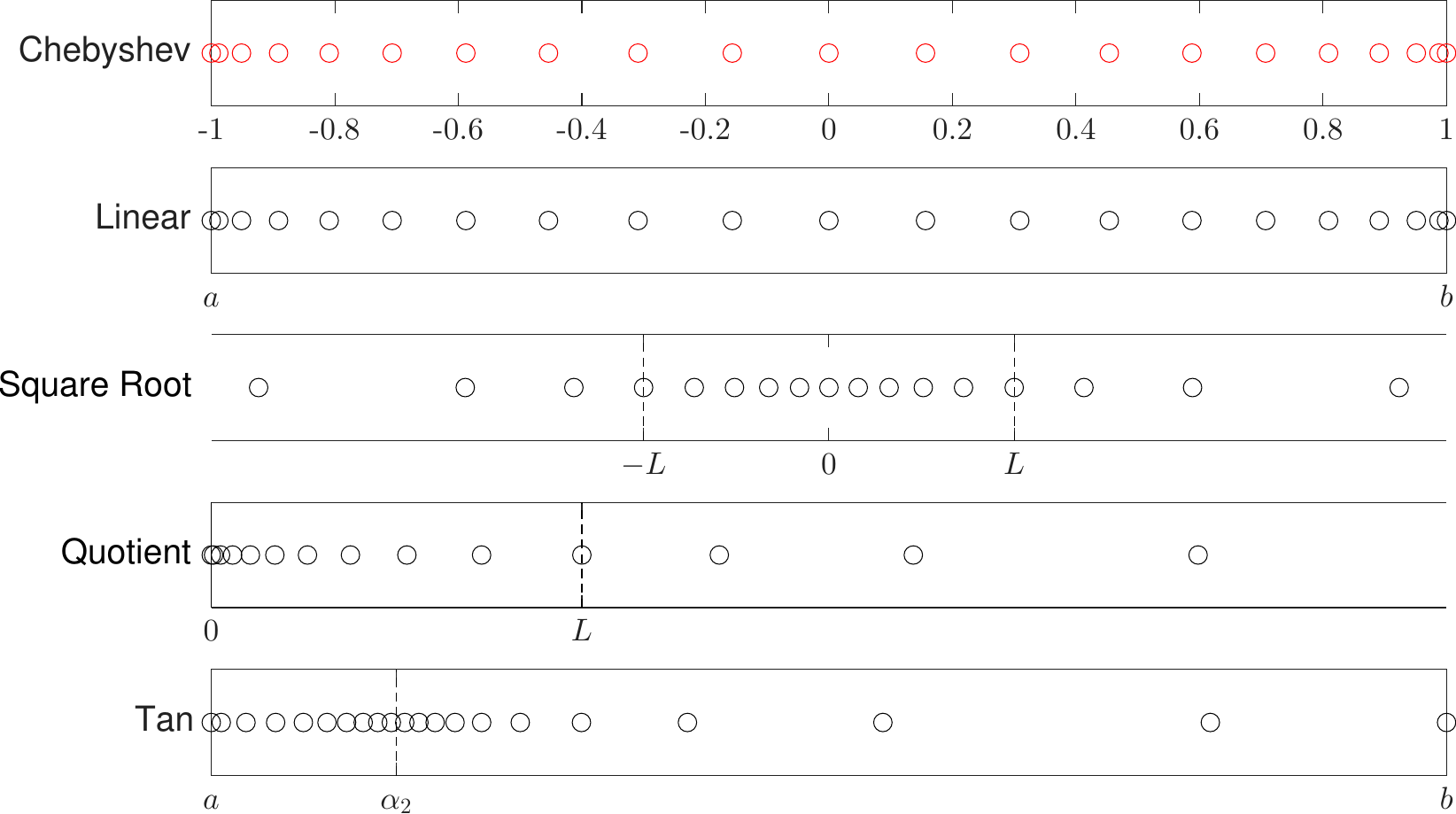}
\caption{An example of the effect of the maps described in Section~\ref{sec:1DMaps}.  The top plot shows the original 
Chebyshev grid, with the remaining rows illustrating the locations
of the collocation points under the various maps.  Note that the Square Root map takes points to the full interval $[-\infty,\infty]$, but we have cut off the domain.  Similarly,
the Quotient map that takes points to $[0,\infty]$ has been cut off 
to the right.}
\label{fig:1Ddomains}
\end{figure}

In this work, we utilise the following maps, $P$ and $C$, examples of which can be seen in Figure~\ref{fig:1Ddomains}:
\begin{enumerate}
    \item \textbf{Linear}: The standard linear map between $[-1,1]$ and $[a,b]$:
    \[
        P_L(x) = y_L(x) = a + (x+1)(b-a)/2, \quad
        J_{P_L}(x) = (b-a)/2,
        \quad
        C_L(y) = -1 + 2(y-a)/(b-a).
    \]

    \item \textbf{Linear01}: The standard linear map between $[0,1]$ and $[a,b]$:
    \[
        P_{L01}(x) = y_{L01}(x) = a + x*(b-a), \quad
        J_{P_{L01}}(x) = (b-a), \quad
        C_{L01}(y) = (y-a)/(b-a).
    \]

    Note that, in practice, we only ever use $[a,b] = [0,2\pi]$.

    \item \textbf{Square Root}: A map between $[-1,1]$ and $[-\infty,\infty]$ with a single parameter $L_{SR}$, such that approximately half the interval is mapped to $[-L_{SR},L_{SR}]$:
    \[
        P_{SR}(x) = y_{SR}(x) = \frac{L_{SR}x}{(1-x^2)^{1/2}}, \quad
        J_{P_{SR}}(x) = \frac{L_{SR}}{(1-x^2)^{3/2}}, \quad
        C_{SR}(y) = \frac{y}{(L_{SR}^2+y^2)^{1/2}}.
    \]
    \item \textbf{Quotient}: A map between $[-1,1]$ and $[a,\infty]$ with a single parameter $L_Q$, such that approximately half the interval is mapped to $[a,a+L_Q]$:
    \begin{equation}
        P_Q(x) = y_Q(x) = a + \frac{L_Q(1+x)}{(1-x)}, \quad
        J_{P_Q}(x) = \frac{2L_Q}{(1-x)^2}, \quad
        C_Q(y) = \frac{(y-a)-L_Q}{(y-a)+L_Q}.
        \label{eqn:quotient}
    \end{equation}

    \item \textbf{Tan}: A map between $[-1,1]$ and $[a,b]$ that allows the user to control the clustering of points around a single value, $\alpha_2$, through the specification of another parameter, $\alpha_1$.  When $\alpha _1 \gg 1$ the points are clustered around $\alpha_2$.
    
    This is defined in three stages:  
    
    First we find it useful to set the values of $\alpha_1$ and $\alpha_2$ in the physical space.  We then need their equivalents in the computational space $\tilde{\alpha}_1 = \alpha_1/J_{P_L}$, $\tilde{\alpha}_2 = C_L(\alpha_2)$.

    Secondly, we apply a map which takes $[-1,1]$ to itself:
    \[
        \tilde{P}(x) = \tilde{y}(x) = \tilde\alpha_2 + \tan(X)/\tilde\alpha_1, \quad
        J_{\tilde{P}}(x) = \lambda \sec^2(X)/\tilde\alpha_1,
    \]
 
    where we define $X$ and $\lambda$ through
    \begin{align*}
        \kappa &= \tan^{-1}\big(\tilde\alpha_1(1+\tilde\alpha_2)\big) /  
                \tan^{-1}\big(\tilde\alpha_1(1-\tilde\alpha_2)\big), \\
        s_0 &= (\kappa -1)/(\kappa + 1), \\
        \lambda &= \tan^{-1}\big(\tilde\alpha_1(1-\tilde\alpha_2)\big) / (1-s_0), \\
        X &= \lambda (x-s_0).
    \end{align*}
    
    Finally, we apply the standard linear map to $\tilde{y}$:
    \[
        P_{T}(x) = y_T(x) = P_L \big(\tilde{P}(x) \big), \quad
        J_{P_{T}}(x) = J_{P_L}(x) J_{\tilde{P}}(x), \quad
        C_T(y) = \tilde{C}\big( C_L(y) \big).
    \]

    In order to compute $C_T$, we first map the physical points into the computational domain using $C_L$ and then apply the inverse of $\tilde{P}$, $\tilde{C}$ with
    \[
        \tilde{C}(y) = s_0 + \tan^{-1}\big(\tilde{\alpha}_1 
        (y - \tilde{\alpha}_2)\big)/\lambda.
    \]

We note that, if $\alpha_2 = a$ (which is what we will use), then the Tan map is conceptually similar to a Quotient-like map on a finite interval, where points are clustered around $a$.  It is, in principle, possible to use such a map instead of the Tan map, but in practice we found it to produce results that are around one order of magnitude less accurate for a fixed computational cost, we thus also include the Tan map.

\end{enumerate}

\subsection{2D domains}
\label{sec:2Ddomains}

Our approach to 2D domains is underpinned by the 1D domains above.  For many of the required domains, it is sufficient to take a tensor product of two 1D domains.  There are, however, some cases where a more careful discretisation is required.  We consider one, representative example, which we call MoonSlice.  An example of which is shown in Appendix~\ref{app:2Ddomains}.  It can be thought of as a horizontal slice of a disc, once the intersection with another disk has been removed.  The key idea here is that the vertical grid can be fixed (spanning the whole height of the domain), but the horizontal grid varies at each vertical collocation point.  This is relatively straightforward to implement as one can use a Linear map in each case, and the 2D Jacobian can be formed in a straightforward way from the 1D Jacobians.

\begin{figure}
    \includegraphics[width=0.46\textwidth]{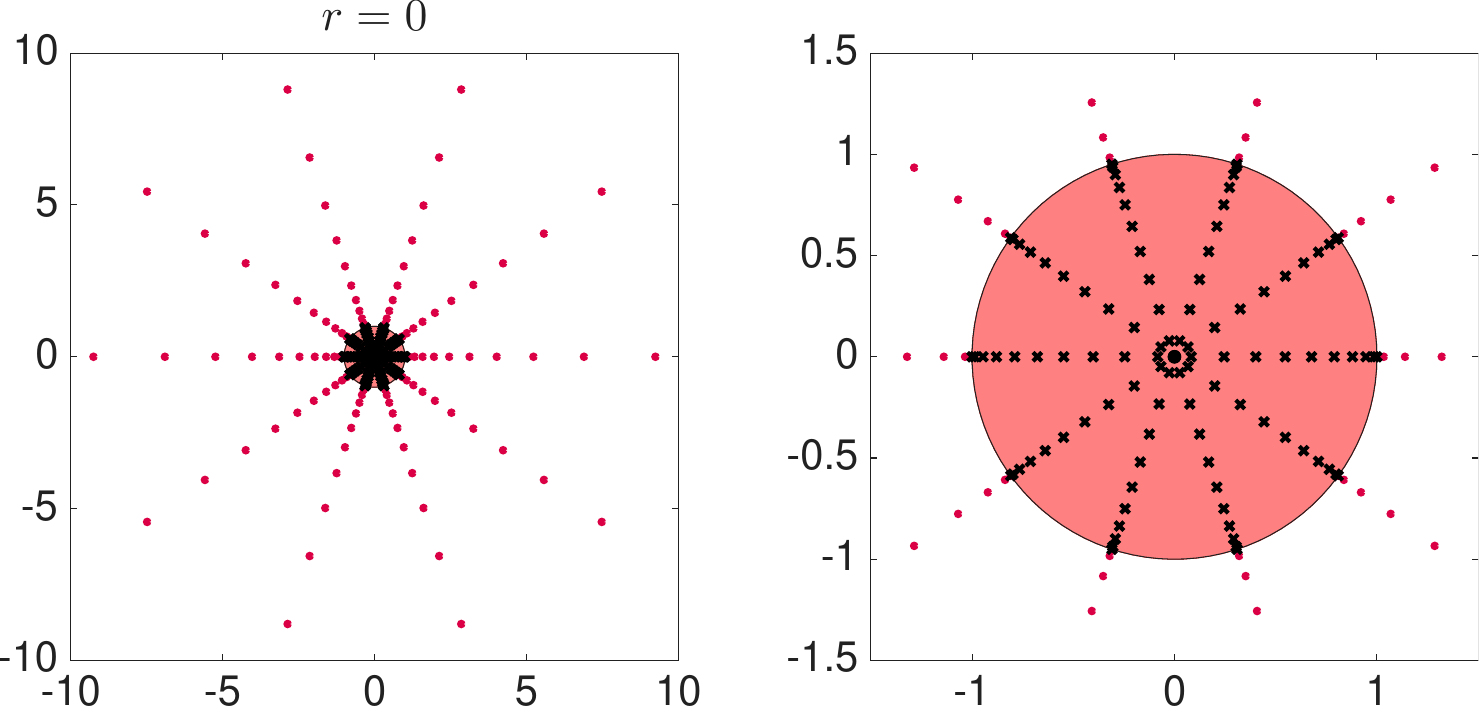}
    \hfill
    \includegraphics[width=0.46\textwidth]{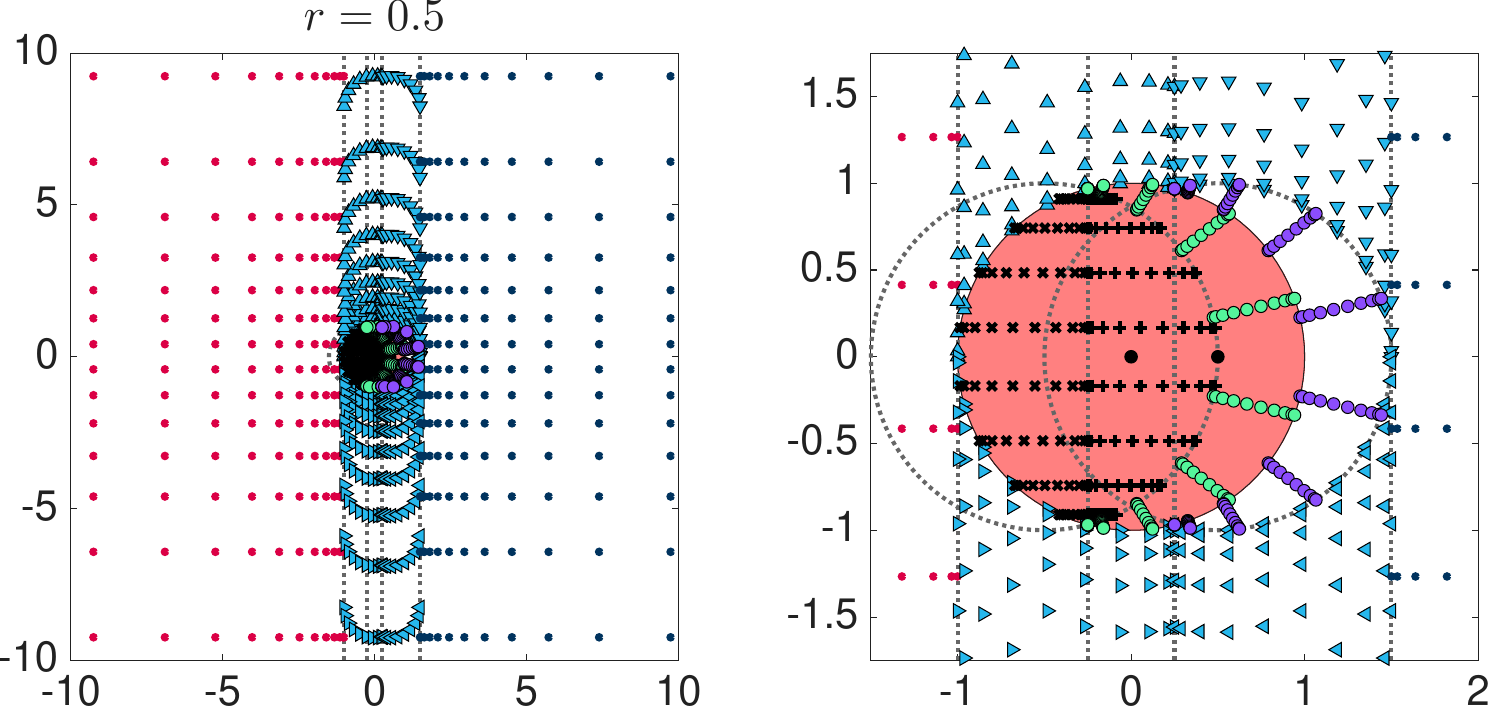}\\[2mm]
    \includegraphics[width=0.46\textwidth]{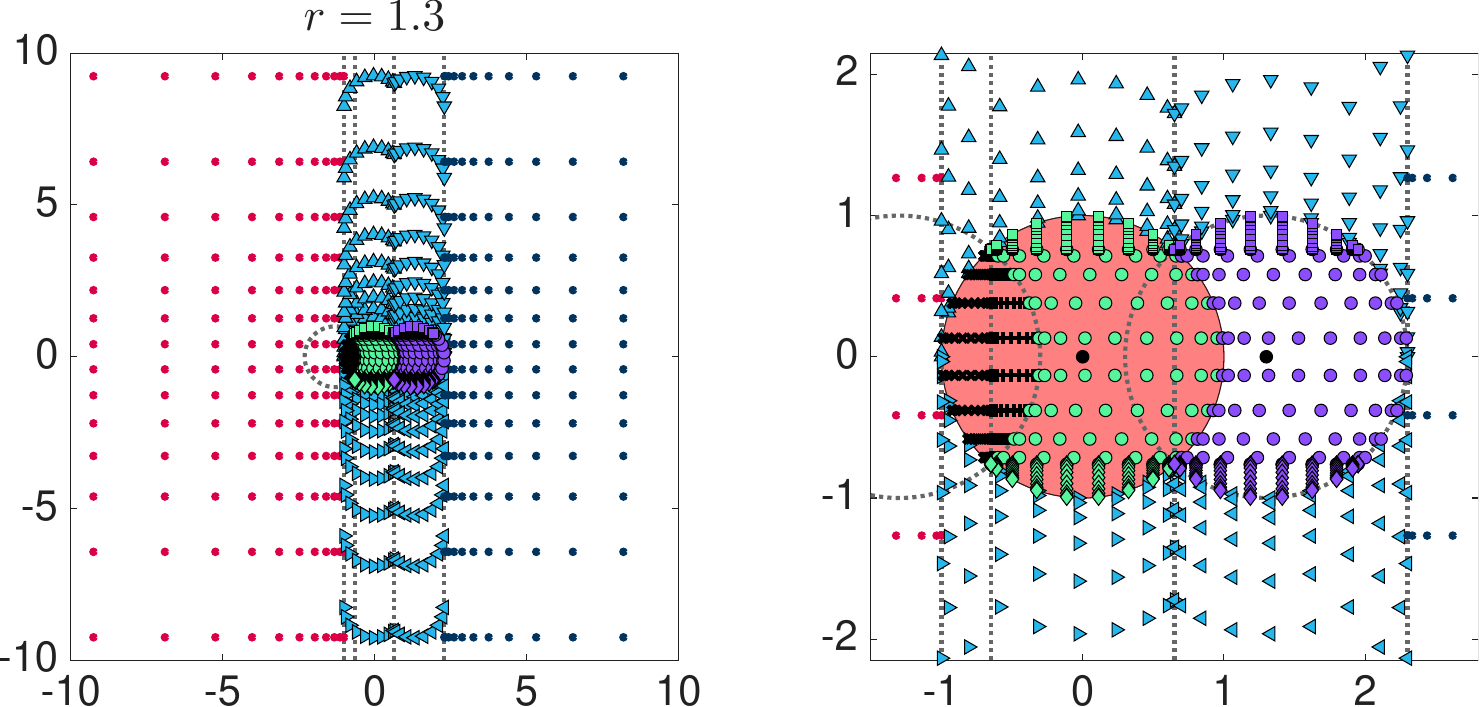}
    \hfill
    \includegraphics[width=0.46\textwidth]{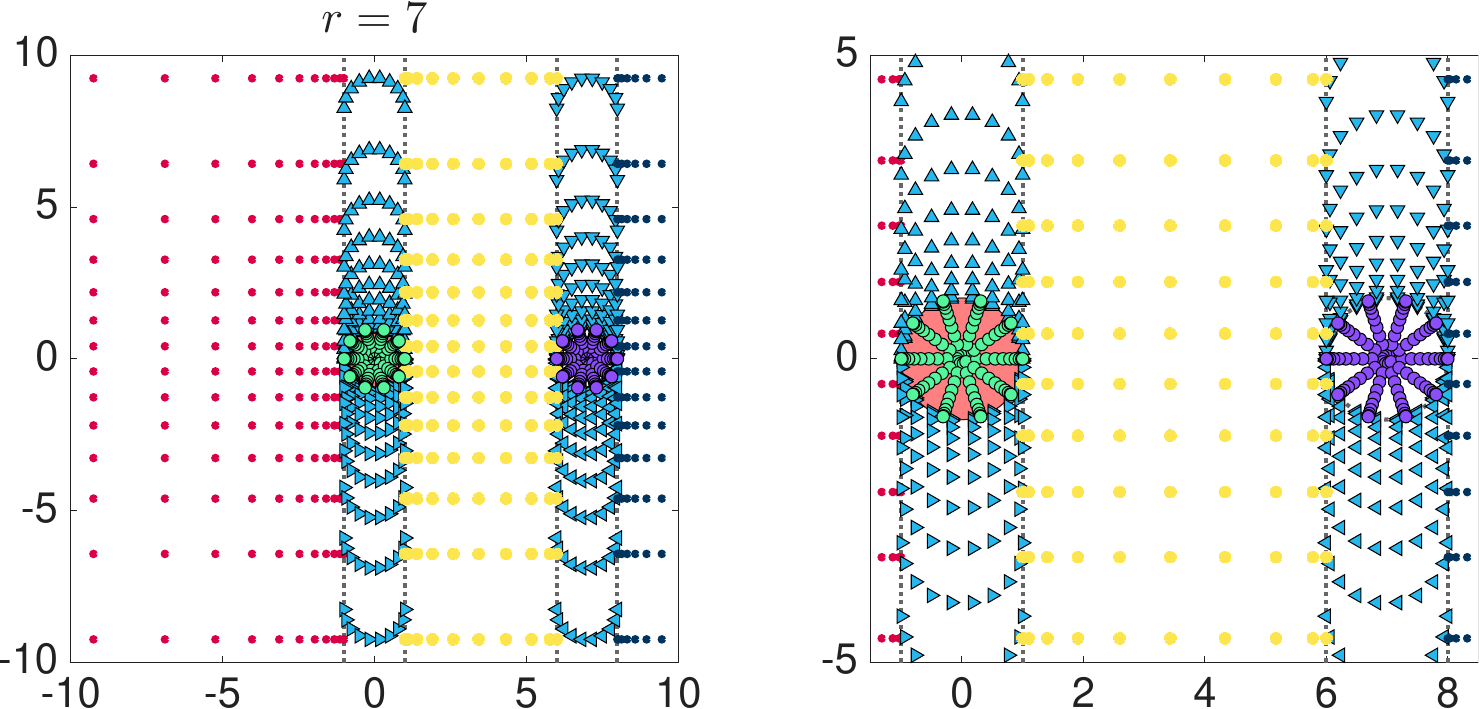}
    \caption{Representative decompositions of the plane with the corresponding collocation points, for disks of diameter $1$.  Each pair of plots corresponds to a different separation of the disk centres; the first disk is fixed at the origin and the second lies at $(r,0)$.  The left-hand plot shows a relatively large part of the plane, with the plot on the right zoomed to show the collocation points in more detail. The disk at the origin is coloured in red.  Collocation points are shown with coloured symbols.  Some helpful lines are indicted by grey dots.}
    \label{fig:2Ddomains}
\end{figure}

Given the domains in Appendix~\ref{app:2Ddomains}, it is then possible to decompose $\R^2$ into the domains in Table~\ref{tab:subdoms}.  Typical examples are shown in Figure~\ref{fig:2Ddomains}. We note that, in order to increase numerical accuracy and robustness, and to minimise the number of domains required, we have further subdivided some of the domains.

\subsection{3D domains}
\label{sec:3Ddomains}

Due to the symmetry of the system, all of the 3D domains we use can be treated as rotations of 2D domains around an axis in the plane of the domain.  In order to avoid multiple coverings of the domain, in a few cases it is necessary to `cut in half' domains from Section~\ref{sec:2Ddomains}.  These are shown in Appendix~\ref{app:2Ddomains}.
Supposing that these domains are located in the $y_1$--$y_2$ plane, almost all of the 3D domains are produced by forming a tensor product with a Fourier grid in $y_3$ on $[0,2\pi]$.  This corresponds to `rotating' the 2D shape around the third axis.  The exceptions are the full cuboids of finite or infinite extent, which are defined using a standard tensor product of Cartesian 1D grids.

\section{Solution of the self-consistent equations} \label{sec:self-consistent}

We are now in a position to accurately compute the integral in the rewritten bulk OZ~\eqref{eqn:ozbulkshift} for an arbitrary particle separation $r$, in either 2D (for disks) or 3D (for spheres).  The remaining question is how one can accurately and efficiently compute the solution of \eqref{eqn:ozbulkshift} given an appropriate closure scheme.

\subsection{Discretisation of the 1D domain} \label{sec:1DDomain}

Recall that, in the homogeneous case, $h$ and $c$ are defined on the positive, half-infinite interval $[0,\infty]$.
As previously described, in most approaches (such as those based on equispaced grids), the domain must be restricted to $[0,L]$ for some sufficiently large $L$.  In this context, `sufficiently large' generally means that both $h$ and $c$ are essentially constant (in fact, constantly zero) for $r>L$.  Here, we choose to once again use a pseudospectral approach for the 1D domain, since this allows us to treat both the finite domain as well as the half-infinite interval.  The latter is advantageous as it removes any effects from the (somewhat arbitrary) domain truncation.  However, it does not actually reduce the number of parameters on which the numerical solution depends as one must choose how to map the finite interval onto the half-infinite interval; typically this includes choosing at least one parameter, e.g., $L_Q$ in the Quotient map.

As with the 2D and 3D domains, there is a natural decomposition of the 1D line such that $h$ and $c$ are smooth on the subdomains. 
We construct a 1D domain by splitting the half-infinite interval at $d$.  We then construct the part of the domain in one of two ways, depending on whether it is finite or half-infinite.  In both cases, the interval $[0,d]$ is defined through a Linear map.  For the infinite case we use a Quotient map for $[d,\infty]$.  For the finite case we use a Tan map for $[d,L]$; we have found that the added flexibility of concentrating the grid around a point is beneficial when compared to using the standard Linear map.

We note one technical detail of this decomposition: there are two points at $d$, one in each interval.  This has no effect on integration (since the single point has measure zero), but does require some care when interpolating.  In particular, if one requires the value of $h$ or $c$ precisely at $d$ (e.g., to compute the reduced pressure) then it is important to evaluate this on the correct interval (namely, the outer interval for $h$ and the inner interval for $c$).

\subsection{Iterative solution of the OZ equation} \label{sec:iterative}

Given that we have fixed a discretised 1D domain, such as one of those defined in Section~\ref{sec:1DDomain}, the next task is to solve the discretised OZ equation~\eqref{eqn:ozbulk_discrete}. 
Previous studies have employed Picard iteration (either directly \cite{broyles} or incorporating information from several previous iterations \cite{ng}), Newton's method 
\cite{zerah} or Picard/Newton hybrids \cite{labik,gillan}.
In the following we describe three different iterative methods -- Picard, Newton and a promising new approach --  noting that this is far from an exhaustive list of options.

Suppose that we have initial guesses, $\nvec{h}_0$ and $\nvec{c}_0$, for $\nvec{h}$ and $\nvec{c}$, respectively, defined on the discretized domain.  It is worth noting that there are two situations we are interested in:
\begin{enumerate}
    \item $\nvec{h}_0$ and $\nvec{c}_0$ are `arbitrary', perhaps including some knowledge of the relevant physics, but we do not expect them to be particularly close to the exact solution.
    \item $\nvec{h}_0$ and $\nvec{c}_0$ are close to the exact solution.  This may arise, e.g., because we are using continuation in a parameter in which the solution changes smoothly, or in time-dependent problems where we expect $h$ and $c$ to vary only a small amount at the next time step.
\end{enumerate}
For the first case we require a robust approach (Picard iteration, which is very standard in the literature) that is relatively insensitive to the initial guess.  For the second, we will present a scheme which, in cases where we have a good initial guess, has proven to be significantly faster (which we call \emph{implicit linearised Picard} (LIP), and believe to be novel). We note that although the LIP scheme is motivated by starting close to the correct solution, in practice we have found it to be highly effective for a wide range of initial guesses.

For a general, self-consistent equation in a vector $\y$,
$
    \y = F(\y),
$
Picard iteration (with a `mixing parameter' $\alpha$) is an iterative scheme.  Given the current approximation to a solution at step $n$, denoted by $\y^n$,  the next approximation is constructed through:
\begin{enumerate}
    \item Construct $\y^{\rm{new}}$ which corresponds to $F$ evaluated at $\y^n$:
    \[
        \y^{\rm{new}} = F(\y^n).
    \]
    \item Construct $\y^{n+1}$ by mixing $\y^n$ and $\y^{\rm{new}}$:
    \[
        \y^{n+1} = (1-\alpha) \y^n + \alpha \y^{\rm{new}}.
    \]
\end{enumerate}
A typical Picard scheme starts with an initial guess $\y^0$ and a given, fixed, mixing parameter, $\alpha$.  It then iterates the above steps until either $\y^{n+1}$ and $\y^{n}$ differ by less than a given tolerance (measured in some norm), or a maximum number of iteration steps has been reached.
Some care needs to be taken in choosing $\alpha$.  If it is chosen too large then the Picard scheme may not converge, while if it is chosen too small then the number of iterations required to converge to a given tolerance may be prohibitively large.

To motivate our second method (linearised iterative Picard) we restrict ourselves to a self-consistent equation in two variables, however the method can be generalised:
\begin{equation} \label{eqn:y1Picard}
\begin{aligned}
	\y_1 &= F_1(\y_1,\y_2), \\
	\y_2 &= F_2(\y_1,\y_2).
\end{aligned}
\end{equation}
Note that the discretised OZ equation~\eqref{eqn:ozbulk_discrete} can be written in this form, where $\y_1 = \nvec{h}$ and $\y_2 = \nvec{c}$. We may write a corresponding set of ODEs by considering $\y_i$ to be a function of time:
\begin{equation} \label{eqn:bilinearODE1}
\begin{aligned}
	\dv{\y_1}{t} &= F_1(\y_1,\y_2) - \y_1 \coloneq R_1(\y_1,\y_2), \\
	\dv{\y_2}{t} &= F_2(\y_1,\y_2) - \y_2 \coloneq R_2(\y_1,\y_2).
\end{aligned}
\end{equation}
In particular,~\eqref{eqn:y1Picard} is satisfied by $\y_1$ and $\y_2$ if and only if these are also equilibrium solutions of~\eqref{eqn:bilinearODE1}, nullifying the residuals $R_i$. We now discretise~\eqref{eqn:bilinearODE1} with explicit Euler in time, and consider a single step,
\begin{equation} \label{eqn:eulerpicard}
\begin{aligned}
	\y_1' - \y_1 &=   \dt R_1(\y_1,\y_2), \\
	\y_2' - \y_2 &=   \dt R_2(\y_1,\y_2),
\end{aligned}
\end{equation}
where $\y_i'$ denotes an update of $\y_i$, and $\dt$ is the step size.  Taking $\dt = \alpha$, this is precisely equivalent to the Picard scheme with mixing parameter $\alpha$.  We note the usual stability issues of explicit Euler~\cite{Iserles2009Book}, which may require a small step size; correspondingly, the Picard approach will require a small mixing parameter.

Motivated by this, one may think to apply a more robust numerical scheme in time, such as implicit Euler.  However, this results in another set of nonlinear, implicit equations which must be solved, namely
\begin{equation}
\begin{aligned}
	\y_1' - \y_1 &=   \dt R_1(\y_1',\y_2'), \\
	\y_2' - \y_2 &=   \dt R_2(\y_1',\y_2').
\end{aligned}
\end{equation}
If $\y_1$ and $\y_2$ are sufficiently close to a fixed point of \eqref{eqn:bilinearODE1}, then we may linearise the right-hand side.
To be precise, we assume that the residuals $R_i$ have a particular form:
\begin{equation}
\begin{aligned}
    R_1(\y_1,\y_2) &= f_1(\y_2) + P_1(\y_1,\y_2), \\
	R_2(\y_1,\y_2) &= f_2(\y_2) + P_2(\y_1,\y_2),
\end{aligned}
\end{equation}
where $P_1$ is linear in its first argument, $P_2$ in its second, and $f_i$ are arbitrary functions.
We choose to linearise the above by replacing $\y_2'$ with $\y_2$ in $R_1$, and $\y_1'$ with $\y_1$ in $R_2$, noting that other linearisation choices are possible.
This leads to $P_1(\y_1',\y_2')\approx \tilde{P}_1[\y_2]\y_1'$ and $P_2(\y_1',\y_2')\approx \tilde{P}_2[\y_1]\y_2'$, where e.g. $\tilde{P}_1[\y_2]$ is a linear operator acting to its right, parametrised by $\y_2$.
We therefore obtain
\begin{equation} \label{eqn:lipform1}
	\begin{aligned}
		\left[\dt \tilde{P}_1[\y_2] - \id\right] \y_1' &= -\y_1 -\dt f_1(\y_2),\\
		\left[\dt \tilde{P}_2[\y_1] - \id\right] \y_2' &= -\y_2 -\dt f_2(\y_1),
	\end{aligned}
\end{equation}
or equivalently
\begin{equation} \label{eqn:lipform2}
	\begin{aligned}
		\left[\dt \tilde{P}_1[\y_2] - \id\right](\y_1' - \y_1) &= 
		-\dt R_1(\y_1,\y_2),	\\
		\left[\dt \tilde{P}_2[\y_1] - \id\right](\y_2' - \y_2) &= 
		-\dt R_2(\y_1,\y_2),
	\end{aligned}
\end{equation}
where $\id$ is the identity operator. Due to our assumptions, these are linear equations in $\y_i'$ and we may solve them using any suitable algorithm, such as Cholesky or LU decomposition~\cite{Trefethen2022Book}.  
The solutions for $\y_i'$ become the next approximation in the iterative scheme, and once again setting $\dt = \alpha$ makes the analogy to Picard iteration clearer.

In our case, we can express the OZ relation as
\begin{equation} \label{eqn:system_discrete}
	\nvec{h} - \nvec{c} - \rhotch = 0, 
\end{equation}
where e.g., $\nvec{h}$ is the vector of collocated values of the correlation function $h$ and we use the notation $\nvec{A}\times_n \nvec{b}$ to denote a tensor contraction between an order-$N$ tensor $\nvec{A}$ (with $N \geq n$) and a vector $\nvec{b}$ such that $(\nvec{A} \times_n \nvec{b})_{i_1 \dots i_{n-1} i_{n+1} \dots i_N} = \sum_{i_n} \nvec{A}_{i_1 \dots i_N} \nvec{b}_{i_n}$.

To place this in fixed-point form as above, we need only set $\y_1=\nvec{c}$, $\y_2=\nvec{h}$, $P_1(\nvec{c},\nvec{h}) = -\nvec{c}-\rhotch$, $P_2(\nvec{c},\nvec{h})=-\nvec{h}+\rhotch$, and $f_i$ both as the identity function.
Making the substitution into \eqref{eqn:eulerpicard}, our Picard iteration then reads
\begin{equation} \label{eqn:picard}
	\begin{aligned}
		\nvec{c}'-\nvec{c} &= \alpha\left(
		\nvec{h} - \nvec{c} - \rhotch\right), \\
		\nvec{h}'-\nvec{h} &= \alpha\left(
		\nvec{c} - \nvec{h} + \rhotch \right).
	\end{aligned}
\end{equation}
As in \eqref{eqn:ozwithdomains}, the above equalities each hold on only one side of two subdomains $r\in[0,d]$ and $[d,\infty]$, with the closure holding elsewhere in order to make-up a fully determined system.
For ease of presentation, we have left this implicit.

When it comes to the implicit linearised Picard scheme \eqref{eqn:lipform2}, we are aided by the fact that $\rhotch$ is linear in both $\nvec{c}$ and $\nvec{h}$, so we can relax some of the assumptions about proximity to a fixed point.
One update of the LIP scheme then reads
\begin{equation}
	\label{eqn:lip}
	\begin{aligned}
		\left[\alpha\rho \nvec{T}\times_3 \nvec{h} + (1+\alpha)\id\right](\nvec{c}'-\nvec{c})
		&= \alpha \left(\nvec{h} - \nvec{c} - \rhotch\right), \\
		\left[\alpha\rho \nvec{T}\times_2 \nvec{c} - (1+\alpha)\id\right](\nvec{h}'-\nvec{h})
		&= \alpha \left(\nvec{h} - \nvec{c} - \rhotch\right),
	\end{aligned}
\end{equation}
where each of the operators on the left-hand side will be recognised as a matrix, so that each iteration involves the solution of two systems of linear equations.

Finally, it is certainly possible to solve \eqref{eqn:system_discrete} with a canonical Newton method, and in fact the Jacobian is trivial due to the bilinearity of the OZ and the locality of the closure $G$,
\begin{equation}\label{eqn:newton}
	\begin{aligned}
		(\rho \nvec{T}\times_3 \nvec{h} + \id)(\nvec{c}'-\nvec{c}) +
		(\rho \nvec{T}\times_2 \nvec{c} - \id)(\nvec{h}'-\nvec{h}) &= 
		- \rhotch, \\
		\nabla_\nvec{c} G(\nvec{c},\nvec{h})(\nvec{c}'-\nvec{c}) +
		\nabla_\nvec{h} G(\nvec{c},\nvec{h})(\nvec{h}'-\nvec{h}) &=
		G(\nvec{c},\nvec{h}),
	\end{aligned}
\end{equation}
which results in a linear system that can be solved for $\nvec{c}'$ and $\nvec{h}'$.
We present this here to indicate the ease with which one might implement a familiar method, but as we shall explain in Section \ref{sec:validation_iterative}, we did not find it suitable for the particular investigations in this work.

\section{Numerical validation} \label{sec:validation}

Having presented all aspects of our numerical method, we are now in a position to validate and test its performance.
We begin by explaining the benchmarks to which we can compare, and appropriate error metrics. 
This is followed by an investigation of the geometric parameters described in Section \ref{sec:discretisation}, and of the iterative solution schemes presented in the previous section.
We conclude the performance testing by demonstrating the need for infinite domains in situations where the correlation functions are known to be long-ranged.

\subsection{Initial comparison with MSA+HCY analytic solution}

To begin with, we demonstrate the agreement between our numerical solution of the OZ equation for a model with a HCY interparticle potential using the MSA closure \eqref{eqn:msa} (MSA+HCY) in 3D, which has a known exact solution due to Waisman \cite{Waisman1973,Hoye1977}.
Explicitly, this potential is of the form
\begin{align}\label{HCY potential}
	\phi(r) = 
	\begin{cases} 
		\infty, & r < d \\
		-\frac{\beta}{r}e^{-z(r-d)}, & r\ge d
	\end{cases},
\end{align}
where $\beta$ is the usual inverse temperature, and we fix the decay-constant $z$ equal to 2 throughout this work.
The nature of the analytical solution \cite{Kahl1989,Cummings1979,Cummings1979a} to this model is such that obtaining explicit values of the pair correlation function at large separations is prohibitively tedious.
As such, we make a pointwise comparison of numerical and analytic solutions up to $r=6$, for three representative test cases in the free parameters density $\rho$ and inverse temperature $\beta$.
As an unremarkable case, which we shall refer to as case A, we take a low density and high temperature $\rho=0.316, \beta=0.2$.
For a high-density case, B, we take $\rho=0.7$ and $\beta=1.113$. 
Thirdly, we take a case C with $\rho=0.316$ and $\beta=1.113$ which, as we will demonstrate later in this paper, is close to the critical point in the density-temperature plane for this model. 
These cases are summarised in table \ref{tab:testcases} for later reference.

\begin{table}
	\centering
	\sisetup{table-text-alignment=left}
	\begin{tabular}{ll
    		S[table-number-alignment=left, table-format=1.3]
			S[table-number-alignment=left, table-format=1.3]
            }
		Name & Description & {Density} & {Inverse temperature} \\
		\hline
		A & Typical case & 0.316 & 0.2 \\
		B & High density & 0.7 & 1.113 \\
		C & Near-critical & 0.316 & 1.113 \\
		\hline
	\end{tabular}
	\caption{The three test cases used throughout Section \ref{sec:validation}.}
	\label{tab:testcases}
\end{table}

\begin{figure}
	\centering
	\includegraphics{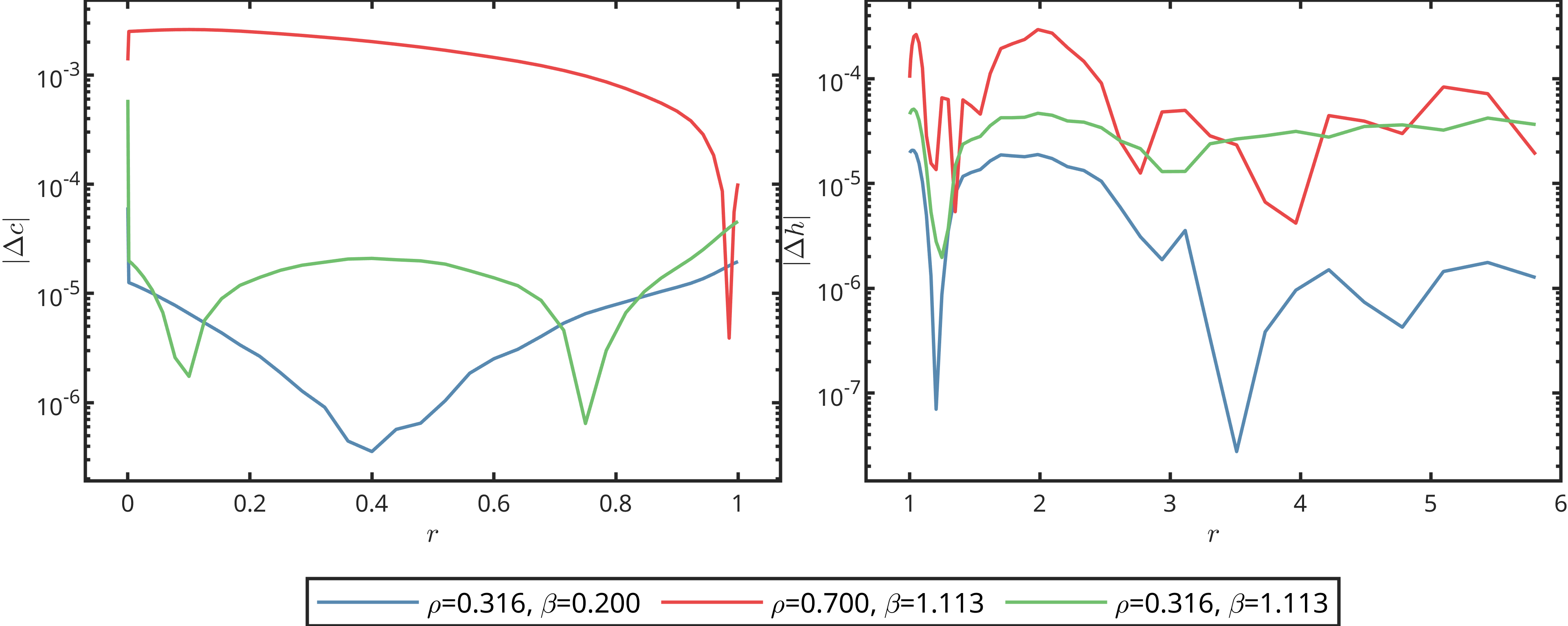}
	\caption{Absolute pointwise error versus the Waisman exact solution for the three test cases, A in blue, B in red, C in green.}
	\label{fig:pointwise_error}
\end{figure}

The results of this exercise, in figure \ref{fig:pointwise_error}, confirm that our method is accurate for the range of separations considered, although close to the critical point the pair correlation function becomes increasingly long-ranged and pointwise comparison for $r \leq 6$ is insufficient for complete validation.
As an alternative, we can create a pure test of the integration and interpolation procedures we use to build the tensor by measuring the residuals of the discrete system of equations \eqref{eqn:ozbulk_discrete} with the exact solution substituted in.
Using the bilinearity of the OZ relation, as in Section \ref{sec:iterative}, we calculate the pair correlation function $\nvec{h}$ consistent with the exact direct correlation function $\nvec{c}$, over the whole domain $[0,\infty]$, then compute the normed residual over the interval $[0,d]$, where the closure $\nvec{h}=-1$ applies.
We call this error measure of our numerical machinery the \emph{exact consistency} in what follows.

As an alternative that can measure the quality of our iterative schemes and avoid the truncation of the exact $h(r)$, we can use a detail of the Waisman exact solution to the MSA+HCY model.
The reduced compressibility of the solution
\begin{equation}
	\ichired = \left(\pdv{\beta p}{\rho}\right)_\beta
	\label{eqn:redcompressibility}
\end{equation}
at any point $(\rho,\beta)$ is in some sense a primary quantity of the analytical solution process, and thus independent of the evaluation of $h(r)$ at long range.
This thermodynamic quantity can be related to the direct correlation function (using an integration measure appropriate for three dimensions) via
\begin{equation} 
	\label{eqn:compressibility_direct}
	\ichired = 1 - \rho \int_0^\infty \dd{r} 4 \pi r^2 c(r)
\end{equation}
and to the pair correlation function via
\begin{equation}
	\label{eqn:compressibility_pair}
	\ichired = \left(1 + \rho \int_0^\infty \dd{r} 4\pi r^2 h(r) \right)^{-1}
\end{equation}
and so it is a convenient scalar measure of the global accuracy of our numerical solutions.
We shall henceforth refer to the relative difference in the inverse compressibility as the \emph{exact error}.
For the three cases shown in figure \ref{fig:pointwise_error}, the inverse compressibility and the exact error are shown in table \ref{tab:introresults}. 
We see that the inverse compressibility is significantly reduced for case C, which is close to the critical point, and that this also leads to the largest relative error, despite the fact that the residuals shown in figure \ref{fig:pointwise_error} are generally larger for case B, at high density.
Case A, at low density and high temperature, illustrates that the method is very accurate for such an `easy' case.

\begin{table}
	\centering
	\sisetup{table-text-alignment=left}
	\begin{tabular}{l
			S[table-number-alignment=left, table-format=1.3]
			S[table-number-alignment=left, table-format=1.3]
			S[table-number-alignment=center, table-format=1.1e1]
			S[table-number-alignment=center, table-format=1.1e1]
			}
			
		Case & {Density} & {Inverse temperature} & {$\ichired$} & {$\abs*{\Delta\ichired}/\ichired$} \\
		\hline
		A & 0.316 	& 0.2 	& 2.9		& 3.6e-6 \\
		B & 0.7 	& 1.113 & 7.4		& 3.7e-4 \\
		C & 0.316 	& 1.113 & 8.9e-5	& 7.6e-2 \\
		\hline
	\end{tabular}
	\caption{Numerical errors for the compressibility and correlation functions, which are shown in figure \ref{fig:pointwise_error}.  Here, $\abs*{\Delta\ichired}/\ichired$ is the relative error in the inverse compressibility, which we refer to as the \emph{exact error} for brevity.}
	\label{tab:introresults}
\end{table}

\subsection{Collocation domain parameters}

The collocation of the correlation functions on the one-dimensional interval $[0,\infty]$, and onto the four subdomains of $\mathbb{R}^D$ described in Section \ref{sec:discretisation}, leads to four free numerical parameters that we can vary.
There are $\nod$ points used to discretise $[0,d]$ with an affine mapping of the Chebyshev-Gauss-Lobatto (CGL) points \eqref{eqn:cglpoints}, and $\ndl$ used to discretise $[d,\infty]$ with a quotient map \eqref{eqn:quotient} of the CGL points.
The map introduces a further free parameter $\ldinf$ (which corresponds to $L_Q$ in \eqref{eqn:quotient}) used to set the concentration of the grid points.

In order to reduce the number of parameters, each of the subdomains ($\clg,\cll,\cgl,\cgg$) are discretised with a small integer multiple (typically, 1, 2, or 3) of $\ndl$ points.  The results are insensitive to the exact choices for the number of points used.

\begin{figure}
    \centering
	\includegraphics{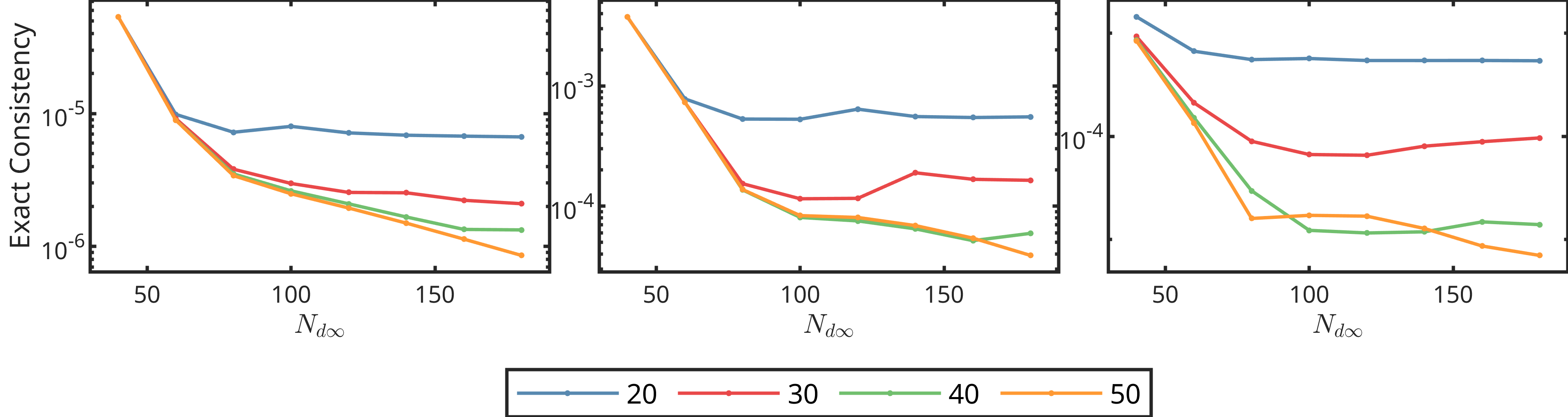}
    \caption{Convergence of the consistency in the number of grid points, both $\ndl$ (x-axis) and $\ndisc$ (colour). 
    	Shown from right to left are the three test cases A, B and C.
    	Note that varying $\nod$ in the range from 20 to 50 was found to have no significant effect; we have set it at 40 here.
    	Meanwhile, $\ldinf$ is fixed to 6. 
    	The variation in the same dataset along the $\ldinf$ direction is shown in figure \ref{fig:ldinf_convergence}.}
    \label{fig:n0d_ndisc_convergence}
\end{figure}

\begin{figure}
	\centering
\includegraphics{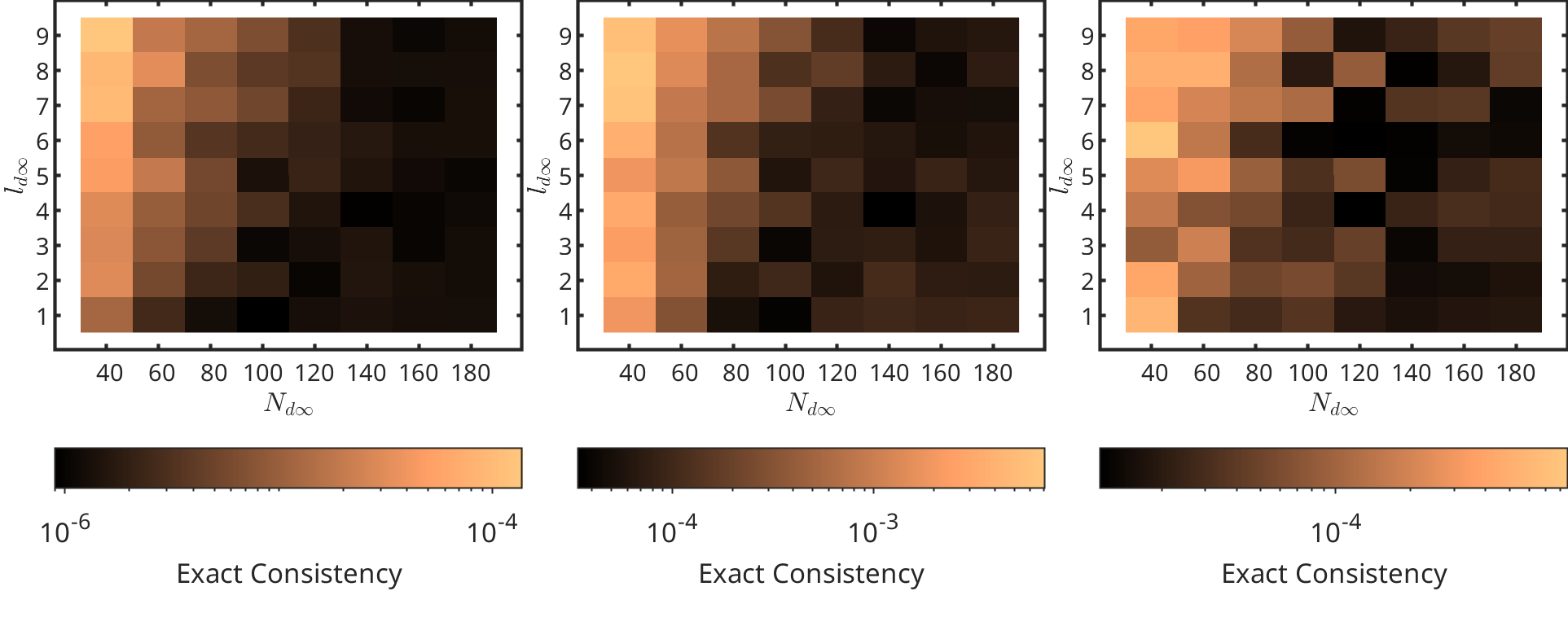}
	\caption{
		The consistency of the Waisman exact solution under our numerical system of equations, as it varies with the number of points, and the exact mapping of them in $[d,\infty]$, for the three test cases (A) left, (B) centre, (C) right. Note the difference in colour scales.
		Observe that one can obtain slightly better results with the fewest points by picking the concentration length to be between 4 and 6. 
		With more grid resolution, reasonable choices for this parameter barely affect the result.}
	\label{fig:ldinf_convergence}
\end{figure}

In practice, we have found that choosing $\nod$ to be anything higher than around 20 is sufficient in all cases. 
This is unsurprising when one inspects the general form of the direct correlation function on $[0,d]$, indeed it is smooth and monotonic, and thus easily-represented by a high-order polynomial basis.
The variation in accuracy of the three cases of interest for a range of $\ndl$ and $\ndisc$ is shown in figure \ref{fig:n0d_ndisc_convergence}, and for a range of $\ndl$ and $\ldinf$ in figure \ref{fig:ldinf_convergence}.
The main conclusions are as follows:
First, the number of points $\ndl$ used to represent the tail of the pair correlation function $h(r)$ is the greatest determinant of the numerical error; in this work we avoid using any fewer than 80 points.
Second, the discretisation of the $\mathbb{R}^D$ subdomains is most impactful for long-tailed critical solutions. Increasing $\ndisc$ entails only a one-time startup cost, as explained in Section \ref{sec:tensor}, therefore we set it to at least 40 throughout this work.
Finally, the effect of the quotient map parameter $\ldinf$ is noticeable but has no obvious pattern. 
Values of 4 to 6 appear marginally better than others tested, we thus empirically choose to remain in that range.

\subsection{Iterative schemes} \label{sec:validation_iterative}

To test our hypothesis proposed in Section \ref{sec:iterative}, that an implicit fixed-point iteration can be more stable than standard Picard iteration, we solved for the three test cases with a range of mixing parameters, which we consider to be analogous to the time step of a differential equation integrator.
Both the Picard iteration and our linearised implicit Picard method were started with the same initial condition, which we shall briefly describe for completeness.
Our aims in choosing an initial condition for the correlation functions were simplicity, and a desire not to bias the solver toward unphysical solutions, for example those with negative compressibility, $\chired$.
To this end, we pick both correlation functions to be exponentially-decaying with unit length-scale,
\begin{equation}
	\begin{aligned}
		c(r) = c_0 e^{d-r} & \quad r<d, \\
		h(r) = -c_0 e^{d-r} & \quad r>d,
    \end{aligned}
	\label{eqn:ic}
\end{equation}
and fix $c_0<0$ for a given density and temperature so that the initial compressibility $\ichired=1+2b\rho$ for some positive constant $b$. 
This is intended to match the high-temperature, low-density form from the van der Waals equation of state.

\begin{figure}
	\centering
\includegraphics{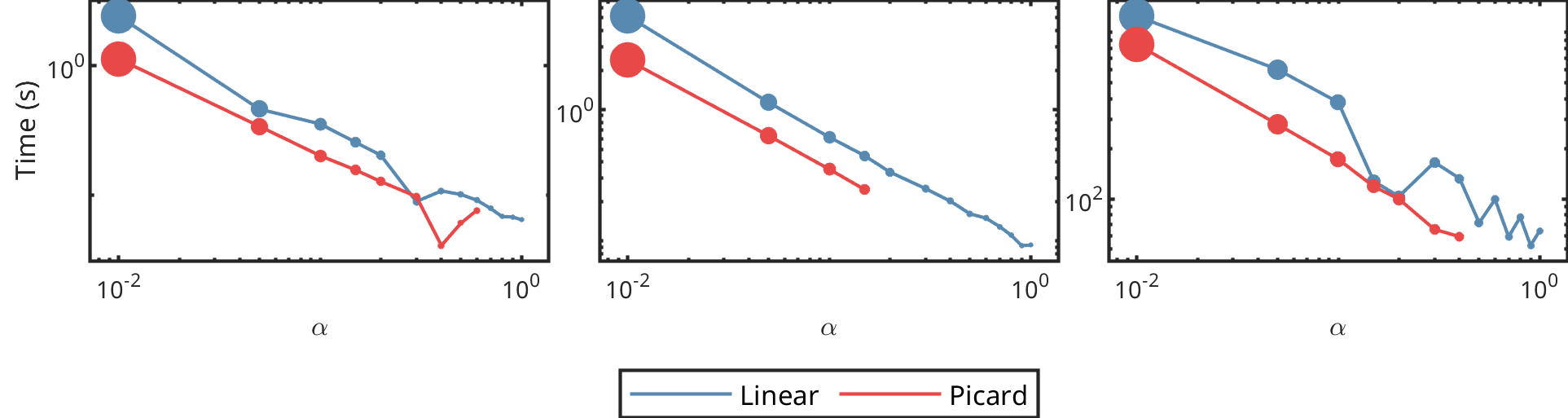}
	\caption{Convergence times of the Picard and linearised implicit Picard for the three test cases (A) left, (B) centre, (C) right.
		Cases where the solver converged to an incorrect solution are blanked-out.
		The chief conclusions are as follows.
		The LIP scheme is about half as fast as standard Picard with the same mixing parameter, but is  stable for much higher mixing parameters, and in some cases such as high densities, this allows it to out perform Picard.
		In all plots, the size of each dot indicates the number of iterations needed to converge to a tolerance of \num{1e-10}, relative only to data on the same plot.
		Note that the ideal scaling of time-to-convergence with mixing parameter is $t\propto 1/\alpha$.
		%Anderson acceleration is very often the fastest scheme, but it is not robust and there is little discernible pattern to which combinations of mixing parameter and window length lead to a correct solution.
		The tests shown here used a numerical domain characterised by $\nod=40,\ndl=80,\ndisc=40,\ldinf=5$.
	}
	\label{fig:iterconvergence}
\end{figure}

The results of this test are shown in figure \ref{fig:iterconvergence}. 
In general, in situations where both methods converge, we found that they do so in approximately the same number of iterations, so we plot only the time to convergence.
Typically in testing, one iteration of the LIP solver takes twice the time of a Picard update. 
Inspecting \eqref{eqn:picard} and \eqref{eqn:lip}, the reason for this is quite clear: the computational effort is dominated by the order-3, order-1 tensor contraction, with complexity $\mathcal{O}((\nod+\ndl)^3)$.
In the Picard case \eqref{eqn:picard}, since we are free to rearrange the order of the double contraction $(\nvec{T}\times_3\nvec{h})\times_2\nvec{c}$, we need only to perform a single order-3, order-1 contraction, whereas in the LIP case \eqref{eqn:lip}, we must calculate both $\nvec{T}\times_3\nvec{h}$ and $\nvec{T}\times_3\nvec{c}$, doubling the effort. 
Nonetheless, the results in the figure demonstrate that LIP is more stable than Picard with respect to increased mixing (step-size). 
So, especially at high densities, this can allow a LIP solver to outperform standard Picard.
The increased stability of LIP also removes the need to hand-tune the mixing parameter for a particular problem, which can be a time-consuming but necessary element of Picard iteration: the cost of not doing so is a failure to converge, or wasteful computation.
Given these promising results, we feel implicit techniques for linearisable fixed-point problems, and the analogy with differential equation integration is an interesting avenue for future study.

It should be mentioned also that, in the preparation of this work, we experimented with Newton and other gradient-based iterative solvers, and with acceleration schemes for fixed-point iterations such as Anderson acceleration \cite{Saad2025}.
While these methods often yielded a notable speed-up, we found they frequently converged to spurious, obviously unphysical solutions.
Given the high level of accuracy required for investigations near the critical point, the robustness of standard Picard iteration was more highly valued for our purposes here.

\subsection{Finite domains}

\begin{figure}
	\centering
\includegraphics{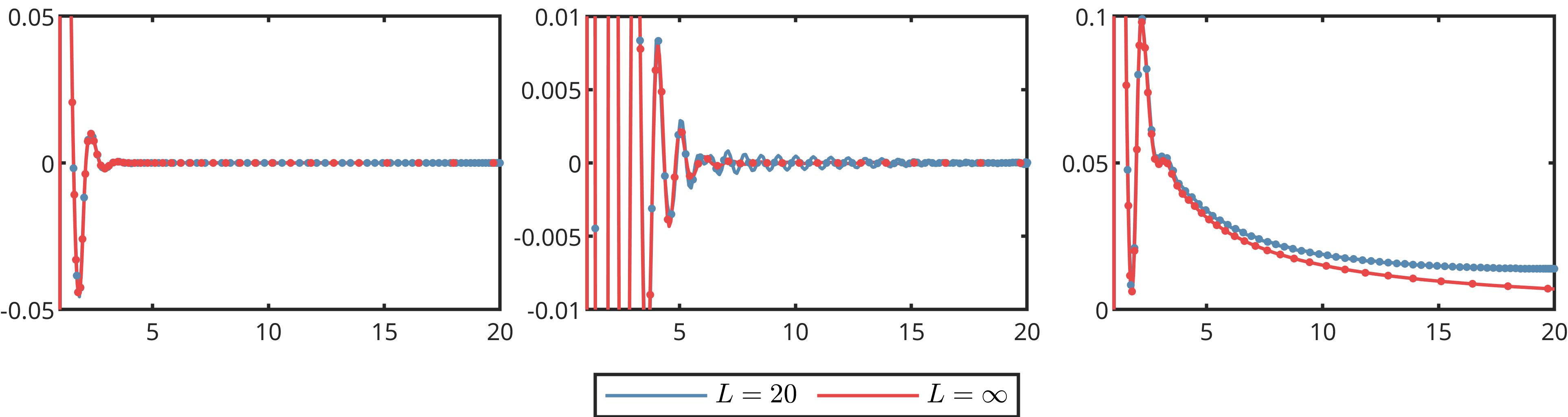}
	\caption{Example profiles of the correlation function $h(r)$ for both a finite and an infinite domain, in blue and red, respectively.
		The agreement for the low density, high temperature case A (left) is undeniable, whereas in the critical case C (right), the long-ranged nature of the physical solution is incompatible with a finite domain, yielding clearly inconsistent results.
		For the high-density case B (centre), the oscillatory correlation function has a relatively large gradient (if not magnitude) at $r=L$, which leads to numerical oscillations in the polynomial interpolant, that one may think of as artificial, back-mirrored packing oscillations.
		Both tests shown were conducted with $\ndl=80$ and $\ndisc=40$. The mapping parameter for the infinite case was $\ldinf=5$.
        }
	\label{fig:finiteprofiles}
\end{figure}

\begin{figure}
	\centering
\includegraphics{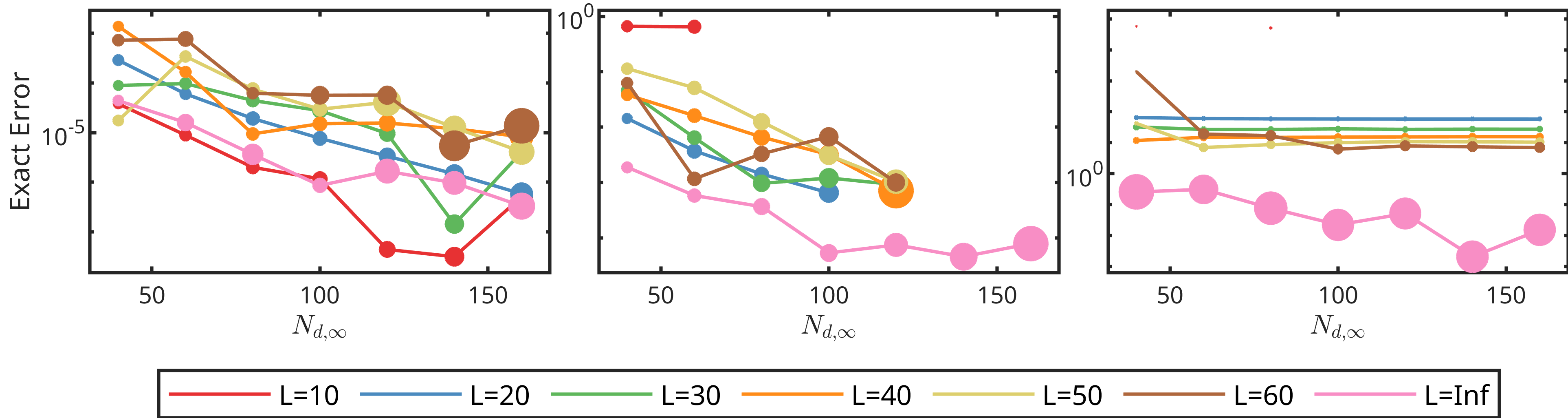}
	\caption{Quality of the solution for test cases on finite domains of increasing extent, and with added grid points. 
		Shown from right to left are the three test cases A, B and C.
		One sees that in a typical case like A all choices perform well, with $L=10$ and $L=\infty$ yielding the best results. 
		In the high-density case B, $L=10$ is insufficient to capture the nature of the solution, and all the finite $L$ calculations fail to converge beyond some number of grid points. 
		Our hypothesis is that this is due to the oscillatory nature of the high-density solution being both incompatible with finite domains, and better-represented with finer grids.
		The final plot demonstrates that none of the finite calculations are adequate to represent solutions near the critical point.
		This is an effect that improves a little with increasingly large finite domains, but is unchanged by resolution.
		Note that varying the mapping parameters described in \ref{sec:1DMaps} had only a minor effect.}
	\label{fig:finitel}
\end{figure}

An important distinction of our method compared to previous work is the freedom to use an infinite domain.
If the domain is truncated, say at $r=L$, one needs to make an assumption about the behaviour of $h$ beyond that point, for $r>L$, as a boundary condition.
As a neutral choice, to examine the effect of truncating the correlation functions, we assume that $h(r)$ is constant for $r>L$. 
Representative solution profiles for the three test cases on an infinite domain and a finite domain with $L=20$ are shown in figure \ref{fig:finiteprofiles}.
One can see that the assumption of constant $h$ is sound for low density and high temperature (case A on the left), where it decays quickly to zero, but totally incorrect near the critical point (case C on the right).
For high density, represented by case B (centre), the magnitude of the correlation can be small but with oscillations persisting to large separations, so the boundary condition is poor in this case too.
The non-zero gradient requires an oscillatory polynomial to satisfy the boundary condition in addition to interpolating the collocated solution.
These oscillations can be thought of as an artifact of a wall in the two-body, but not the one-body domain (which ought to be identical in bulk), leading to packing. 
Figure \ref{fig:finitel} substantiates this for a wider range of domain sizes and resolutions, and reinforces that an infinite domain is crucial for calculations near the critical point, unless one has access to a more-informed boundary condition \cite{Belloni1993}.

\section{Critical properties of the MSA for the HCY model}

\subsection{Review} \label{sec:exponentreview}

At the critical point the total correlation function develops a slow, power-law decay for large values of $r$, characterized by a universal critical 
exponent $\eta$, according to
\begin{equation}
h(r)\thicksim r^{-(D-2)-\eta},    
\end{equation}
where the density and temperature are fixed to their values at the critical point, $\rho=\rho_{\text{crit}}$ and $T=T_{\text{crit}}$.
In addition to this structural exponent we will consider two further thermodynamic exponents, $\gamma$ and $\delta$, characterizing the divergence of the reduced isothermal compressibility. These are defined according to the relations
\begin{align}
\label{eqn:chiexponents}
\chi_{\text{red}} \thicksim
  \begin{cases}
    |t|^{-\gamma}, & \rho=\rho_{\text{crit}},\\
    |\Delta \rho|^{1-\delta}, & T=T_{\text{crit}},
  \end{cases}
\end{align}
where the (dimensionless) reduced isothermal compressibility is given by $\chi_{\text{red}}=k_{\text{B}}T\rho \chi_{T}$ and where $t=T/T_{\text{crit}}-1$ and $\Delta \rho=\rho/\rho_{\text{crit}}-1$. 
The density difference $\Delta\rho$ is the (scaled) difference between the coexisting liquid and gas densities in the phase-separated fluid and serves as an order-parameter for the phase transition. 
Within mean-field theory these exponents take the values $\eta=0$, $\gamma=1$ and 
$\delta=3$, independently of the spatial dimension. 
In 3D the MSA predicts $\eta=0$, $\gamma=2$ and 
$\delta=5$ and the exact Ising model values are $\eta=0.036$, $\gamma=1.237$ and 
$\delta=4.78$. 

The critical exponents are not all independent quantities. 
Indeed, the Widom scaling relation,
\begin{equation}\label{widom}
\beta = \gamma/(\delta -1),   
\end{equation}
allows us to predict the value of the coexistence curve 
exponent $\beta$ from knowledge of $\gamma$ and $\delta$.
{Therefore the exponent $\beta$ provides a connection between $t$ and $\Delta \rho$, according to
\begin{equation}\label{beta_def}
 t \sim \Delta \rho^{1/\beta}.
\end{equation}
This is convenient, since 
accurate numerical determination of the curvature of the binodal at the critical point 
is a very demanding task. (Note that sufficiently close to the critical point the curvature of the spinodal will match that of the binodal, so the exponent $\beta$ could potentially be useful in fitting the spinodal curve around the critical point.)
In 3D the Widom relation predicts the following values for $\beta$: $1/2$ (mean-field), 
$1/2$ (MSA) and $0.327$ (Ising). 

Another relation between the critical exponents is the Josephson hyperscaling relation,
\begin{equation}\label{josephson}
\eta = 2 - D\frac{\delta-1}{\delta+1}.    
\end{equation}
In contrast to the Widom relation, this involves explicitly the dimensionality, $D$, of the system. It implies that relation \eqref{josephson} will generally not hold for mean-field theory, since its predictions for the critical exponents are independent of $D$.  
The hyperscaling relation is valid for mean-field theory only at the upper critical dimension $D=4$. 
As a quick check to show that \eqref{josephson} applies to the exact exponents: inputting the 3D Ising value $\delta=4.78$ into \eqref{josephson} yields $\eta=0.038$, consistent with expectations. 
Similarly, inputting $\delta=5$ into equation \eqref{josephson} yields $\eta=0$, as expected, thus demonstrating that the MSA also satisfies hyperscaling.
(Note that using the mean-field exponent $\delta=3$ in \eqref{josephson} leads to the inconsistent result $\eta=0.5\ne 0$.)

In 2D the mean-field exponents remain unchanged and the MSA values (which are presently unknown) will be determined in Section \ref{sec:exponents2d}. The exact 2D Ising exponents are $\eta=1/4$, $\gamma=7/4$, $\delta=15$ and $\beta=1/8$.
We note that the exact $\delta$-exponent deviates very significantly from its mean-field value (i.e. 15 vs. 3).

In numerical work it is standard practice to define the \textit{effective} exponents, according to
\begin{align}
\delta_{\text{eff}} &= - \frac{d \ln\left( \chi_{\text{red}} \right)}{d \ln\left( \Delta \rho \right)} +1, \label{eqn:delta_eff} \\
\gamma_{\text{eff}} &= - \frac{d \ln\left( \chi_{\text{red}} \right)}{d \ln\left( t \right)}, \label{eqn:gammaeff} \\
\eta_{\text{eff}} &= - \frac{d \ln\left( h(r) \right)}{d \ln\left( r \right)} -D +2. \label{eqn:etaeff}
\end{align}
As the thermodynamic parameters are tuned towards the critical point the effective exponents eventually plateau to their true critical values. 
An additional benefit of studying the effective exponents is that it enables identification of the extent of the `critical region' about the critical point, namely the 
zone of thermodynamic parameter-space within which the correct asymptotic behavior sets in.

\subsection{Critical exponent determination} \label{sec:exponents3d}

\begin{figure}
	\centering
\includegraphics{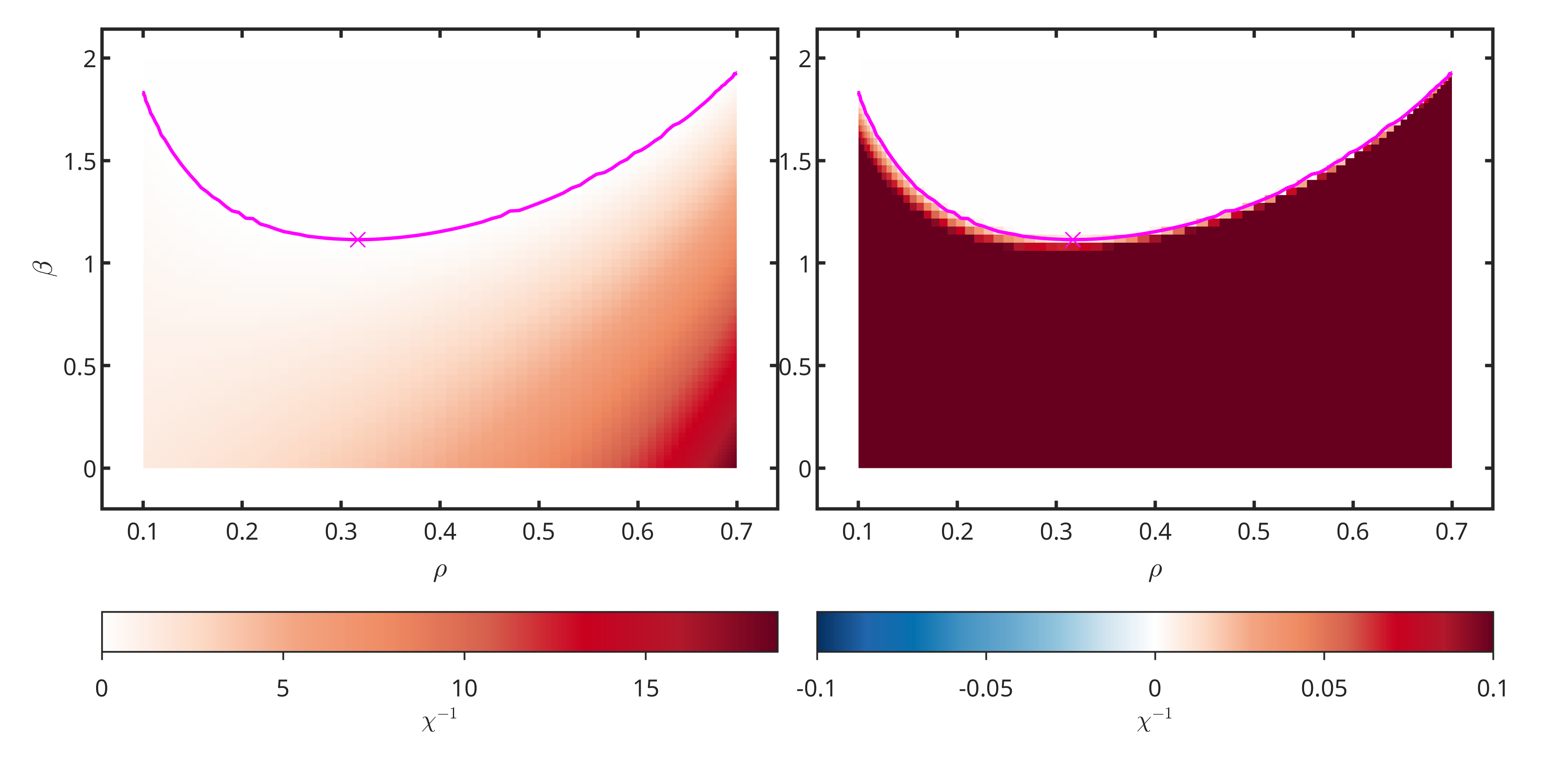}
	\caption{The inverse compressibility of the MSA+HCY model in 3D, over a subset of the density-temperature plane, including the critical point.
	In the right panel, we clip the colour range of the plot around $\ichired=0$, to highlight the spinodal curve.
    }
	\label{fig:ichi3d}
\end{figure}

\begin{figure}
	\centering
	\includegraphics{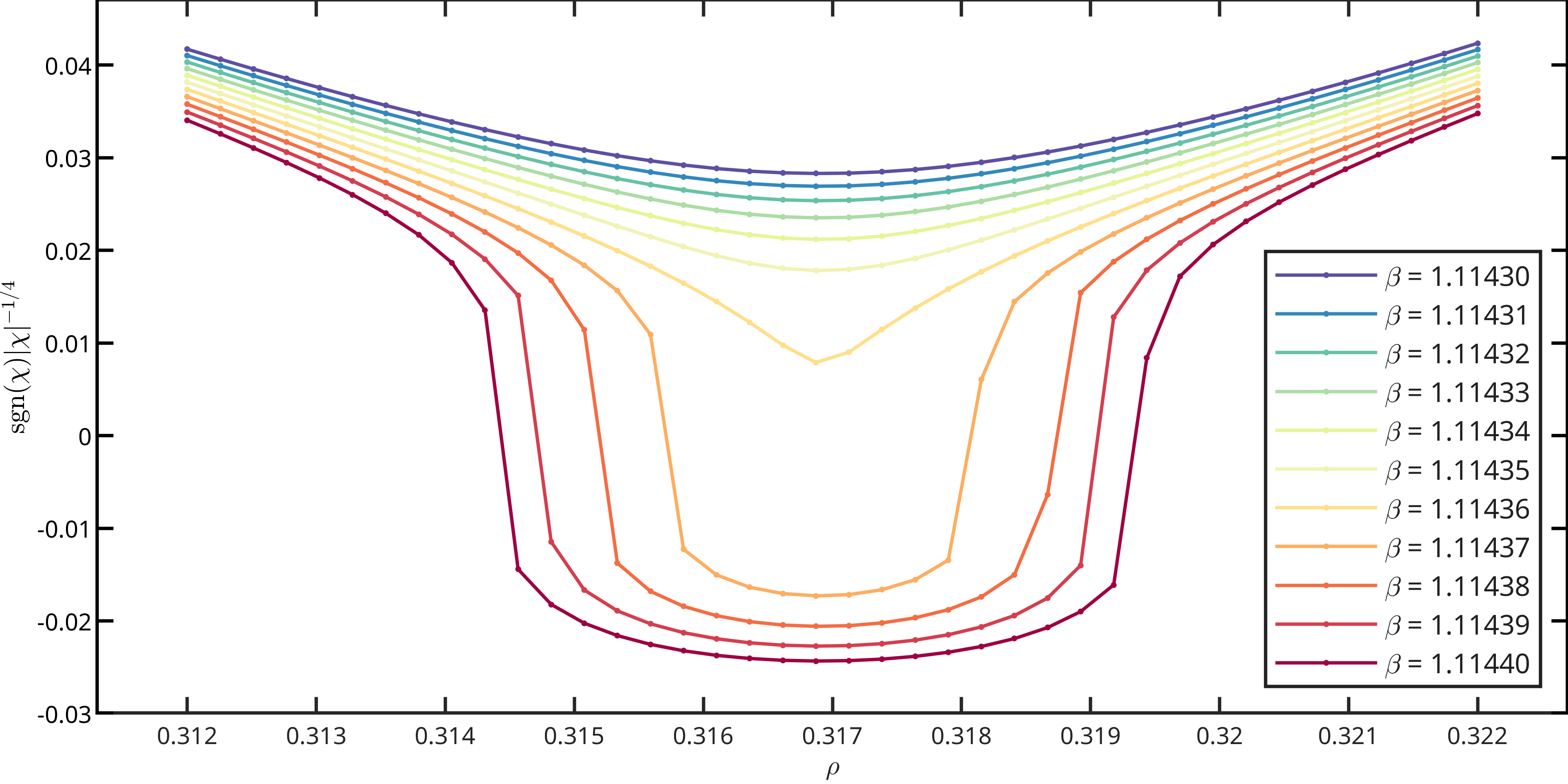}

	\caption{Compressibility isotherms near the critical point in 3D, for a range of temperatures. 
	As the temperature decreases toward the critical point a cusp begins to appear, defining the asymptotic critical regime where the exact solution indicates that $\ichired \sim\abs{\Delta\rho}^{4}$.
	Beyond the critical temperature, the compressibility is negative.
	Tests shown here were conducted with $\nod=40,\ndl=80,\ndisc=40,\ldinf=5$ and Picard iteration with mixing parameter of $\alpha=0.3$.
	}
	\label{fig:ichiisotherms3d}
\end{figure}

For the purpose of identifying the spinodal region and the critical point of the HCY+MSA model, we work with the reduced compressibility defined in equations \eqref{eqn:redcompressibility} and \eqref{eqn:compressibility_direct}.
As the spinodal region is approached from the exterior, one expects the inverse compressibility to evolve continuously, vanishing on the spinodal curve and becoming negative within the spinodal region.
The shape of the compressibility surface in density and temperature near the critical point is defined by \eqref{eqn:chiexponents}.
Therefore, once we have numerically solved the OZ equation and obtained estimates for both correlation functions, it is a simple matter to deploy Gauss-Chebyshev quadrature as described in Section \ref{sec:pseudospectral} to evaluate the compressibility via \eqref{eqn:compressibility_direct}.
Since the density parametrises the OZ equation, and the temperature (or inverse temperature $\beta$) the interparticle potential and thus the closure, the phase diagram of this model can be built as a (inverse) compressibility surface in $(\rho,\beta)$ space.
Carrying this out, we obtain results as shown in figure \ref{fig:ichi3d}.
The spinodal curve thus revealed has a recognisable form, and defines a critical point at the minimal $\beta$.
Focussing around this region of interest, to accurately locate the critical point, we take isothermal samples of the compressibility, and look for the first density and temperature at which negative compressibilities occur.
This process is illustrated in figure \ref{fig:ichiisotherms3d}, and for the 3D case we estimate the critical point to be at $(\rho_c,\beta_c)$=\critpointthree, which is in excellent agreement with the exact predictions \cite{brader_critical}.

It is then a matter of estimating the critical exponents using expressions \eqref{eqn:delta_eff} to \eqref{eqn:etaeff}. 
The exponent $\eta$, which describes the decay of the pair correlation function, is in principle the simplest to estimate, since it requires only one single solution of the OZ relation, exactly at the critical point.
Determining the exponent is then a matter of analysing the gradient, $\pdv*{\ln h}{\ln r}$, numerically. 
Because our numerical estimate of the correlation function is represented as a Chebyshev polynomial (composed with the quotient map \eqref{eqn:quotient}), we could evaluate this derivative using the usual pseudospectral differentiation matrix \cite{Trefethen,Canuto2006Book}.
In practice, however, $h$ may be negative during the oscillatory behaviour at small separations or at the largest separations, where its magnitude is small and thus subject to numerical noise.
The logarithm of $h$ is unsuitable in such cases and, because pseudospectral differentiation is global, this data cannot be excluded.
Instead, we take a subset of the separations $r$ over which we expect that the correlation has its asymptotic form and is fully converged, interpolate onto a fresh collocation grid and evaluate the pseudospectral derivative. 
As a point of comparison, we also calculate a least-squares straight line fit over this sub-interval.
This process is illustrated for our estimated critical point in figure \ref{fig:eta3d}, where the right panel makes clear that both routes to the gradient are consistent.
Based upon the pseudospectral derivative, we estimate that $\eta$ lies in the range \numrange{-0.016}{0.01}, with the linear fit pointing to a best-estimate of \num{-0.006}, consistent with the exact value, $\eta=0$, for the MSA+HCY model in 3D.

The process to estimate the exponent $\gamma$ describing the change in compressibility with temperature is similar in spirit, but it requires a range of compressibilities and thus several solutions to the OZ relation, at differing temperatures. 
However, so long as these test points lie on a line in the phase plane, ending at the critical point, we are free to choose their exact location, thus it makes sense to pick $\beta_i$ such that $\ln t_i = \ln\abs{\beta_c/\beta_i - 1}$ are an affine transformation of the CGL points.
In this way, once we solve the OZ relation at all the test temperatures to determine $\chi^{-1}_{\text{red},i}$, we can evaluate $\gamma_\text{eff}=\pdv*{\ln\ichired}{\ln t}$ via pseudospectral differentiation.
This is illustrated in figure \ref{fig:gamma3d} for a succession of temperatures up to our critical point estimate, where we find a maximal gradient of $\gamma=1.988$.
We can place a bound on this estimate by performing the same test at temperatures either side of our best estimate of the critical temperature, as in figure \ref{fig:ichiisotherms3d}. 
In this case, testing at $\beta=1.11435$ and $1.11437$ yields estimates of 1.974 and 2.003, respectively. Therefore, our final estimate for the exponent is \num{\gammathree}, which is consistent with the exact value of $\gamma=2$ for the 3D MSA+HCY model.

There are a slew of critical exponents one may wish to estimate for a given model, but fortunately they are not independent, and just two estimates are needed to uniquely determine the rest \cite{Binney}.
We have chosen $\eta$ and $\gamma$ as the two from which to start, for the following reasons:
As mentioned, $\eta$ is a particularly convenient choice, since it only relies on a single solution of the OZ relation.
The reason to choose $\gamma$ over $\delta$, which would require a very similar practical process, is that exponents with greater magnitudes are more difficult to estimate numerically.
Indeed, if we assume that good statistics require our independent variable ($t$, $\Delta\rho$, etc.) to range over several orders-of-magnitude and that there is some threshold for $\ichired$ below which critical phenomena dominate, then the required solving tolerance scales exponentially with the magnitude of the exponent.
Concretely, if $x$ is our independent variable, $\ichired\propto x^\alpha$, the onset of criticality is at $\ichired\approx X$, and we require that $\min{x}/\max{x} < 1/C$, then the smallest inverse compressibility we must calculate is at most $X/C^{\alpha}$.
For example, if we take from figure \ref{fig:ichiisotherms3d} that $X\approx \num{1e-8}$, $C>2$, then for the exponent $\delta=5$, we would require an inverse compressibility as small as \num{1e-16}, which is not feasible with our method or, we'd venture, any other relying on 64 bit floating point arithmetic.

\begin{figure}
	\centering
\includegraphics{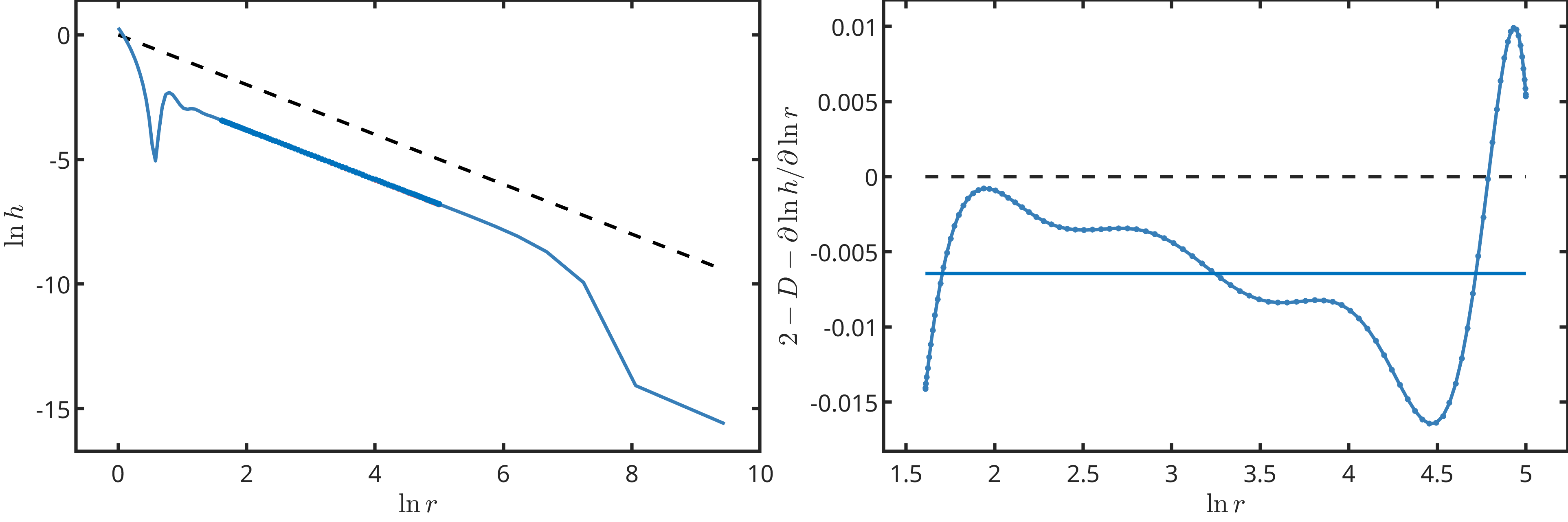}
	\caption{Illustration of the process used to calculate the exponent $\eta$ describing the spatial decay of the pair correlation function at the critical point, in the 3D case.
	We use both a pseudospectral derivative of the collocated solution for $h(r)$, and a least-squares linear fit, using a subset of the data indicated with dots in the left panel.
	From this derivative we estimate the exponent according to the definition \eqref{eqn:etaeff}, the results of which are shown in the right panel.
	Of course, for the linear fit, this is a constant value, but the pseudospectral derivative helps indicate the validity of this assumption over the subset of $r$ used.
	The guidelines included in each panel are intended to indicate the theoretical prediction for the gradient of $-1$, from $\eta=0$.}
	\label{fig:eta3d}
\end{figure}

\begin{figure}
	\centering
	\includegraphics{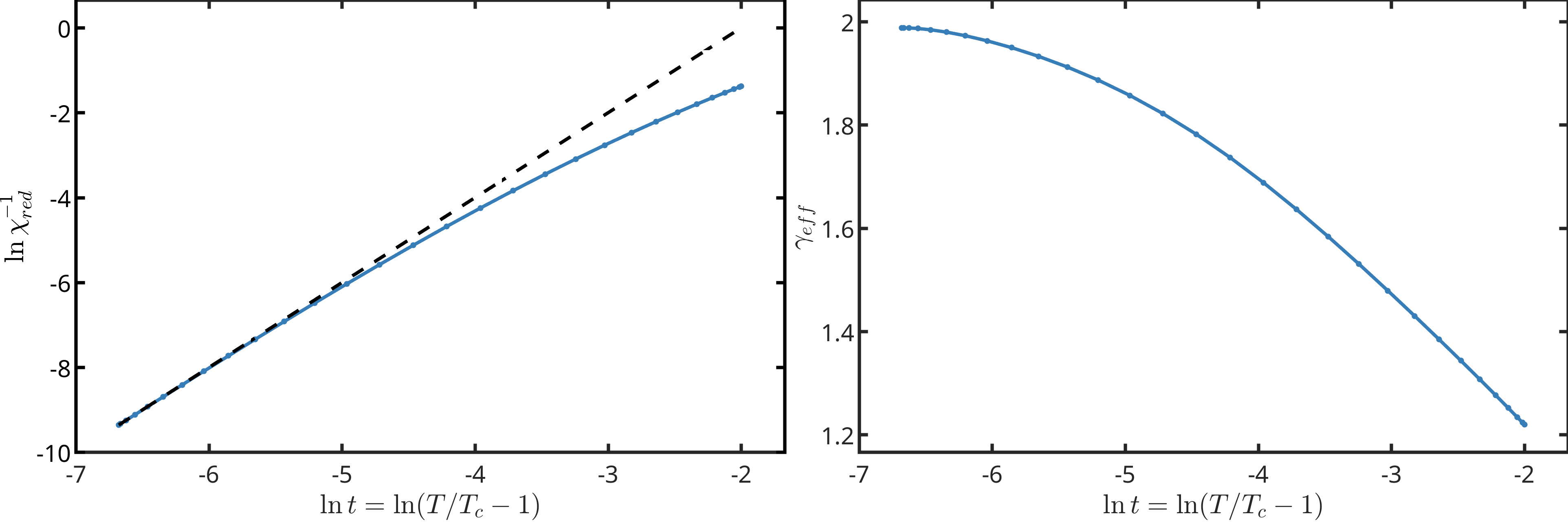}
	\caption{Illustration of the process used to calculate the exponent $\gamma$ describing the change in the compressibility with temperature in the vicinity of the critical point, in the 3D case.
	Since we have the freedom to choose the temperatures at which to sample the compressibility, we pick them such that $\ln t_i$ lie on an affinely-mapped grid of CGL points.
	The effective exponent \eqref{eqn:gammaeff} is then simply calculated using pseudospectral differentiation, and is shown in the right panel.
	As the temperature is decreased, the gradient of $\ichired$ tends toward a maximum, close to the theoretical value of $\gamma=2$ (which we indicate with a guideline in the left panel).}
	\label{fig:gamma3d}
\end{figure}

\subsection{Two-dimensional results} \label{sec:exponents2d}

With the speed afforded by our numerical method, it is a simple matter to compute the compressibility landscape for the MSA+HCY model also in 2D.
This result is shown in figure \ref{fig:ichi2d}, where the spinodal curve is clearly revealed by the sign change of $\ichired$.
Once again we can estimate the location of the critical point by computing isotherms of the compressibility near the apex of the spinodal. These results are shown in figure \ref{fig:ichiisotherms2d} and suggest that the critical point is located at $(\rho_c,\beta_c)=\critpointtwo$.

Now that we have estimates for the spinodals in both 3D and 2D, see figures \ref{fig:ichi3d} and \ref{fig:ichi2d}, respectively, we are able to compare them qualitatively.
We observe that the critical point in 2D lies at much higher values of $\beta$ than that in 3D. 
	This is not surprising, since phase separation is largely determined by the `integrated strength' of the attractive part of the interparticle interaction potential (i.e. its spatial integral) and this is naturally larger in 3D than in 2D at equal values of $\beta$. 
	Moreover, we note that the curvature of the 2D spinodal clearly differs from that in 3D.

Our procedure for estimating the critical exponents is unchanged, and is outlined for $\eta$ and $\gamma$ by figures \ref{fig:eta2d} and \ref{fig:gamma2d}, respectively, although there are some elements worth commenting on.
Firstly, the power law trend is less cleanly exhibited in 2D than 3D; we attribute this to the slower decay of the correlation function, as $r^{-\eta}$ rather than $r^{-1}$, and the fact that our iterative numerical method converges latest at the greatest separations.
Secondly, and more positively, the precision with which we estimate $\gamma$ in 2D is far better than any of the other exponents tested for. 
We believe that this is partly due to the aforementioned relationship between the magnitude of the exponent and the required tolerance, but this is not a full explanation as we also found calculations at the smallest compressibilities more reliable in 2D than 3D.

Our estimates for the 2D exponents are collected, along with those for the 3D model, in table \ref{tab:exponents}, where we also calculate the exponents $\delta$ and $\beta$ according to the scaling relations presented in Section \ref{sec:exponentreview}.
We point out that our uncertainty intervals for all the exponents of the 3D MSA+HCY model contain the exact values, and these differ from the estimates for the 2D model: illustrating that the MSA+HCY model is non-mean-field.
In fact, our best estimates for the exponents $\eta$ and $\gamma$ differ from the exact values by 0.6\% and 0.8\%, respectively.
Interestingly, we also find that the general trend of exponents increasing or decreasing as the dimensionality is reduced to be the same in all cases as the Ising model exponents, to which more research effort has been dedicated.

\begin{figure}[pos=h]
	\centering
\includegraphics{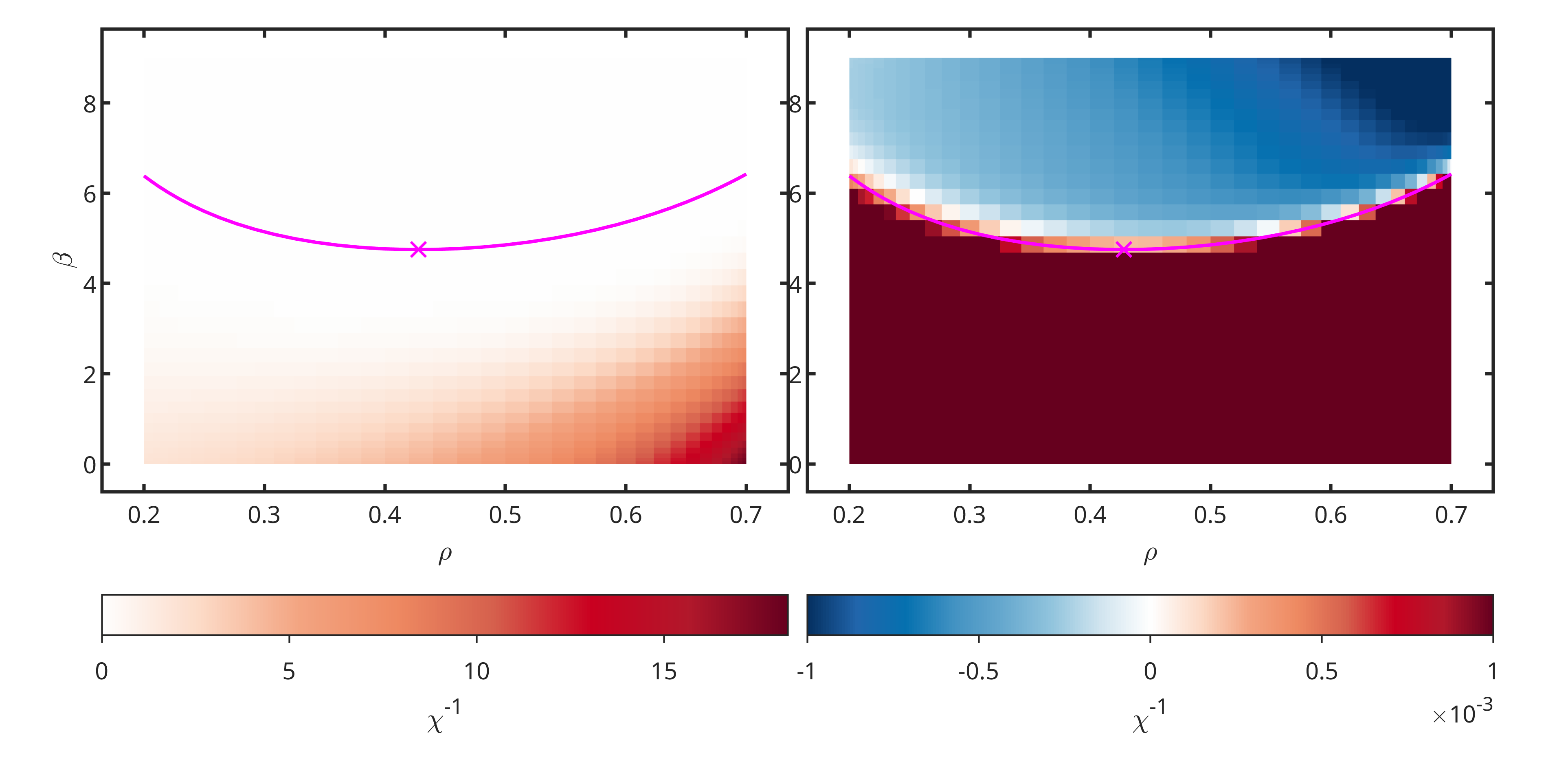}
	\caption{Inverse compressibility over the density-temperature plane for the HCY+MSA model in 2D, with full and clipped dynamic range. 
		Overlaid is the numerically-estimated spinodal and critical point, using the compressibility (albeit with higher-resolution evaluation than shown).
		The critical temperature is notably higher than for the equivalent potential in 3D.}
	\label{fig:ichi2d}
\end{figure}

\begin{figure}
	\centering
	\includegraphics{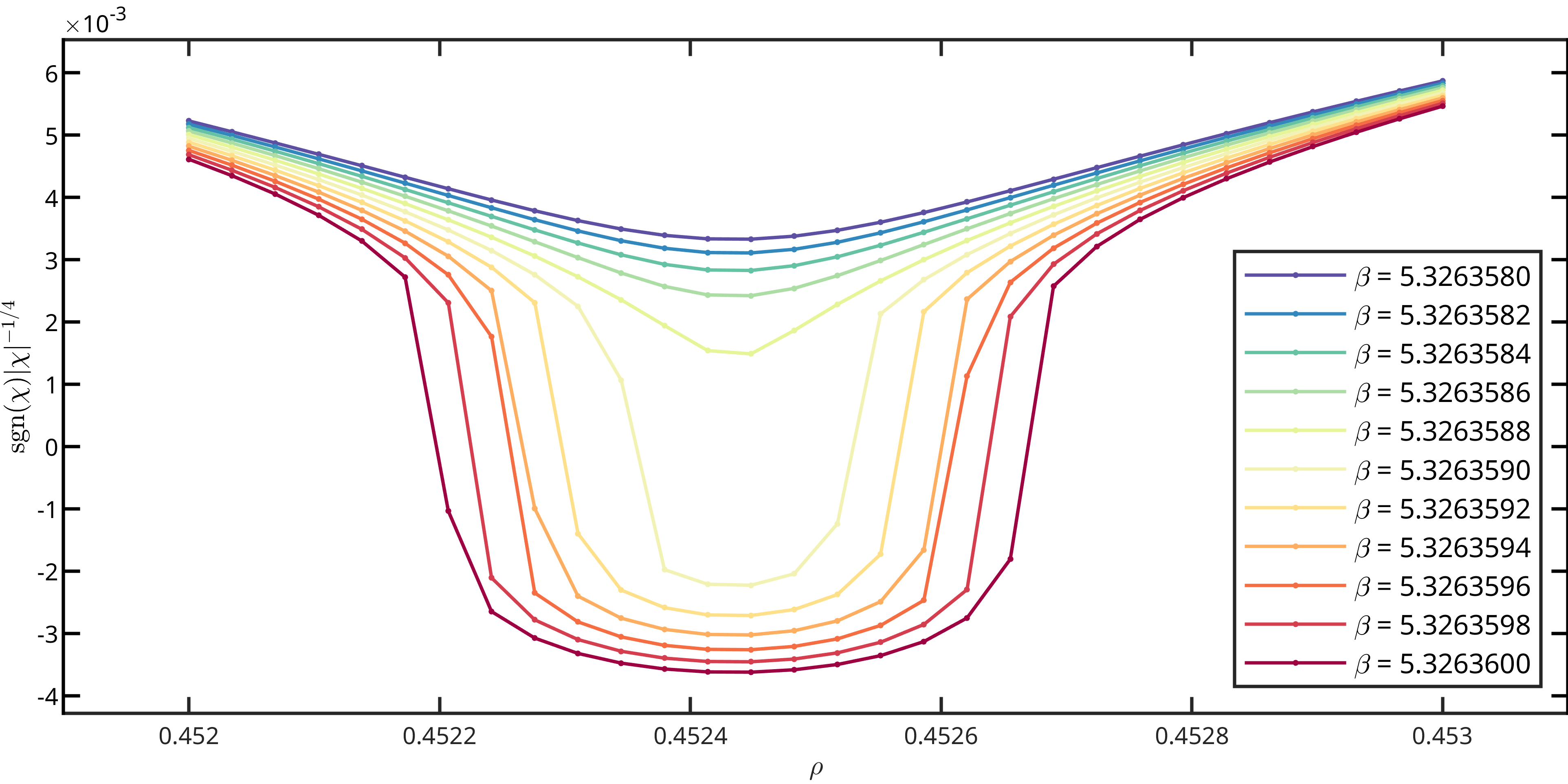}
	\caption{Compressibility isotherms near the critical point in 2D, for a range of temperatures, as in figure \ref{fig:ichiisotherms3d}.
		We have scaled the compressibility using the same $1/4$ power in 2D: there is no theoretical justification for this in two dimensions, but it allows comparison between the two plots.
}
	\label{fig:ichiisotherms2d}
\end{figure}

\begin{figure}[pos=h]
	\centering
\includegraphics{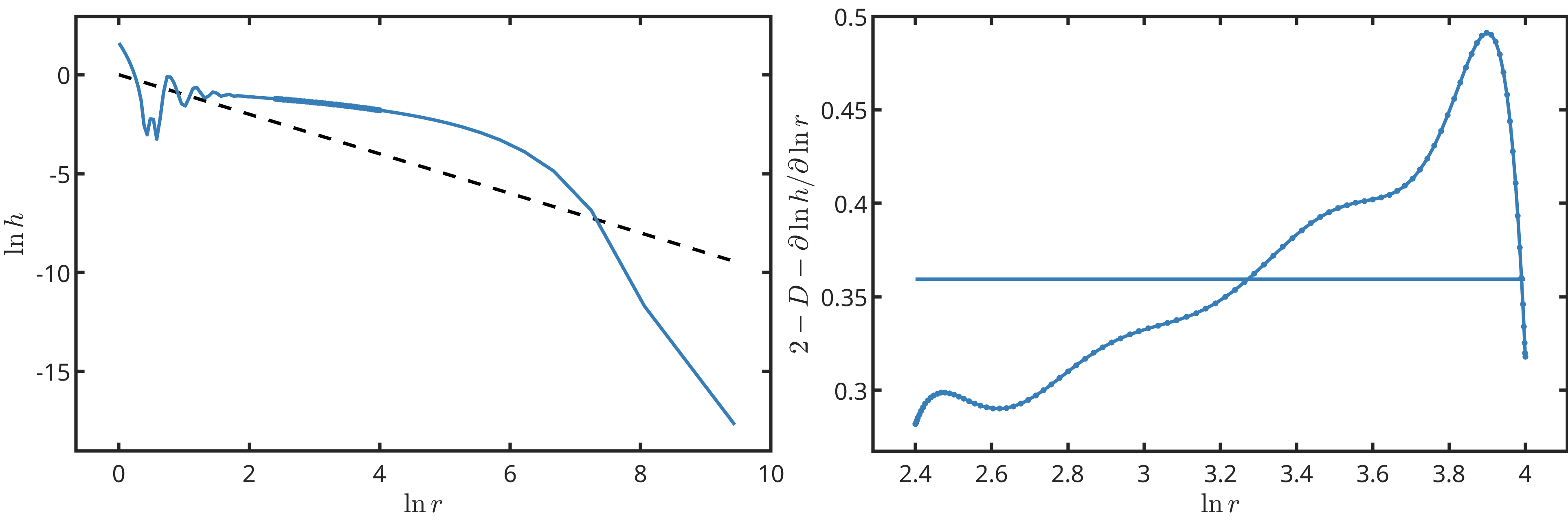}
	\caption{$\eta$-determination in 2D. 
	Note that we are not able to recover such a clean power law relationship between $h$ and $r$ as in 3D, and so the slope in the right panel has something of a trend with increasing $r$, we have accordingly estimated $\eta$ with a wide uncertainty.
	To provide a point of comparison with the 3D data, we include the same gradient $-1$ guideline.}
	\label{fig:eta2d}
\end{figure}

\begin{figure}[pos=h]
	\centering
	\includegraphics{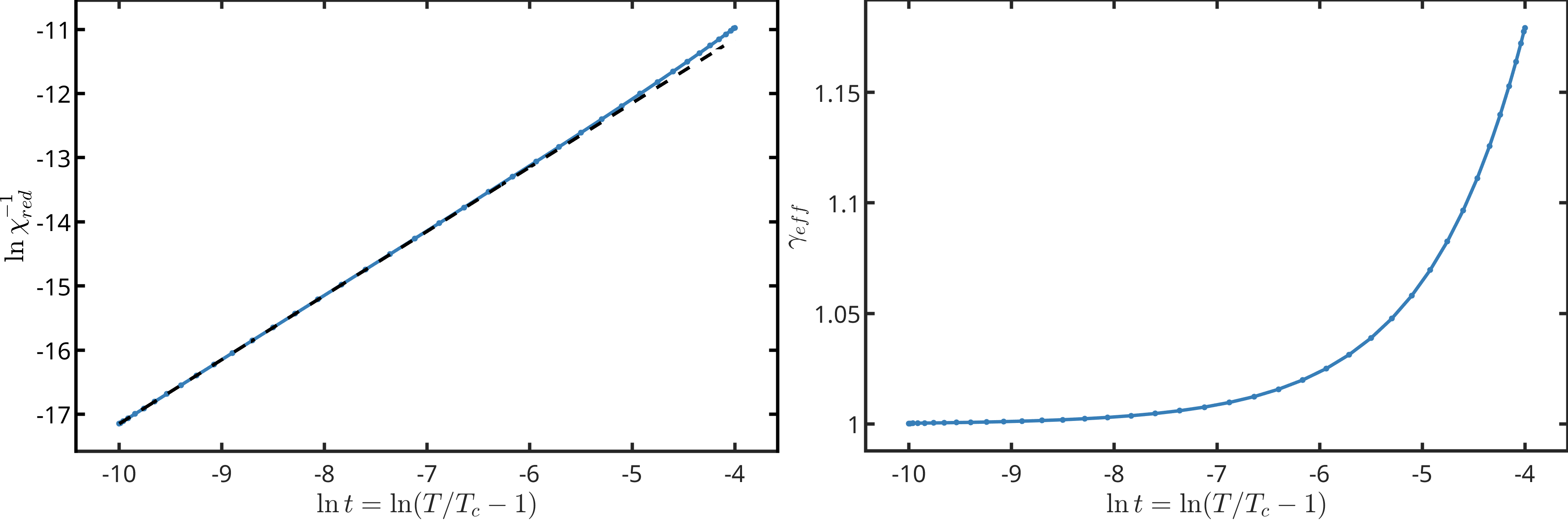}
	\caption{$\gamma$-determination for 2D.
	In contrast to the case of $\eta$, for this exponent we were able to extract a clean trend of the effective exponent as temperature decreases toward the critical point. 
	Our data seem to point strongly toward an estimate of $\gamma=1$ in 2D,
	and we include a guideline with this gradient in the left panel to compare.}
	\label{fig:gamma2d}
\end{figure}

\begin{table} 
	\centering
	\sisetup{
		table-text-alignment=left,
		table-number-alignment=center,
		table-format=-1.4(1),
		table-align-uncertainty=true,
		separate-uncertainty=true,
		table-auto-round}
	\begin{tabular}{l|c|S[table-format=1.4]cS|cS}
			%Exponent & {Mean field} & {Ising 3D} & {Exact MSA+HCY 3D} & {Best estimate 3D} & 
			%{Ising 2D} & {Best estimate 2D} \\
			Exponent & {Mean field} & {Ising 3D} & {Exact MSA 3D} & {Best estimate MSA 3D} & 
			{Ising 2D} & {Best estimate MSA 2D} \\
			\hline
			$\eta$	& 0	& 0.036	& 0	& \etathree		& 1/4	& \etatwo	\\
			$\gamma$& 1	& 1.237	& 2	& \gammathree	& 7/4	& \gammatwo	\\
			$\delta$& 3	& 4.78	& 5	& \deltathree	& 15	& \deltatwo	\\
			$\beta$	&1/2& 0.327	&1/2& \betathree	& 1/8	& 0.11(0.05) 	\\
		\hline	
	\end{tabular}
	\label{tab:exponents}
	\caption{
	Our estimates of the critical exponents of the MSA+HCY model in two and three dimensions, presented alongside mean field and Ising model exponents, and those from the Waisman exact solution for context.
	The exponents $\eta$ and $\gamma$ were estimated directly, as described in Sections \ref{sec:exponents3d} and \ref{sec:exponents2d}, while $\delta$ and $\beta$ (not to be confused with the inverse temperature) were calculated according to the scaling relations \eqref{josephson} and \eqref{widom}.
	}
\end{table}

\section{Conclusions and outlook} \label{sec:conclusions}

In this paper we have developed, tested and implemented a novel method of solving the bulk OZ equation rapidly and with high accuracy using a combination of 
pseudospectral methods and (linearized implicit)  Picard iteration. 

Since the OZ equation lies at the heart of many liquid-state theories and plays an important role in equilibrium statistical mechanics, the methods we have developed 
will be very useful for further studies in a variety of different contexts, ranging from studies of liquid microstructure and packing to critical phenomena and 
phase transitions. 
For the present work we have focused on a specific model system, namely the HCY model, which presents all of the generic features of a realistic liquid while using a minimal set 
of parameters. Moreover, this model has the advantage that in 3D it can be solved analytically within the MSA approximation, which provides a very useful benchmark for testing 
numerical schemes. 
We demonstrated that our numerical method can reproduce important features of the analytical solution (the critical exponents $\gamma$ and $\eta$) with excellent accuracy.  
This gave us confidence to then make new predictions for the (previously unknown) critical behaviour of the MSA in 2D. 
When reducing the dimensionality from 3D to 2D, we found that the critical exponents change, thus making clear that the MSA is not a mean-field theory. In addition, the manner in which the exponents vary as a function of $D$ is in line with physical expectations based on experience with the Ising model (the HCY critical behaviour falls within the Ising universality class). 

The MSA assumes that the direct correlation function, $c$, decays asymptotically (for $r\rightarrow\infty$) in the same way as the interparticle pair potential. 
However, even though this approximation is generally successful, it is known that, close to the critical point, $c$ should develop a long-range tail. 
Thus, numerical methods capable of treating infinite domains can be anticipated to become even more essential for investigating the properties of future, improved integral equation closures \cite{Janssen}.
In this regard, since integral equation theories combined with pseudospectral schemes can handle truly infinite domains, they have a clear advantage over stochastic particle-based simulation methods that rely on simulation boxes and thus are always subject to finite-size effects. Also, at present, machine learning and AI-based approaches train on finite size simulation data, and thus the status of their predictions for larger system sizes remains unclear.  

We should draw the reader's attention to the fact that, in this work, the bulk OZ equation was numerically solved without any prior analytic reformulation.
However, in the case of odd dimensions, the bulk OZ equation can be usefully reformulated as a pair of coupled 
integral equations using a Wiener-Hopf factorization, first introduced by Baxter \cite{baxter_factorization1} and, later, in a more obscure fashion by Wertheim \cite{baxter_factorization2}. 
Indeed, the fact that certain models can be exactly solved within the PY and MSA approximations is intimately linked to the existence of a factorized form. 
The Baxter-factorized equations have the advantage that they can be solved in a self-contained way over a finite domain, $[0,L]$, where the only constraint is that the direct correlation function, $c$,
vanishes for separations beyond $L$. 
For approximations, such as the PY or MSA, for which c remains of short range this feature makes the factorized equations very suitable for the study of critical phenomena 
\cite{brader_critical}. 
However, the method is not able to handle the interesting case of 2D and the only way forwards is to confront directly the OZ equation in its original form. 
The numerical methods presented in the present work are thus very useful, since they can well compete with Baxter-based calculations in the case of 3D, but then easily generalize to treat 2D systems at a comparable level of efficiency and accuracy. 
The absence of Baxter-factorized equations in 2D explains why the critical properties of integral equation closures in that dimension have not received serious attention and lack of studies to compare our outputs with.

Furthermore, the present pseudospectral methodology applied to the bulk OZ equation shows good potential for generalization to the inhomogeneous case, namely to systems for which the one-body density varies with spatial position. The efficiency and accuracy of our new methodology will prove very fruitful when extended to this more general case, since solution of the fully inhomogeneous OZ equation using existing numerical schemes is computationally demanding \cite{TschoppSuperadiabatic5, TschoppSuperadiabatic4, TschoppSuperadiabatic6}. Also, going beyond equilibrium, for example using the new superadiabatic DDFT formalism, would greatly benefit from a faster solution of the inhomogeneous OZ equation, that has to be solved at each time-step \cite{TschoppSuperadiabatic, TschoppSuperadiabatic2}.

Finally, we provide an open-source code at \url{https://github.com/JakeSkelton/OZChebClass}, noting that our numerical experiments demonstrate that the approach presented here can be several orders of magnitude faster than other state-of-the-art methods, such as the Fourier-based approach outlined in Section~\ref{sec:numericalChallenges}.

\section*{Acknowledgments}
% \acknowledgments

This work was supported by the Centenary Research Fund via Project No. FC-25-974.

J.S. was supported by the EPSRC Centre for Doctoral Training in Mathematical Modelling, Analysis and Computation (MAC-MIGS) funded by the UK Engineering and Physical Sciences Research Council (EPSRC) grant EP/S023291/1, Heriot-Watt University and the University of Edinburgh.

\bibliographystyle{abbrv}
\bibliography{bibliography.bib,Homogeneous1.bib}

\newpage

\appendix

\section{Real-space tensor formulation} \label{app:tensor}

The tensor $\nvec{T}$ used to encode the real-space formulation of the bulk OZ relation, as 
\begin{equation} \tag{\ref{eqn:ozbulk_discrete}}
	h_i = c_i + \rho\sum_{j,k=1}^{n} T_{ijk} c_j h_k
\end{equation}
is third order and of dimension $\mathbb{R}^{n\times n\times n}$.
Each of the three indices corresponds to a point among the ordered set $R$ used to discretise $[0,L]$, where $L$ is the (possibly infinite) extent of the domain.
The particular set of points we use is a concatenation of two sets of mapped CGL points, as explained in section \ref{sec:1DDomain}, but any set that permits interpolation and numerical integration of ordinates thereon will do.
Each element of $T_{ijk}$ is composed of one integration operator, and two interpolation operators, as follows:
If our entire integration domain is $\Omega=[0,L]^D$, represented by a set of discrete collocation points $\{\s_a\in\Omega | a\in A\}$, each with a corresponding numerical quadrature weight $w_a$, then we can approximate the OZ relation as
\begin{equation} \label{eqn:ozfromint}
	h(\r_i) = c(\r_i) + \rho\sum_{a\in A} 
	w_a c(\s_a) h(\r_i + \s_a) \quad \forall \r_i \in R.
\end{equation}
Now, due to spherical symmetry in the bulk case, we are concerned with the points $\abs*{\s_a}$ and $\abs*{\r_i + \s_a}$, which are not elements of $R$, in general.
In principle, our collocation representation of the unknown functions allows evaluation anywhere within $[0,L]$, but in practice this requires interpolation.
Hence, we define $I(\r_i + \s_a)\in \mathbb{R}^{1\times n}$ to evaluate $h(\r_i + \s_a)$ using a linear combination of $\{h_k | \r_k\in R\}$ (or any other quantity collocated in this way), and likewise 
$J(\s_a)\in \mathbb{R}^{1\times n}$ to represent interpolation from the set $R$ onto $\s_a$, yielding
\begin{equation} \label{eqn:ozfromintandinterp}
	h_i = c_i + \rho\sum_{j,k=1}^{n}\sum_{a\in A} 
		w_a J_j(\s_a) I_k(\r_i + \s_a) c_j h_k
\end{equation}
so that
\begin{equation} \label{eqn:tensorfromintandinterp}
	T_{ijk} = \sum_{a\in A} 
	w_a J_j(\s_a) I_k(\r_i + \s_a).
\end{equation}
In our case, the points, weights, and interpolation operators are chosen to make best use of the accuracy afforded by pseudospectral collocation, and are built from the one-dimensional forms in Section \ref{sec:points_weights_interp} using tensor products.
At this point we should highlight that the contraction in equation \eqref{eqn:tensorfromintandinterp} means that the tensor buries the details of the discretisation of $\Omega$, and once it is assembled, there is no need to keep any of the above integration or interpolation operators in memory: enlarging $A$ only affects the start-up cost.

For a low-order, local discretisation scheme, or if the correlation functions $c$ and $h$ are known to be smooth, one could now proceed to programme the above.
However in the present case, we must represent each of the correlation functions as piecewise smooth, chopped in two by the discontinuity at $r=d$ resulting from the hard-core potential.
This partitions our domain $\Omega$ into four subdomains, as explained in Section \ref{sec:discretisation}, and affects the above formulation because the set of points $\{\s_a | a \in A\}$ used to discretise $\Omega$ will differ with $\r_i$, rendering the integration and interpolation operators functions of $\r_i$.
The sum over all $a \in A$ gets split into four: over the sets $\qty{a \,\middle| \abs*{\s_a} < d, \abs*{\r_i + \s_a} <d}$, 
$\qty{a \,\middle| \abs*{\s_a} < d, \abs*{\r_i + \s_a} > d}$,
$ \qty{a \,\middle| \abs*{\s_a} > d, \abs*{\r_i + \s_a} < d}$, and
$ \qty{a \,\middle| \abs*{\s_a} > d, \abs*{\r_i + \s_a} > d}$. 
These will be recognised as the domains catalogued in table \ref{tab:subdoms}.
To retain smoothness, only data from one piece of a piecewise curve is interpolated in any single operation: if $\abs{\s_a} < d$ then $J_j(\s_a) \neq 0$ if and only if $\r_j < d$.
What this means is that if a summation is only over points in one of the four subdomains, then only data for $c$ from within this subdomain is used. 
This may seem obvious when looking at a form such as equation \eqref{eqn:ozbulkshift}, but it requires special provisions to work with interpolation.
Likewise, if $\abs{\r_i + \s_a} < d$ then $I_k(\r_i + \s_a) \neq 0$ if and only if $\abs{\r_k - \r_i} < d$, meaning a sum over $\s_a$ in one subdomain involves only $h$ data in the translation of that subdomain by $\r_i$, just as explained in Section \ref{sec:discretisation}.
With all this in hand, we can write equation \eqref{eqn:ozfromintandinterp} as
\begin{equation} \label{eqn:ozdiscreteallsubdoms}
\begin{aligned}
		h_i = c_i + \rho\Bigg(&
			\sum_{\qty{a,j 	\middle| \s_a, \r_j 	\in \cll}} 
			\sum_{\qty{k 	\middle| \r_k 		\in \hll}} +
			\sum_{\qty{a,j 	\middle| \s_a, \r_j 	\in \clg}} 
			\sum_{\qty{k 	\middle| \r_k 		\in \hlg}} + \\&
			\sum_{\qty{a,j 	\middle| \s_a, \r_j 	\in \cgl}} 
			\sum_{\qty{k 	\middle| \r_k 		\in \hgl}} +
			\sum_{\qty{a,j 	\middle| \s_a, \r_j 	\in \cgg}} 
			\sum_{\qty{k 	\middle| \r_k 		\in \hgg}}
		\Bigg)
		w_a J_j(\s_a) I_k(\r_i + \s_a) c_j h_k
\end{aligned}
\end{equation}
using a shorthand where each triple sum has an identically-written summand, and implicitly $a\in A$ and $j, k \in 1,\dots,n$.
This expression should be compared with equation \eqref{eqn:ozwithdomains}, and we illustrate the effect of sorted collocation points and the geometric partition on the block structure of the tensor $T_{ijk}$ in figure \ref{fig:oztensorblocks}; each of the four squares in the horizontal plane corresponds to a triple sum in equation \eqref{eqn:ozdiscreteallsubdoms}.

\begin{figure}
	\centering
	\includegraphics{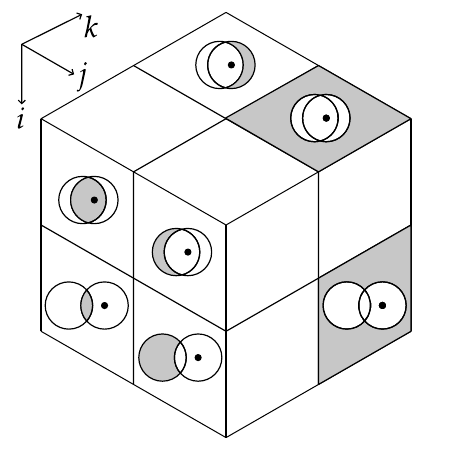}
	\caption{Blocks of the 3-tensor used to discretise the bulk OZ relation in real space, and the corresponding integration subdomains.
	The vertical axis corresponds to the first index (output), that out of the page to the second index (on $c$), and that into the page to the third index (on $h$).
	Each is bisected by the hard-sphere diameter $d$: the upper portion of the vertical axis corresponds to $\r_i<d$, the lower to $\r_i>d$, and likewise for the remaining axes.
	In each pictogram, the dot is the origin $\s=0$, surrounded by a circle of \emph{radius} $d$, and a distance $\r_i$ away from a second circle, within which $\abs{\r_i+\s}<d$.
	The circles are shown a representative distance apart in each block, i.e. of the correct magnitude with respect to $d$, but in reality the value of $i\in 1,\dots,n$ sets the distance, in fine steps, and the intersections are empty for separations exceeding $2d$.}
	\label{fig:oztensorblocks}
\end{figure}

\section{2D Domains}
\label{app:2Ddomains}

In this appendix we show representative examples of each of the 2D domains required to solve the bulk OZ equation using the methodology described in the main text.  The collocation points are shown as green circles, with other points/lines of interest denoted where appropriate.  In each case we give a short description of the domain, as well as the 1D maps used in their construction; see Section ~\ref{sec:1DMaps}.  For each map we give the required parameters; terms denoted by the letter $f$ with subscripts refer to functions that are straightforward to determine geometrically, but too complicated to be informative.  The titles of the figures correspond to the names of the shapes used in the code, \url{https://github.com/JakeSkelton/OZChebClass}, which also provides explicit forms of the functions $f$.  The figures are generated via the `ShapesDocumentation' method included in the code.

\begin{table}[H]
	\begin{tabular}{c p{6cm}}
		\midrule
		\tabfigure{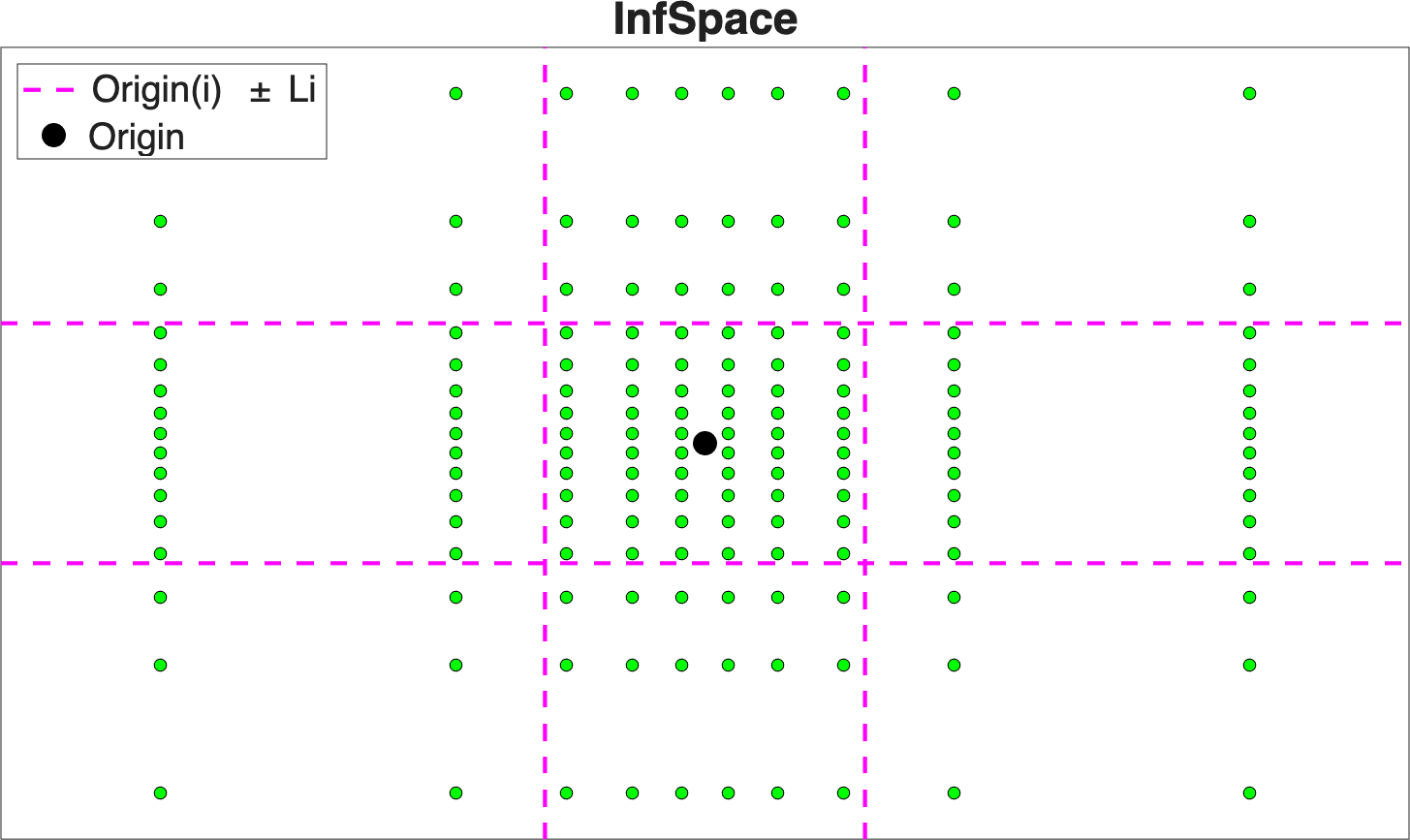}{-0.5\height} & 
		The whole of $\R^2$. \newline
		$\texttt{y}_1$: Square Root (N1, L1, Origin(1)) \newline
		$\texttt{y}_2$: Square Root (N2, L2, Origin(2))
		\\
		\\
		\midrule
		\tabfigure{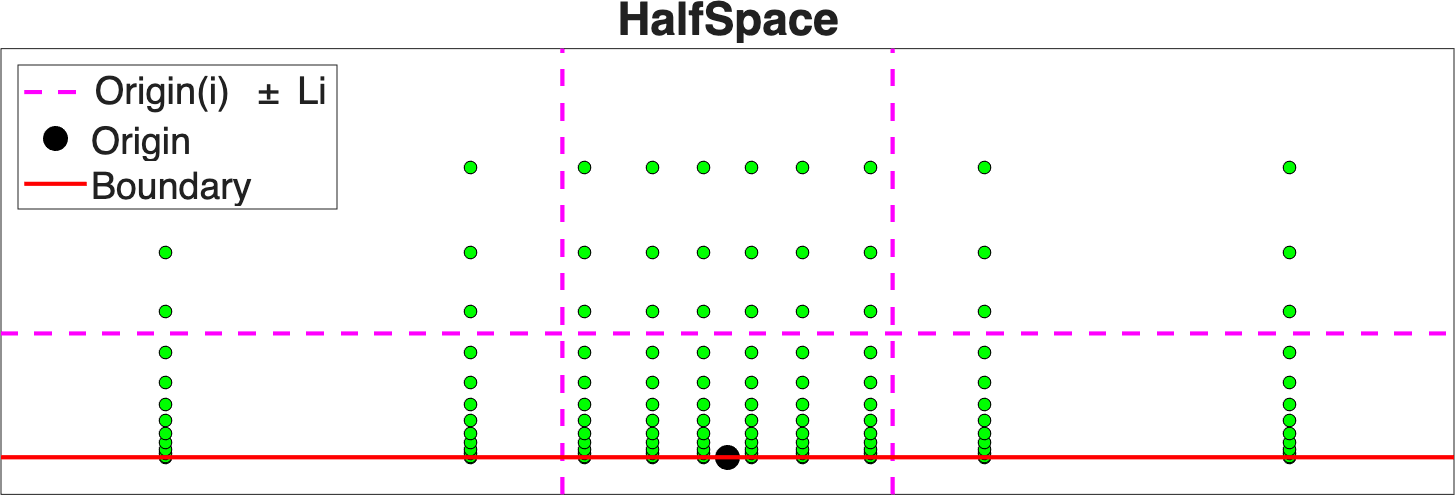}{-0.7\height} &
		The upper half plane with \newline $y_2 \geq$ Origin(2). \newline
		$\texttt{y}_1$: Square Root (N1, L1, Origin(1)) \newline
		$\texttt{y}_2$: Quotient (N2, L2, Origin(2))
		\\
		\\
		\midrule
		
		\tabfigure{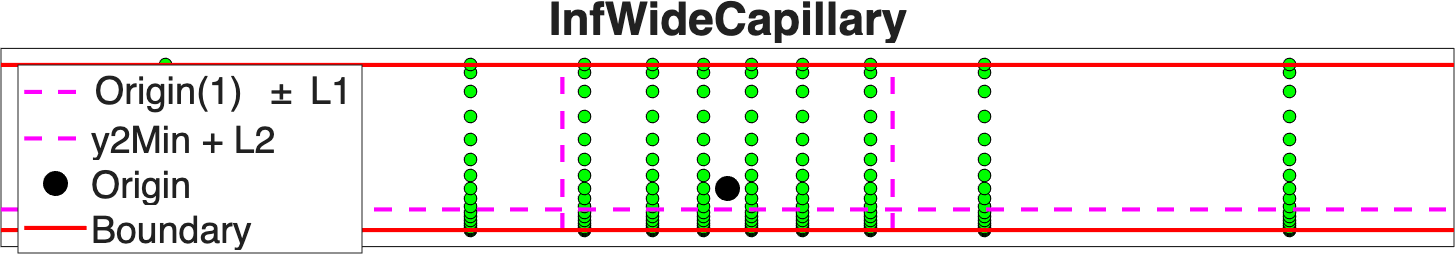}{-0.9\height} &
		An infinite strip between y2Min and y2Max. \newline
		$\texttt{y}_1$: Square Root (N1, L1, Origin(1)) \newline
		$\texttt{y}_2$: Quotient (N2, L2, y2Min, y2Max, \newline \mbox{} \qquad \qquad \qquad Origin(2))
		\\
		\\
		\midrule
		\tabfigure{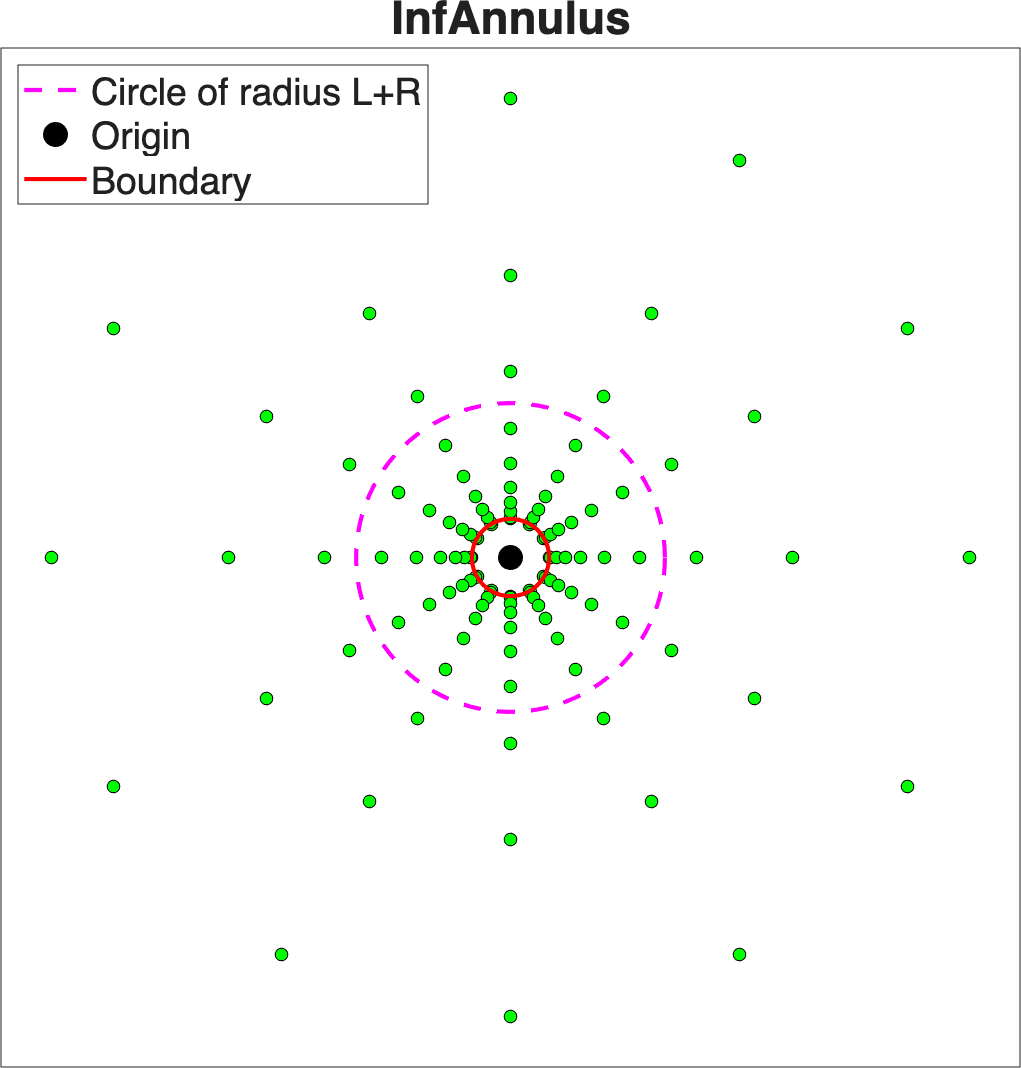}{-0.6\height} & 
		An infinite annulus with inner radius R, centred at Origin, in polar coordinates. \newline
		$\texttt{y}_1$: Quotient (N1, L1, R) \newline
		$\texttt{y}_2$: Linear01 (N2, 0, $2\pi$)
		\\
		\\
		\midrule
	\end{tabular}
\end{table}

\begin{table}[H]
	\begin{tabular}{c p{6cm}}
		\midrule
		\tabfigure{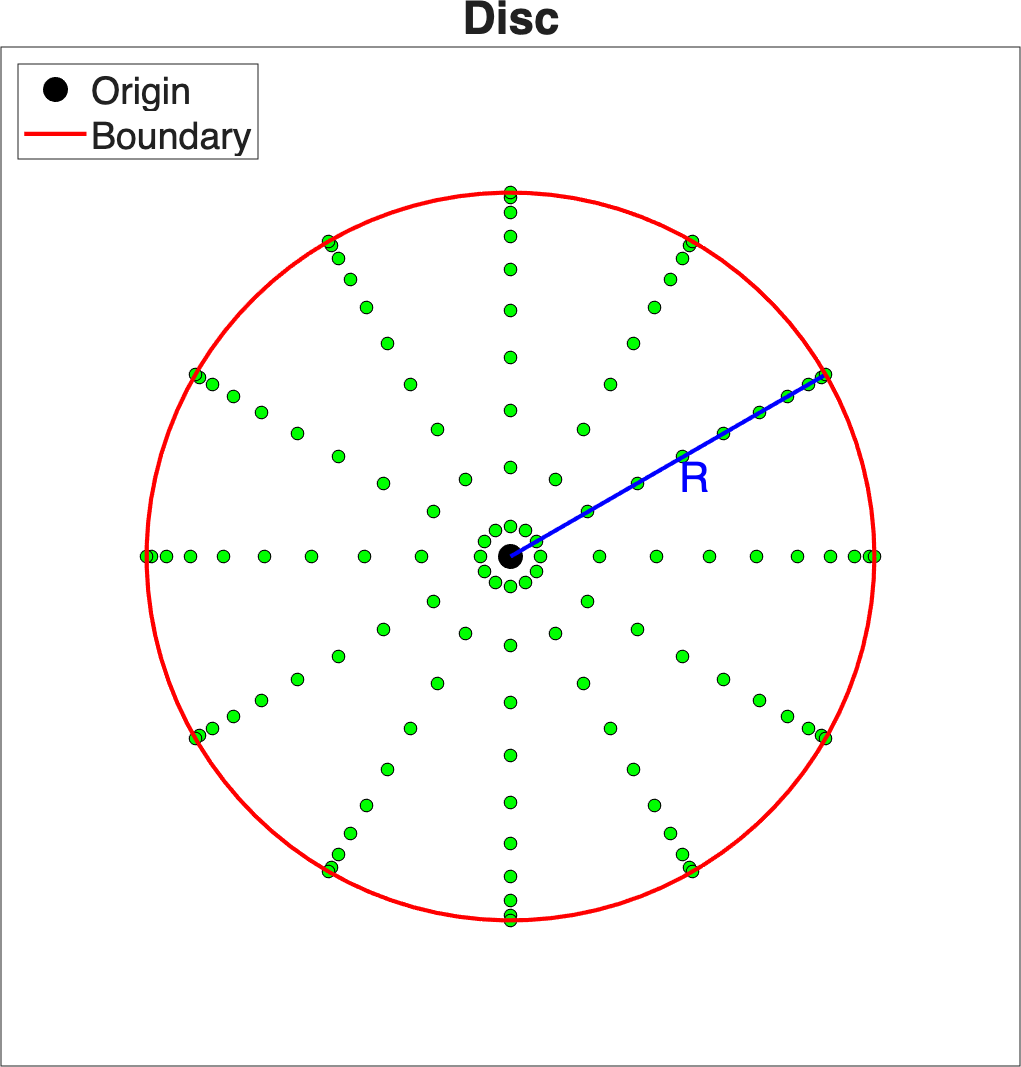}{-0.7\height} &
		A disk of radius R centred at Origin, in polar coordinates. \newline
		$\texttt{y}_1$: Linear (N1, -R, R) \newline
		$\texttt{y}_2$: Linear01 (N2, 0, $2\pi$) \newline
		Note that there are some subtleties when constructing the disc; see, e.g.,~\cite[Chapter 11]{Trefethen}.
		\\
		\\
		\midrule
		
		\tabfigure{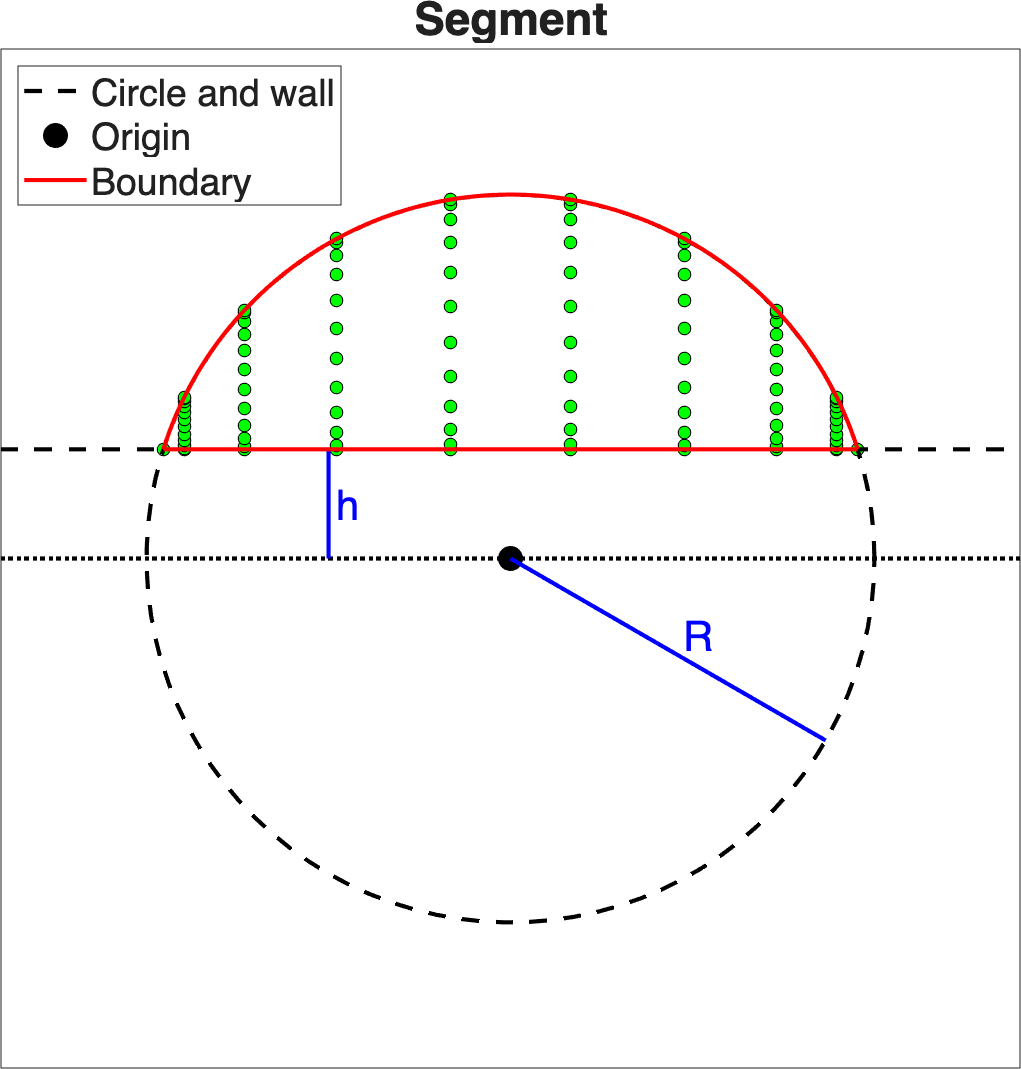}{-0.6\height} & 
		A segment of a disk of radius R, centred at Origin, cut off by a chord h above Origin. \newline
		$\texttt{y}_1$: Linear (N1, $f_{S1}$(h, R)) \newline
		$\texttt{y}_2$: Linear (N2, $f_{S2}$(h, R, y1)) \newline
		\\
		\\
		\midrule
		
		\tabfigure{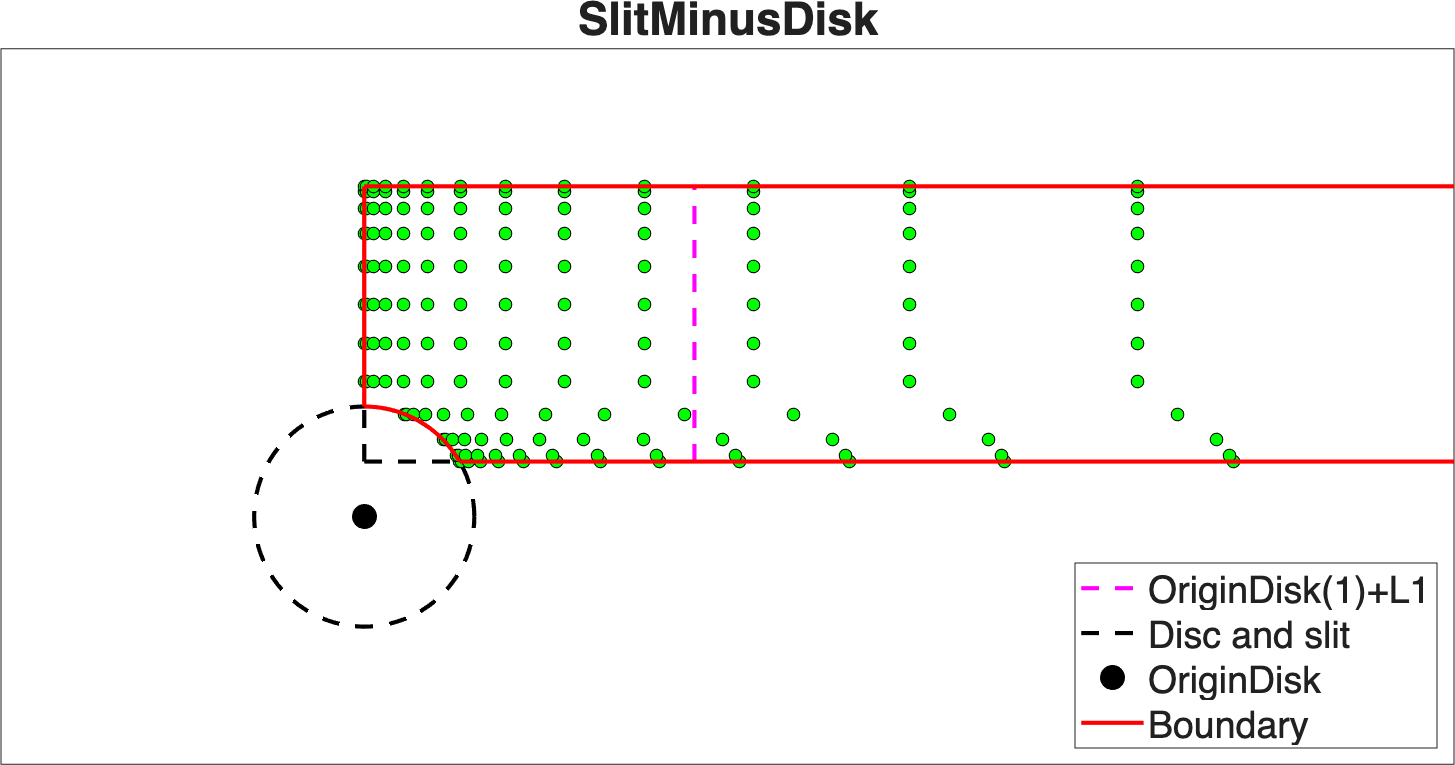}{-0.9\height} &
		Part of a half-infinite slit with upper and lower $y_2$ limits y2Min and y2Max, with the intersection of a disk of radius R, centred at Origin removed.  Note that the $y_1$ coordinate of Origin determines the finite end of the slit.  The distribution of the $y1$ points is determined by L.
		\newline
		$\texttt{y}_1$: Quotient (N1, L, \newline 
		$f_{SMD1}$(y2Min, y2Max, R, Origin, y2)) \newline
		$\texttt{y}_2$: Linear (N2, y2Min, y2Max)\newline
		\\
		\\
		\midrule

	\end{tabular}
\end{table}

\begin{table}[H]
	\begin{tabular}{c p{6cm}}
		\midrule
		
		\tabfigure{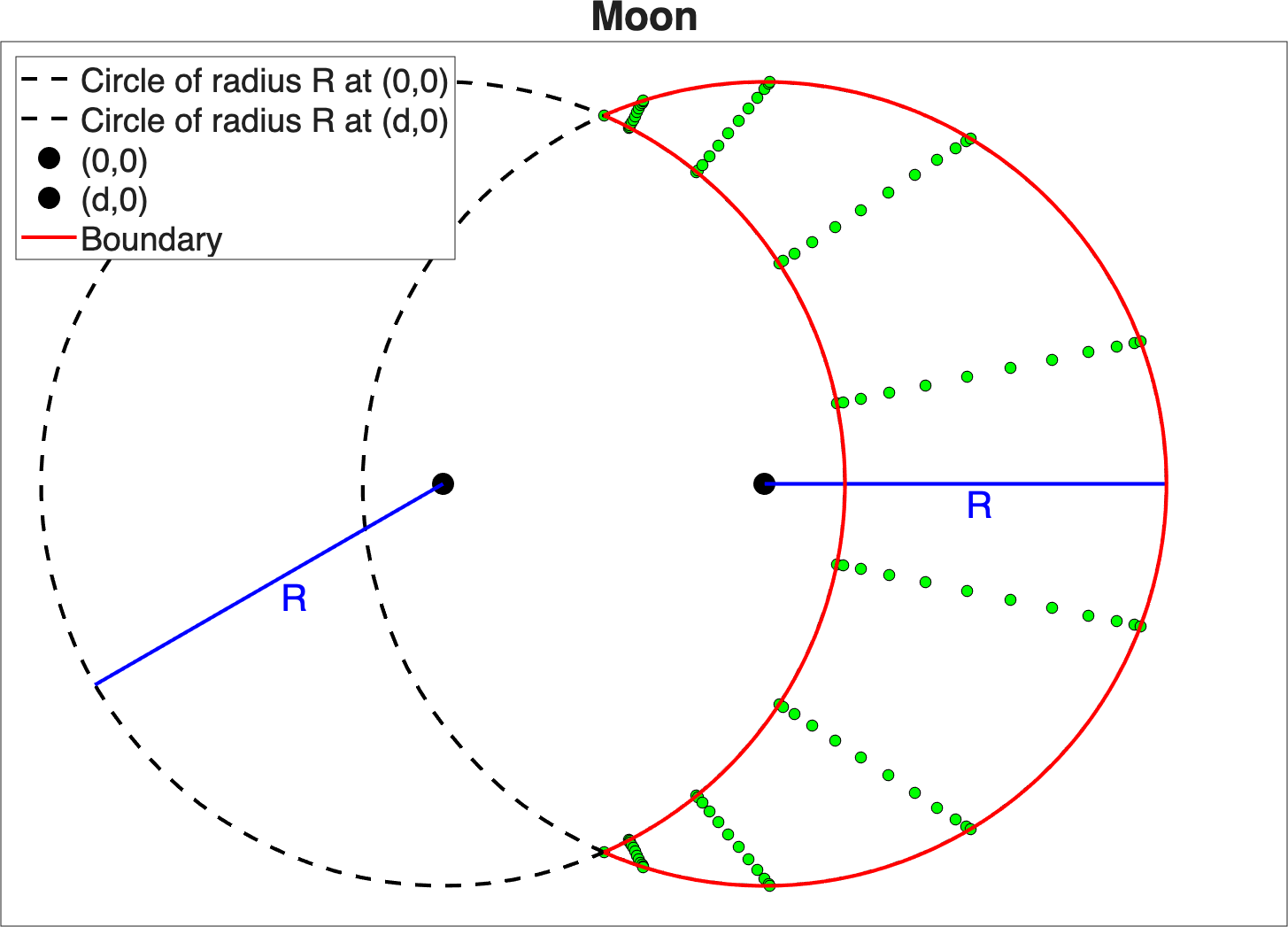}{-0.7\height} & 
		Part of a disk of radius R, centred at (d,0), without the intersection with a disk of radius R, centred at (0,0), in polar coordinates.
		\newline
		$\texttt{y}_1$: Linear (N1, $f_{M1}$(d, R, y2)) \newline
		$\texttt{y}_2$: Linear01 (N2, $f_{M2}$(d, R)) \newline
		\\
		\\
		\midrule
		
		\tabfigure{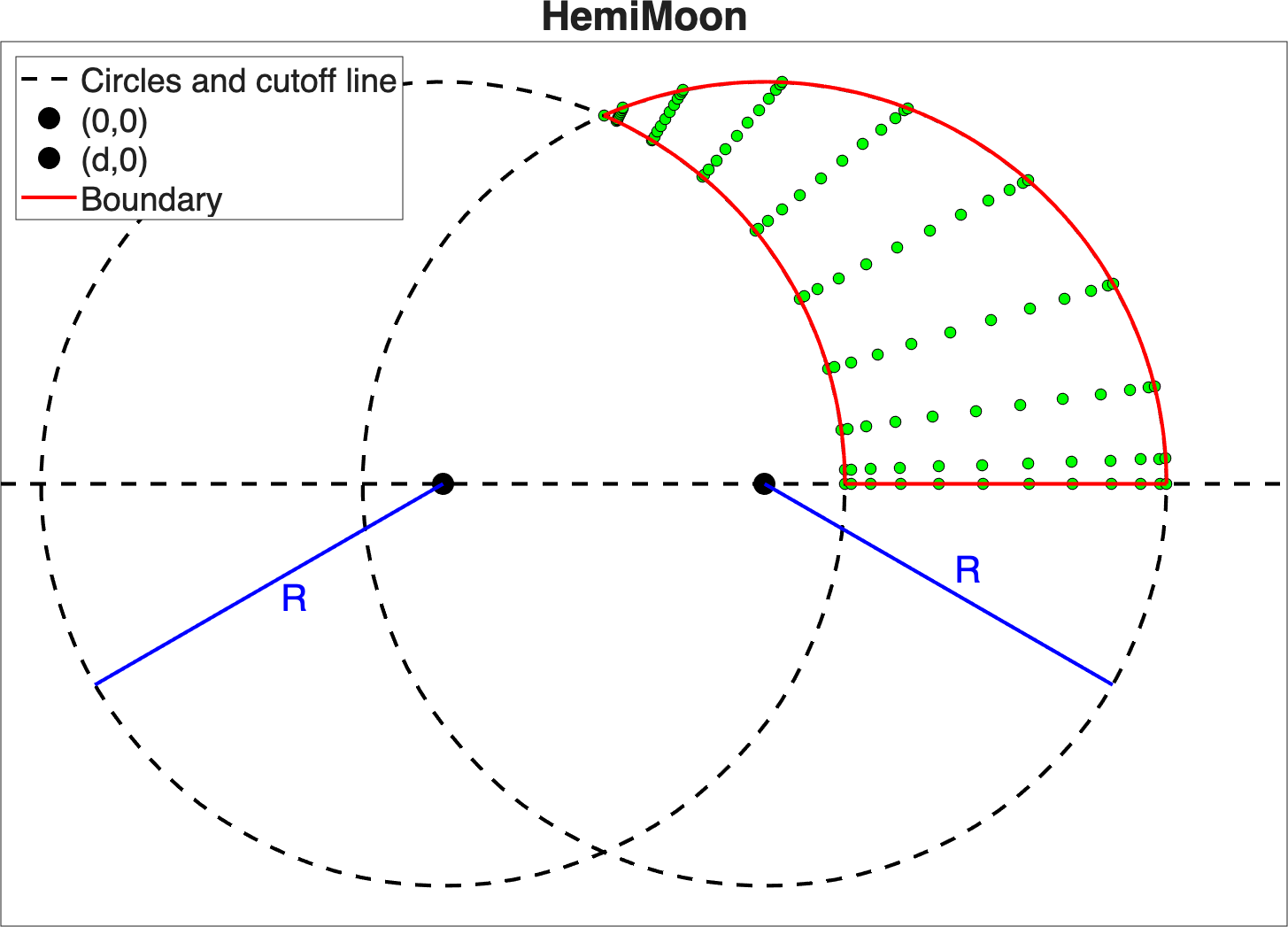}{-0.7\height} &
		Part of a disk of radius R, centred at (d,0), without the intersection with a disk of radius R, centred at (0,0), cut off above the $y_1$ axis, in polar coordinates.
		\newline
		$\texttt{y}_1$: Linear (N1, $f_{HM1}$(d, R, y2)) \newline
		$\texttt{y}_2$: Linear01 (N2, $f_{HM2}$(d, R)) \newline
		\\
		\\
		\midrule

		\tabfigure{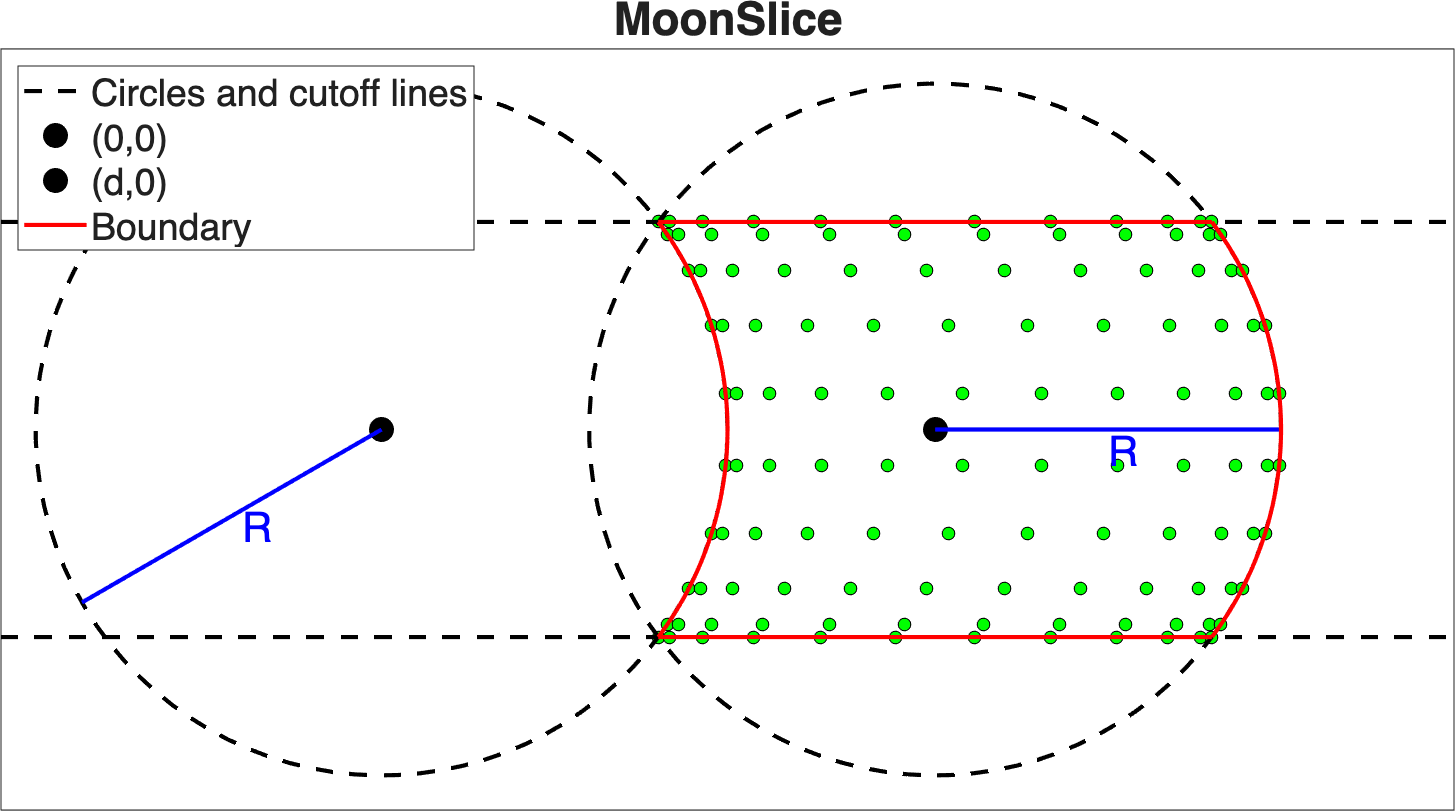}{-0.7\height} &
		Part of a disk of radius R, centred at (d,0), without the intersection with a disk of radius R, centred at (0,0), cut off between the horizontal lines through the intersections of the circles.
		\newline
		$\texttt{y}_1$: Linear (N1, $f_{HM1}$(d, R, y2)) \newline
		$\texttt{y}_2$: Linear (N2, $f_{HM2}$(d, R)) \newline
		\\
		\\
		\midrule

		\tabfigure{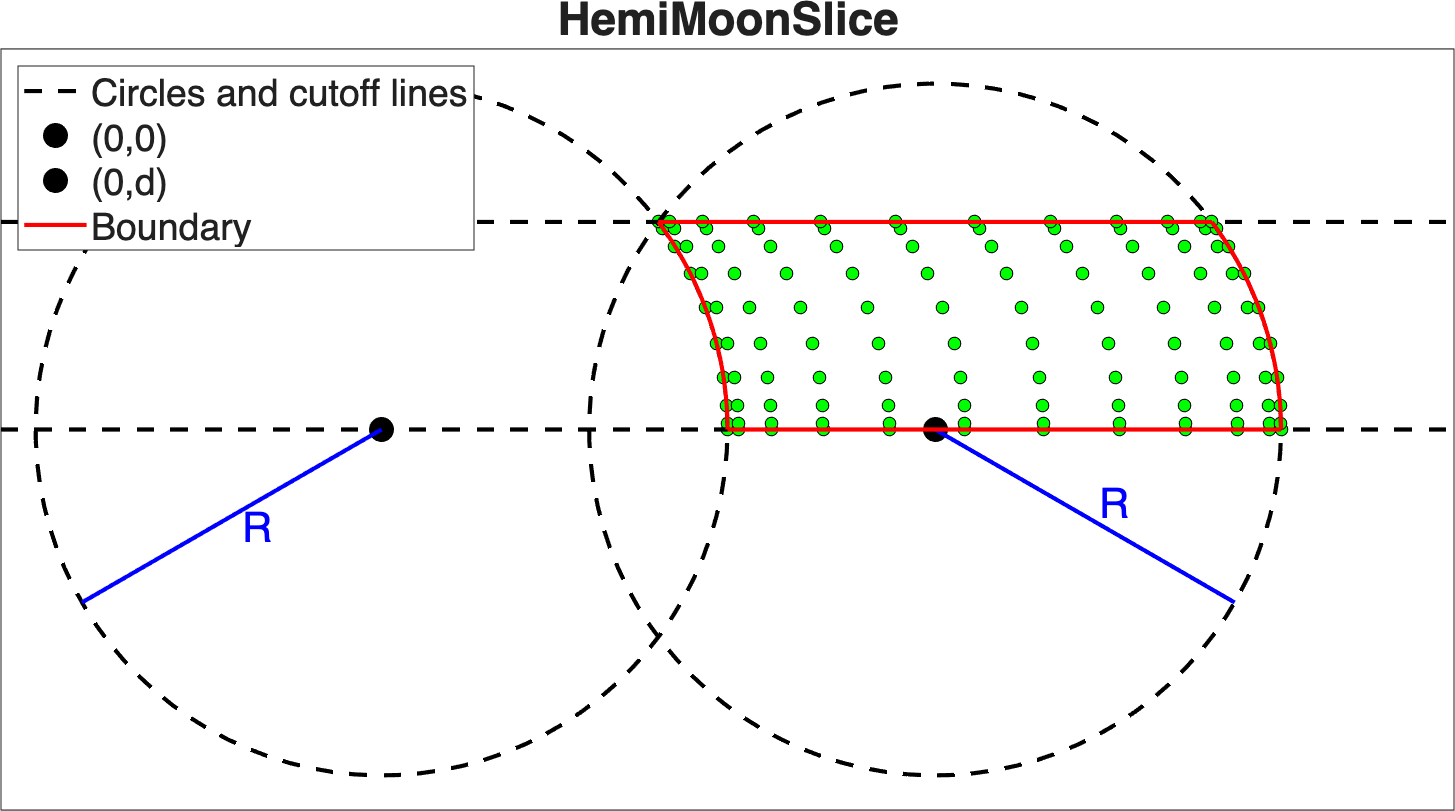}{-0.8\height} &
		Part of a disk of radius R, centred at (d,0), without the intersection with a disk of radius R, centred at (0,0), cut off between the horizontal line through the upper intersection of the circles and the $y_1$ axis.
		\newline
		$\texttt{y}_1$: Linear (N1, $f_{HM1}$(d, R, y2)) \newline
		$\texttt{y}_2$: Linear (N2, $f_{HM2}$(d, R)) \newline
		\\
		\\
		\midrule
		
	\end{tabular}
\end{table}
\end{document}